# Structural Evolution during Reversible Halogen Intercalation into $WTe_2$: Commensurate–Incommensurate $WTe_2I$ and Multistage $WTe_2Br_x$ ($x$ = 0.5, 1.0 and 1.25)

Patrick Schmidt,[a] Carl P. Romao[b] and Hans-Jürgen Meyer*[a]

Halogen intercalation into the layered material tungsten ditelluride ($WTe_2$) provides a unique pathway to tune its structural and electronic properties. In this study, we detail the synthesis and characterization of the new bromine-intercalated phases $WTe_2Br_x$ ($x$ = 0.5, 1.0, and 1.25), and reinvestigate the iodine-intercalated analogue, $WTe_2I$. A defining feature of the bromine system is its rapid and re-versible "breathing" behavior at room temperature, allowing guest molecules to be absorbed or released from the van der Waals gaps under ambient conditions. Structural analysis shows that the bromine-poor phase $WTe_2Br_{0.5}$ crystallizes in the orthorhombic space group *Pmmn*, thereby maintaining a uniform stacking sequence. In contrast, the bromine-rich $WTe_2Br_{1.25}$ phase (space group *Imm*2) adopts an architecture where two distinct types of bromine layers alternate between the host layers. For the iodine system, the compound $WTe_2I$ exhibits both incommensurate and commensurate (3+1)D modulated variants in the superspace group $P2_1/m(\alpha 0\gamma)00$. In the commensurate polytype, the structural modulation locks into a rational vector, $\boldsymbol{q}$ = (1/2, 0, 1/6), which can be described also as a 3D supercell. Electronic structure calculations show $WTe_2Br_{0.5}$ and commensurately modulated $WTe_2I$ to be metals with flat bands at the Fermi energy arising from the intercalation. These findings demonstrate the unusual stability and structural flexibility of anionic intercalation in a transition metal dichal-cogenides.

## Introduction

Layered transition metal dichalcogenides (TMDCs) have attracted intense interest for their unique electronic, magnetic, and topological properties.[1-3] In these materials, strong in-plane covalent bonding and weak interlayer van der Waals forces give rise to a two-dimensional architecture that is highly responsive to external perturbations, such as intercalation or the application of pressure.[2, 4] This open layered structure enables a rich intercalation chemistry in which various guest species can be incorporated. Typically, cations, neutral molecules or organic molecules can be inserted between the layers, thereby modifying the host's structure and properties.[5-16]

$WTe_2$ is a particularly intriguing member of the TMDC family and crystallizes in an orthorhombic $T_d$ phase, where it exhibits semimetallic behavior,[17-19] exceptionally large, non-saturating magnetoresistance,[20, 21] and complex topological properties—ranging from a 2D topological insulator in its monolayer form to a type-II Weyl semimetal in the bulk.[22, 23] It also displays superconductivity at low temperatures[24] and shows a charge density wave (CDW) state[25-32], underscoring its rich and exotic physical properties. The crystal structure (space group $Pmn2_1$) features distorted octahedral coordination of tungsten atoms by tellurium atoms, with zigzag W–W chains running within the *ab*-plane along the *a*-axis and van der Waals gaps separating the $WTe_2$ layers.[33] These structural features render $WTe_2$ highly receptive to intercalation, offering an opportunity to tune its electronic ground state through the controlled insertion of guest species.

Cationic intercalation into the structure of $WTe_2$ has been reported for $A_{0.5}WTe_2$ compounds ($A$ = K, Rb, Cs), which we have previously characterized structurally and electronically, showing that alkali ions act as electron donors.[34] While cationic intercalation into TMDCs is well established, involving alkali metals,[5, 8-11, 13, 35, 36] organometallic species, or molecular donors,[12, 14-16] anionic intercalation remains extremely rare. Halogen molecules ($I_2$, $Br_2$, $Cl_2$), despite their strong oxidizing nature, have largely resisted clean insertion into TMDCs in a structurally defined and stoichiometric manner. Instead, halogens have often been introduced during crystal growth as transport agents and later found in low concentrations or as surface dopants. For example, $Cl_2$ and $I_2$ have been shown to form bound excitonic centers in $MoS_2$, $WS_2$, and $MoSe_2$, being responsible for characteristic luminescence features at low temperatures.[37-48] However, these guest species reside in dilute, disordered environments, without forming ordered intercalation phases or significantly expanding the host lattice. As such, these systems are better classified as halogen-doped or halogen-decorated TMDCs—rather than true intercalation compounds.

[a] Section of Solid State and Theoretical Inorganic Chemistry, Institute of Inorganic Chemistry, Eberhard Karls University Tübingen, Auf der Morgenstelle 18,72076 Tübingen, Germany.

[b] Department of Materials, Faculty of Nuclear Sciences and Physical Engineering, Czech Technical University in Prague, Trojanova 13, Prague 120 00, Czech Republic.

Recently, we have demonstrated that $WTe_2I$ is the first and so far only example of a structurally confirmed anionic halogen-intercalated TMDC.[49, 50] In this compound, planar monolayers of iodine are inserted between the $WTe_2$ layers of the orthorhombic host, expanding the *c*-axis by ~56% and forming a fully stoichiometric intercalate. Importantly, the intercalation process was shown to be topotactically reversible: heating $WTe_2I$ above 100 °C under inert conditions fully deintercalated iodine and regenerated pristine $WTe_2$. These results established $WTe_2I$ as a prototypical system for reversible anionic intercalation in layered semimetals. The crystal structure of $WTe_2I$ was refined an averaged three-dimensional structure from powder X-ray diffraction (PXRD) data, which revealed notably large anisotropic displacement parameters (ADPs) for iodine, already suggesting a positional modulation.

Subsequent single-crystal X-ray diffraction studies indicated the presence of an incommensurately modulated structure. Density functional theory (DFT) calculations supported these findings, indicating significant charge transfer from $WTe_2$ to iodine, a shift of the Fermi level, and soft phonon modes characteristic of structural instability. In this work we present an incommensurate and a commensurate structure, appearing simultaneously for $WTe_2I$, and for better realization we report a supercell of the structure.

Despite many reports of halogen-related doping no other halogen-intercalated TMDC compound has been structurally verified to date. The present work expands this chemistry by demonstrating that bromine, though more volatile and reactive than iodine, can also be intercalated into $WTe_2$ in well-defined stages. We describe the synthesis and reversible intercalation of bromine into $WTe_2$, leading to the formation of stable $WTe_2Br_x$ compounds with distinct structural and electronic signatures, representing only the second known example of an anionic halogen intercalation phase in the TMDC family. Time-resolved powder X-ray diffraction (PXRD) reveals that bromine can be reversibly inserted and removed from the van der Waals gaps, with distinct diffraction patterns for each stoichiometry ($x$ = 0.5, 1.0, and 1.25), highlighting a remarkable "breathing" behavior that enables dynamic control of composition and structure under mild conditions.

## Results and discussion

### Incommensurate and commensurate modulations of $WTe_2I$

As reported previously,[49] $WTe_2I$ forms when $WTe_2$ is reacted with iodine at 40–200 °C. Elemental analyses confirm a stoichiometry corresponding to one iodine per $WTe_2$ ($WTe_2I$) within experimental uncertainty (see Experimental Section). No phase transition of $WTe_2I$ single crystals was detected between 100 K and 270 K.

Single-crystal X-ray diffraction studies, performed on several crystal specimens at 150 K, revealed two closely related (3+1)D modulated structure variants in the superspace group $P2_1/m(\alpha 0\gamma)00$ with almost identical basic lattice parameters (Table 1), in good agreement with the previously reported unit cell.[49] The incommensurate variant is described by the modulation vector $\boldsymbol{q}$ = (0.4487, 0, 0.1617), whereas the commensurate polytype has $\boldsymbol{q}$ = (1/2, 0, 1/6) and can equivalently be described as a 3D supercell. Using the transformation matrix (1 0 -3, 0 1 0, 1 0 3), a primitive $P2_1/m$ supercell with $a$ = 12.1694(1) Å, $b$ = 21.8799(2) Å, $c$ = 12.1726(1) Å and $\beta$ = 117.635(1)° is obtained. For ease of comparison and broader reusability, the commensurate structure was also refined as an explicit 3D superstructure in SHELXL; the resulting model is crystallographically equivalent and is provided as a reference in the deposited data (Table 1). All modulated crystals show pronounced pseudo-merohedral twinning, where the main reflections overlap, while the satellites are well separated in the incommensurate case and partially overlap for the commensurate polytype. The twin law corresponds either to a twofold rotation about *c*-axis or an equivalent mirror perpendicular to it; owing to the metric pseudosymmetry ($\beta \approx 90°$), both descriptions are indistinguishable, and a conventional twofold twin matrix was applied, with the twin fraction of the second domain refined to 0.432(2). Depending on the batch, crystals were found to be fully commensurate or incommensurate (with slightly different incommensurate $\boldsymbol{q}$-vectors), and some specimens contained mixed domains, combining commensurate and incommensurate regions, or even multiple distinct incommensurate modulations alongside the commensurate phase (Figure S1). Although satellites are visible up to $m$ = ±4, refinements were restricted to $m$ = ±2 and carried out with the minimal, physically meaningful set of modulation waves needed to reproduce the observed atomic displacements, thereby avoiding over-parameterization. Refinement statistics are summarized in Table 1.

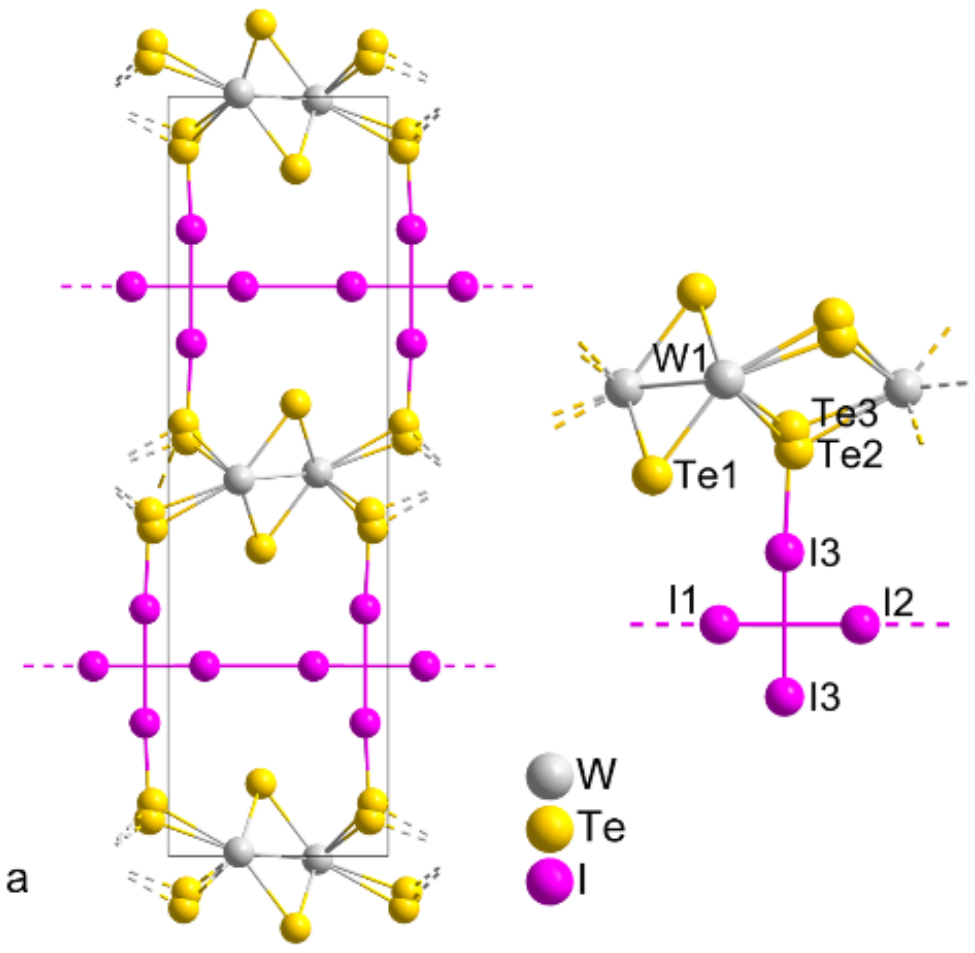

Figure 1. Section of the average incommensurate $WTe_2I$ structure (left) and a smaller section with atom labels (right). Note, the zig-zag chains of tungsten atoms running parallel to the *c*-axis.

The average 3D structure is essentially identical to the previously reported model of $WTe_2I$ (see Figure 1), but requires two additional split sites, Te3 and I3. These minor positions are ~15.5% occupied, while the corresponding main sites (Te2, I1, and I2) are under-occupied accordingly, consistent with a local defect motif rather than a distinct bulk phase. Structurally, the splitting is initiated by a local ~90° rotation of a $I_2$ unit (denoted as I3) out of the iodine-net plane. This reorientation necessitates a concomitant shift of the adjacent tellurium atom. This reorientation is accommodated by a concomitant shift of the adjacent tellurium atom (Te2→Te3); without this shift, the resulting Te2-I3 contact would be unrealistically small and physically prohibitive. By adopting the Te3 position, the system instead establishes a Te3-I3 distance of ~2.77 Å. This value is characteristic of a covalent bond (sum of covalent radii ~2.71 Å), suggesting that the local defect motif is stabilized by a degree of covalent bonding between the host layer and the guest. This localized covalent interaction likely acts as an 'anchor' that pins the iodine guest to the $WTe_2$ framework, thereby stabilizing the defect motif and dictating the specific periodicity of the long-range structural modulation.

Table 1: Crystallographic data and details of the crystal structure refinement of the different $WTe_2I$ modifications. Refinements were performed for (in)commensurately modulated crystal structures and the corresponding supercell. Detailed information on the modulation functions is given in the crystal information file. All crystals were measured with Mo-$K_\alpha$ radiation at 150 K.

| Modification | $WTe_2I$ | | |
|---|---|---|---|
| | incom- | commensurate | supercell |
| CCDC | 2520418 | 2519690 | 2519736 |
| Formula | $WTe_2I$ | | $W_6Te_{12}I_6$ |
| $M$ / g·mol$^{-1}$ | 565.95 | | 3395.7 |
| Crystal system | monoclinic | | |
| (Super)space group | $P2_1/m(\alpha 0\gamma)00$ | | $P2_1/m$ |
| Modulation wave vector $\boldsymbol{q}$ | (0.4487 0 0.1617) | (1/2 0 1/6) | — |
| $m$ max. | ≤ 2 | ≤ 2 | — |
| $t_0$ | 0 | 0 | — |
| $a$ / Å | 6.3200(2) | 6.30257(5) | 12.1694(1) |
| $b$ / Å | 21.8060(6) | 21.8792(3) | 21.87990(2) |
| $c$ / Å | 3.4713(2) | 3.47075(3) | 12.1726(1) |
| $\beta$ / ° | 90.057(3) | 90.0268(7) | 117.635(1) |
| $V$ | 478.40(3) | 478.600(9) | 2871.38(6) |
| $Z$ | 4 | 4 | 4 |
| $\theta$-range / ° | 1.97-38.01 | 1.89- 45.01 | 1.888-30.508 |
| unique reflections/ parameters | 7009 / 193 | 7023 / 152 | 8947 / 294 |
| twvol2 or BASF | 0.432(2) | 0.394(4) | 0.378(2) |
| $R_1$/$wR_2$/GooF (all data) | 4.86 / 7.47 / 1.10 | 6.99 / 14.37 / 1.08 | 0.0433 / 0.0999 / 1.022 |
| $R_1$/$wR_2$: (all main) | 1.65 / 3.80 | 5.18 / 13.33 | — |
| (all 1st order) | 6.92 / | 8.54 / 17.19 | — |
| (all 2nd order) | 12.09 / | 9.13 / 19.37 | — |
| max./min. $\Delta\rho$/e$^-$·10$^{-6}$ pm$^3$ | -1.02 / 1.25 | -1.96 / 1.75 | -4.081 / 4.297 |

Since the commensurate and incommensurate refinements are nearly identical apart from the modulation vector, we focus here on the incommensurate model. The modulation is dominated by the iodine sublattice, where iodine atoms show a pronounced displacive modulation within the planar net (see Figure 2, Figure 4 and Figure 3 right), whereas the $WTe_2$ layers exhibit only small, coupled distortions. Within the iodine layer, the displacive modulation translates into a pronounced spread of I-I distances. Over one modulation period, the shortest and longest distances within the I1/I2 planar net span 2.688(8)–4.018(8) Å (see Figure S2) and the I3-I3 distance remains comparatively long at 3.272(2) Å, highlighting the strongly distorted and non-uniform iodine net. A representative section of the modulated structure at $t$ = 0 is shown in Figure 2.

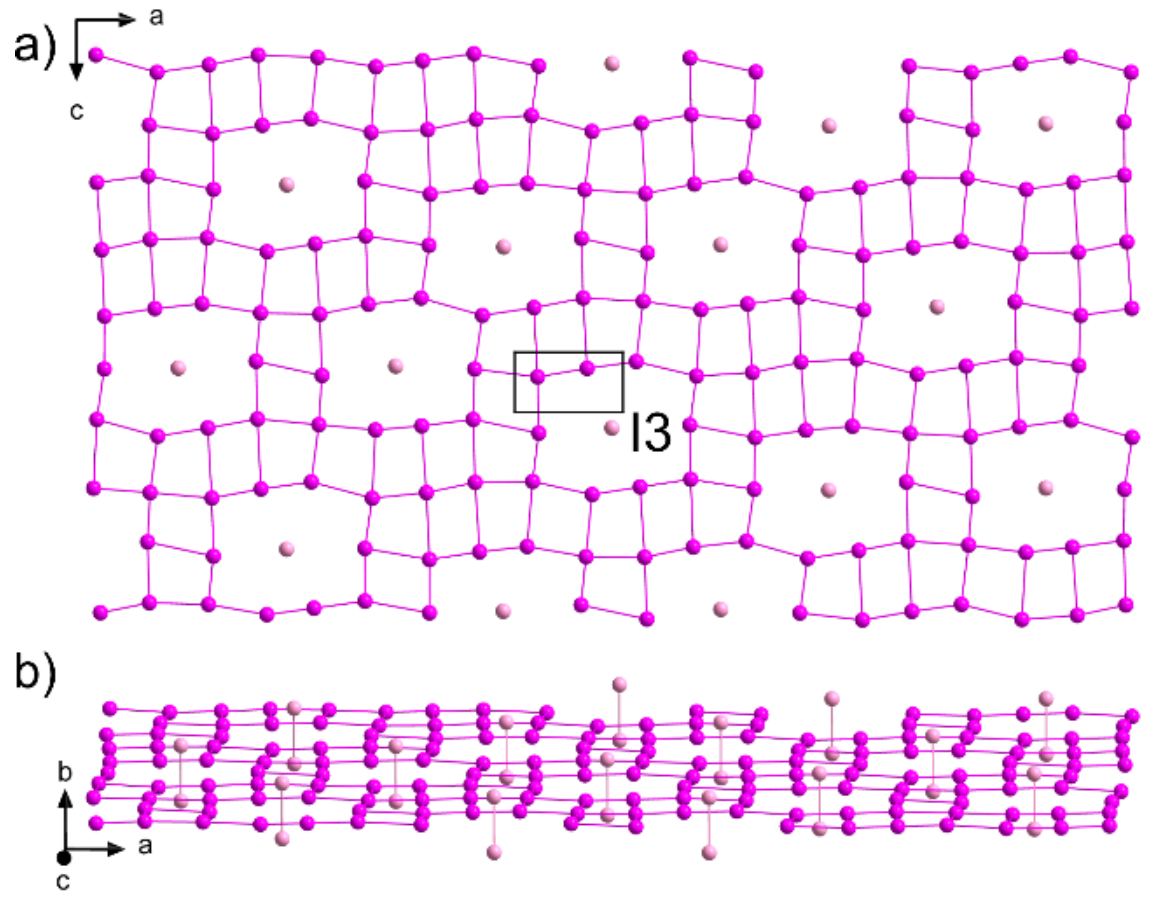

Figure 2. Section of the incommensurate $WTe_2I$ structure a) only showing one iodine layer with $I_2$ dumbbells of I3 being aligned along the $b$-axis, including the projected unit cell; and b) a corresponding view along the $c$-axis, emphasizing the $I_2$ dumbbells of I3. I1 and I2 are displayed purple, and I3 in lighter purple.

The structure solution and refinement indicate that the displacement modulation of the iodine net is most pronounced in the $ac$-plane. Superimposed on this, an occupational modulation is required and was modeled using crenel functions (harmonics, orthogonalized to crenel interval) for the Te2/I2 and Te3/I3 split positions. The crenel intervals are Δ(Te2/I2) = 0.831824 and Δ(Te3/I3) = 0.168176, consistent with the intermittent appearance of the I3 site, which corresponds to a local ~90° out-of-plane reorientation of an $I_2$ unit out of the $ac$-plane that disrupts the wave-like iodine layer. The crenel functions were refined using constraints linking their origins and Δ intervals (Δ(Te2/I1/I2) + Δ(Te3/I3) = 1), ensuring a consistent occupational modulation between the split-site pairs.

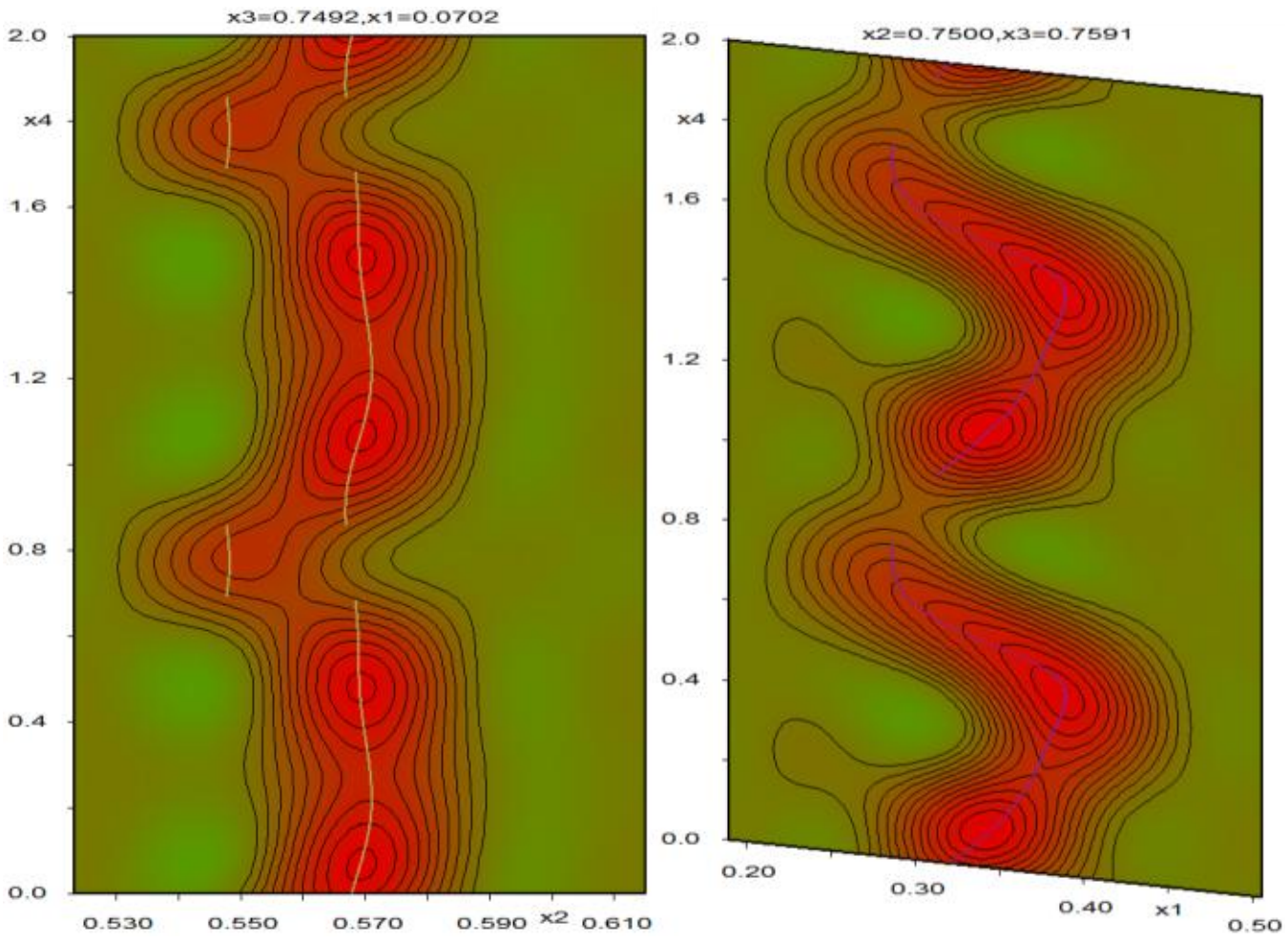

Figure 3. de Wolff sections (Fobs) of $x2$-$x4$ plane for Te2 and Te3 (left) and in the $x1$-$x4$ plane for I1 (right). Refined atomic positions are indicated by lines (Te: yellow; I: purple); the Te2 and Te3 sections are highlighted at $x2$ = 0.57 and $x2$ = 0.55, respectively. The electron density is visualized as a heat map (red = high, green = low).

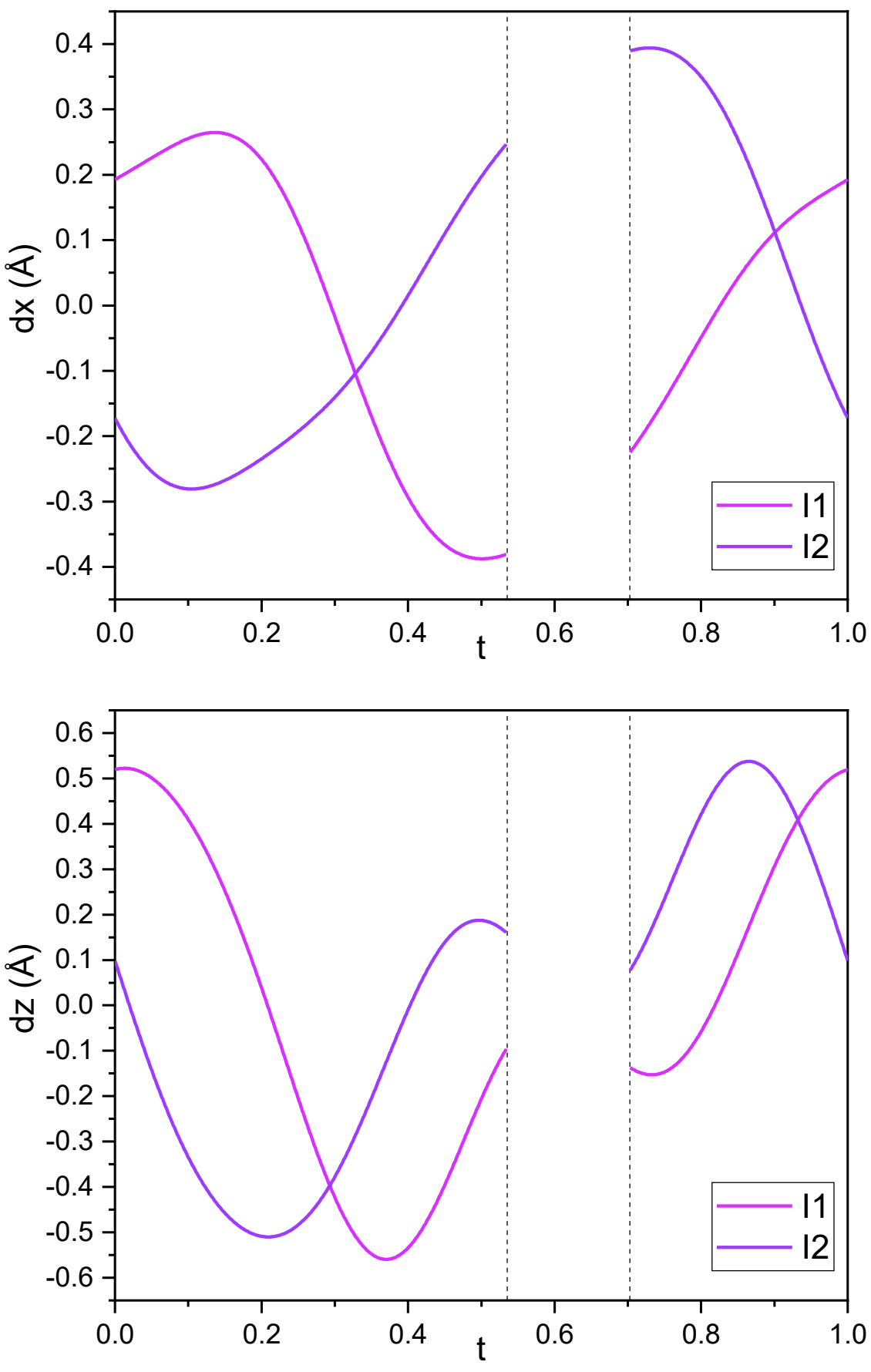

Figure 4. The modulation functions of the deviations from the average positions for I2 and I3 along *x* and *z* directions as a function of internal phase coordinate *t* with crenel limits marked as dashed vertical line.

To accommodate this local motif, Te2 splits into Te2/Te3 and induces subtle but observable distortions in the $WTe_2$ layer. The corresponding de Wolff sections (Fobs) through the (3+1)D Fourier map for Te2/Te3 and I2 is shown in Figure 3.

Although the host-layer modulations are comparatively small, they are clearly resolved. In particular, the position modulation of tungsten atom shows a characteristic saw-tooth modulation in the *x*3–*x*4 section (see Figure 6, Figure 7), resulting in an average W–W distance of 2.86(1) Å. The shortest W–W contact with 2.78(1) Å remains closer to the average, whereas the longest distance with 3.10(2) Å periodically disrupts the chain after segments of roughly ~12 tungsten atoms. This disruption is induced by the I3 atom, which corresponds to the out of plane $I_2$ dumbbell of the iodine layer above and below the projected ac-plane of W chains, shown in Figure 5 and Figure S3.

Apart from these interruptions, the W–W distances vary only modestly with ~0.08 Å (see Figure 7, grey lines). The remaining tellurium sites exhibit similarly small displacements, with amplitudes of approximately $d_{xyz}$(Te1) = ±0.1 Å and $d_{xyz}$(Te2) = ±0.06 Å. Further refinement details, including constraints, parameter correlations, and the complete list of refined variables, are provided in the deposited CIF and in the Supporting Information.

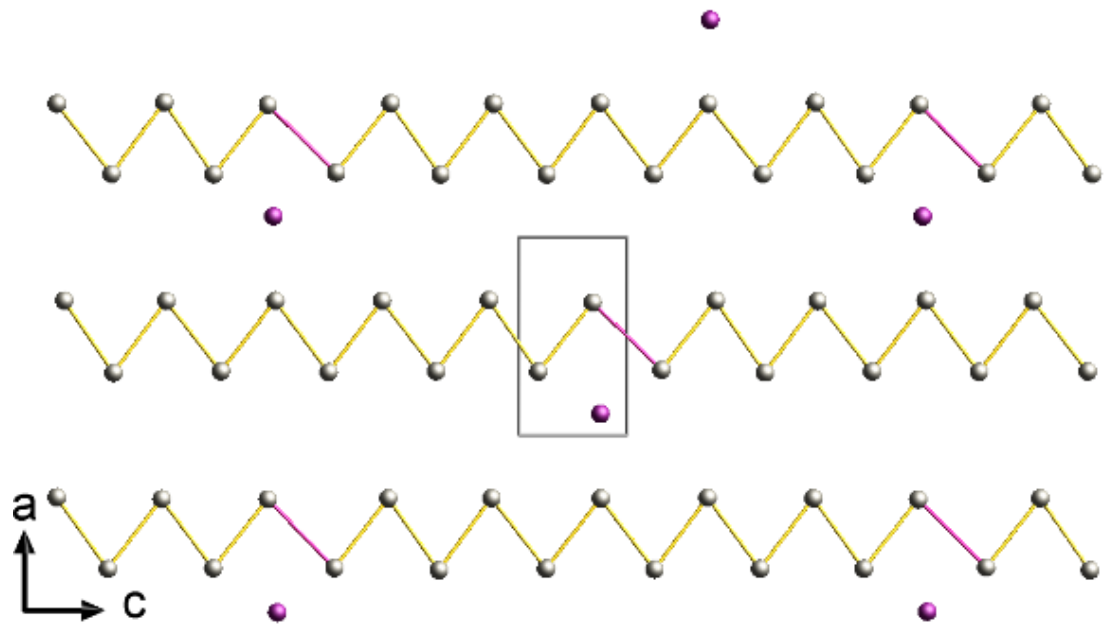

Figure 5. Section of the incommensurate $WTe_2I$ structure viewed along the *b* direction with highlighted unit cell. For clarity, tellurium atoms are omitted and only one interlayer section in the *ac*-plane of the tungsten zigzag chains is shown, together with the $I_2$ dumbbells (I3 sites) aligned along *b* which are directly above and beyond of the *ac*-plane section. Elongated W–W contacts, indicating disruption of the tungsten chain are highlighted by violet bonds (W: grey; I: violet; t = 0.62).

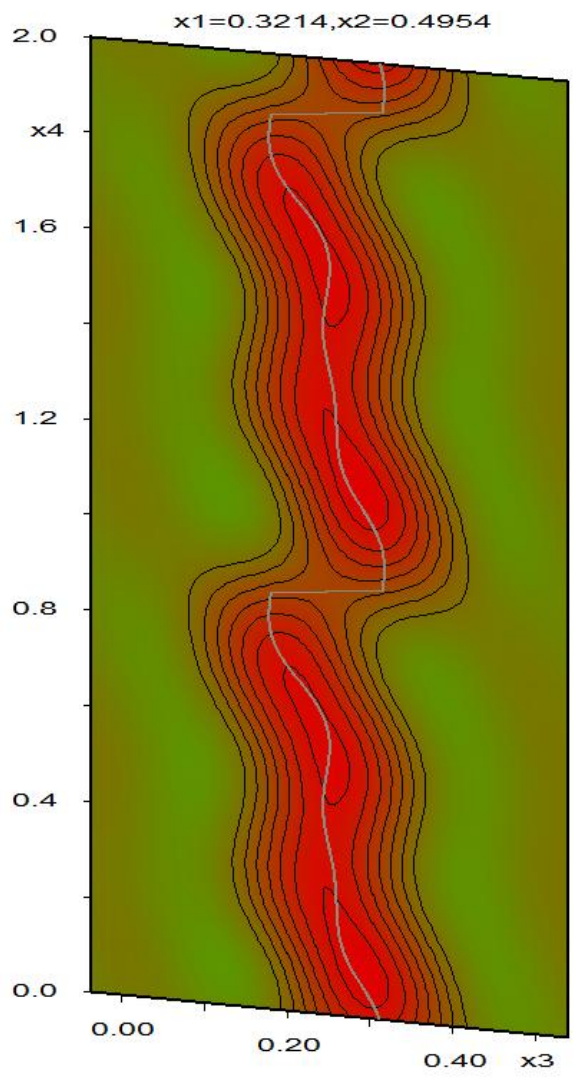

Figure 6. de Wolff sections (Fobs) of x3-x4 plane for W1 in $WTe_2I$. Refined atomic positions are indicated by a grey line. The electron density is visualized as a heat map (red = high, green = low).

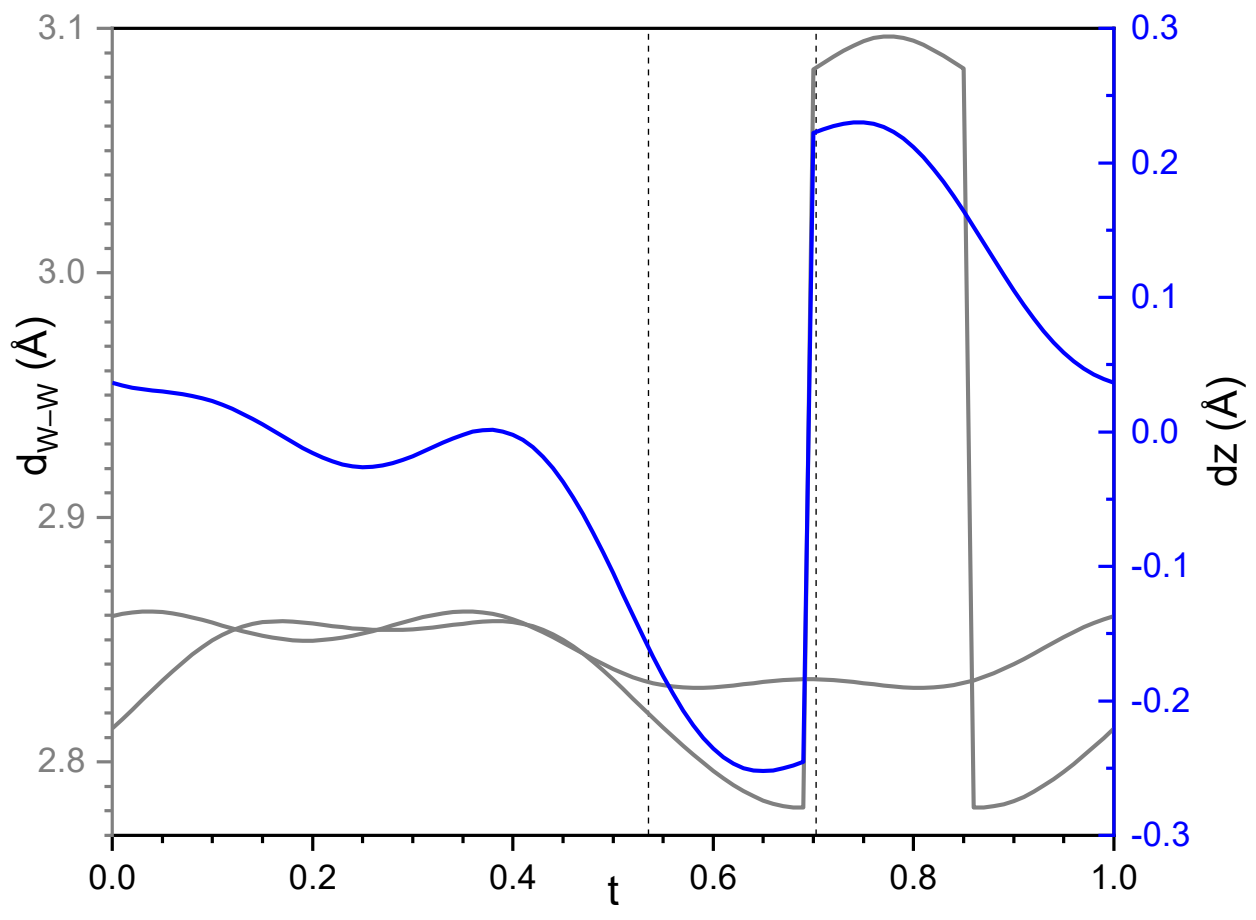

Figure 7. *t* plots of the W-W distances (grey) and positional deviation (blue) of tungsten along *x*3 direction in $WTe_2I$ with crenel limits marked as dashed vertical line.

**Commensurate structure and superstructure model of $WTe_2I$**

In addition to the incommensurate variant, a commensurate polytype is occasionally obtained in which the modulation "locks in" to the rational vector $\boldsymbol{q}$ = (1/2, 0, 1/6) and can therefore be described as a conventional three-dimensional $P2_1/m$ superstructure. Using the transformation matrix (1 0 −3, 0 1 0, 1 0 3) yields a primitive supercell ($a$ ≈ 12.17 Å, $b$ ≈ 21.88 Å, $c$ ≈ 12.17 Å, $\beta$ ≈ 117.64°; $W_6Te_{12}I_6$), which makes the modulation periodicity explicit in real space. In this commensurate description, the iodine net adopts a regular repeating motif that mirrors the rational components of $\boldsymbol{q}$, where along $a$ direction, the rotated defect motif alternates such that every second $I_2$ unit is rotated by ~90° out of the iodine-net plane and appears as the I3 site, while along the $c$ direction the same motif repeats with a six-unit periodicity (Figure 8). The commensurate structure reproduces essentially the same I–I distance distribution, but arranges the shorter and longer contacts into an exactly repeating pattern.

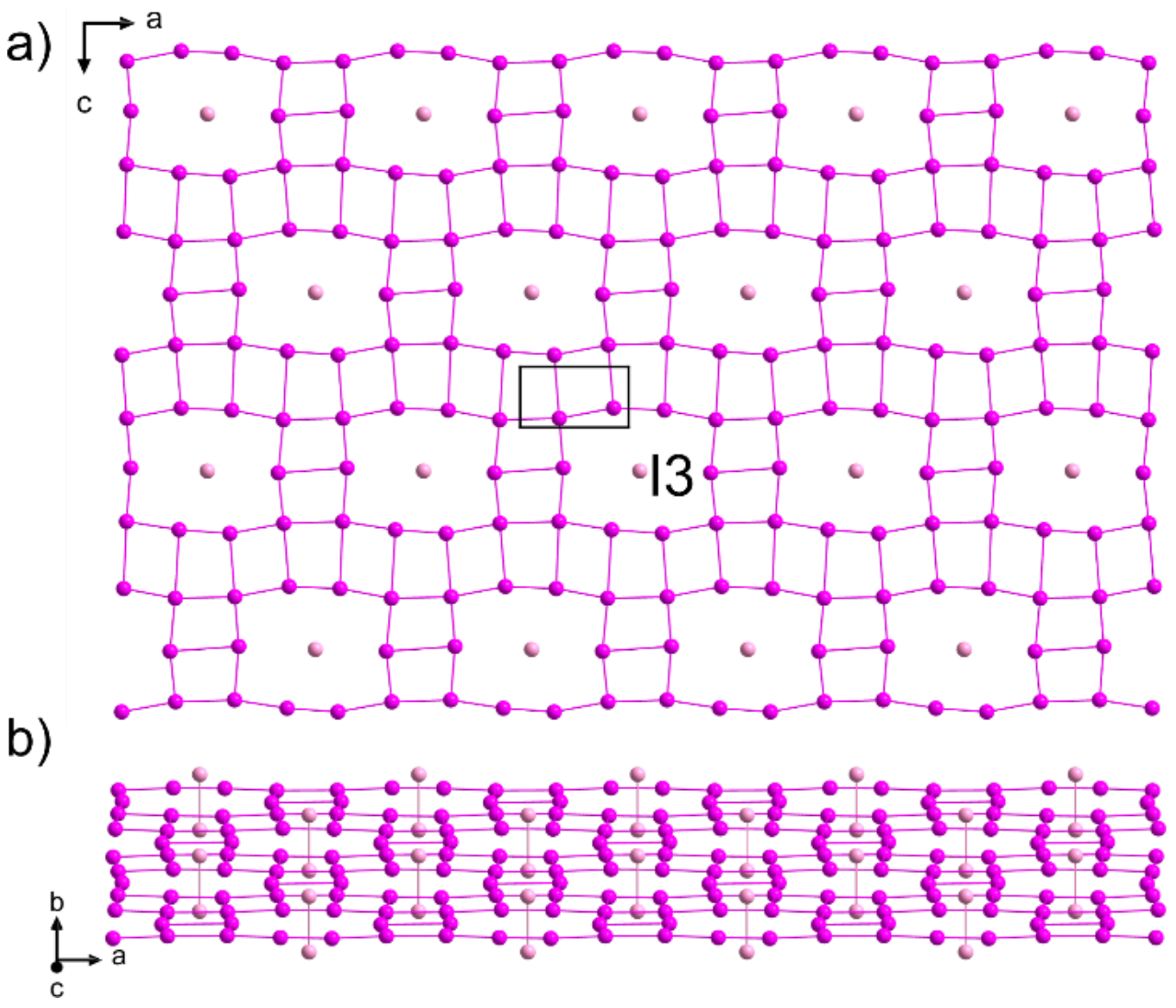

Figure 8. Section of the commensurate $WTe_2I$ structure only showing a) one iodine layer with highlighted I3 units aligned along the $b$-axis with unit cell shown, and b) a corresponding view along the $c$-axis, emphasizing the $I_2$ dumbbells (I1/I2: purple; I3: lighter purple).

Refinement in the commensurate setting is advantageous over a purely 3D average-cell description because the modulation is represented explicitly rather than being "smeared out", typically improving agreement factors and yielding cleaner residual density. Overall, the commensurate model can be viewed as a periodic approximant of the incommensurate structure that facilitates direct real-space visualization of both the iodine-net displacement wave and the ordered occurrence of the I3 defect motif. For ease of comparison and broader reusability, the commensurate structure was also refined as an explicit 3D superstructure in SHELXL; the resulting model is crystallographically equivalent and is provided as a reference in the deposited data (Table 1). Because the diffraction data are affected by twinning and residual modulation, the supercell refinement was improved by modeling the interlayer iodine net as a two-component disordered layer (atom sites Ixa/Ixb; $x$ = 2-11), with coupled occupancies constrained to unity and the b component accounting for ~20% occupancy. The minor b component represents a slightly b-rotated relative alignment of the iodine layer, whereas the out-of-plane iodine positions remain unaffected (see Figure S4).

**Synthesis and "breathing" behavior of $WTe_2Br_x$**

$WTe_2Br_x$ powder phases were prepared by Schlenk techniques by exposing orthorhombic $WTe_2$ to an excess of bromine liquid between 0 and 30 °C. Single crystals were intercalated by bromine vapor at room temperature for 4 days or with liquid bromine at 6 °C for 2 days. Although crystallinity is generally compromised by the multiple phase transitions during bromine uptake and release, the vapor intercalation route consistently afforded crystals of higher quality than direct contact with liquid bromine. Thermal stability is a critical constraint in this system. While intercalation proceeds smoothly even near room temperature, heating $WTe_2$ with excess bromine in a closed vessel to 60 °C leads to decomposition of the host structure and formation of binary halides, for example $TeBr_4$ and $WBr_6$.

A striking feature of this system is a fast and reversible "breathing" behavior, present at room temperature, which involves an uptake and release of guest bromine. The bromine content can be increased up to the maximum stage with $x$ = 1.25 by supplying liquid bromine, and bromine can conversely be removed from the van der Waals gaps (at room temperature) under constant argon flow. The actual bromine content in the structure of $WTe_2Br_x$ is highly temperature sensitive during this de-intercalation step. Evaporation of excess bromine under a constant flow of argon yields $WTe_2Br_{1.25}$ when the temperature is maintained between 0 and 5 °C, whereas the comparatively stable $WTe_2Br_{0.5}$ phase can be obtained at around 25 °C. Starting from $WTe_2Br_{0.5}$, re-intercalation to the bromine-rich stage is particularly rapid and proceeds within minutes.

Elemental analysis by argentometric titration of $WTe_2Br_{0.5}$, and deintercalation of $Br_2$ saturated $WTe_2Br_x$ confirms the variable bromine content and supports three distinct composition regimes in $WTe_2Br_x$ at $x$ = 0.5, 1.0, and 1.25.

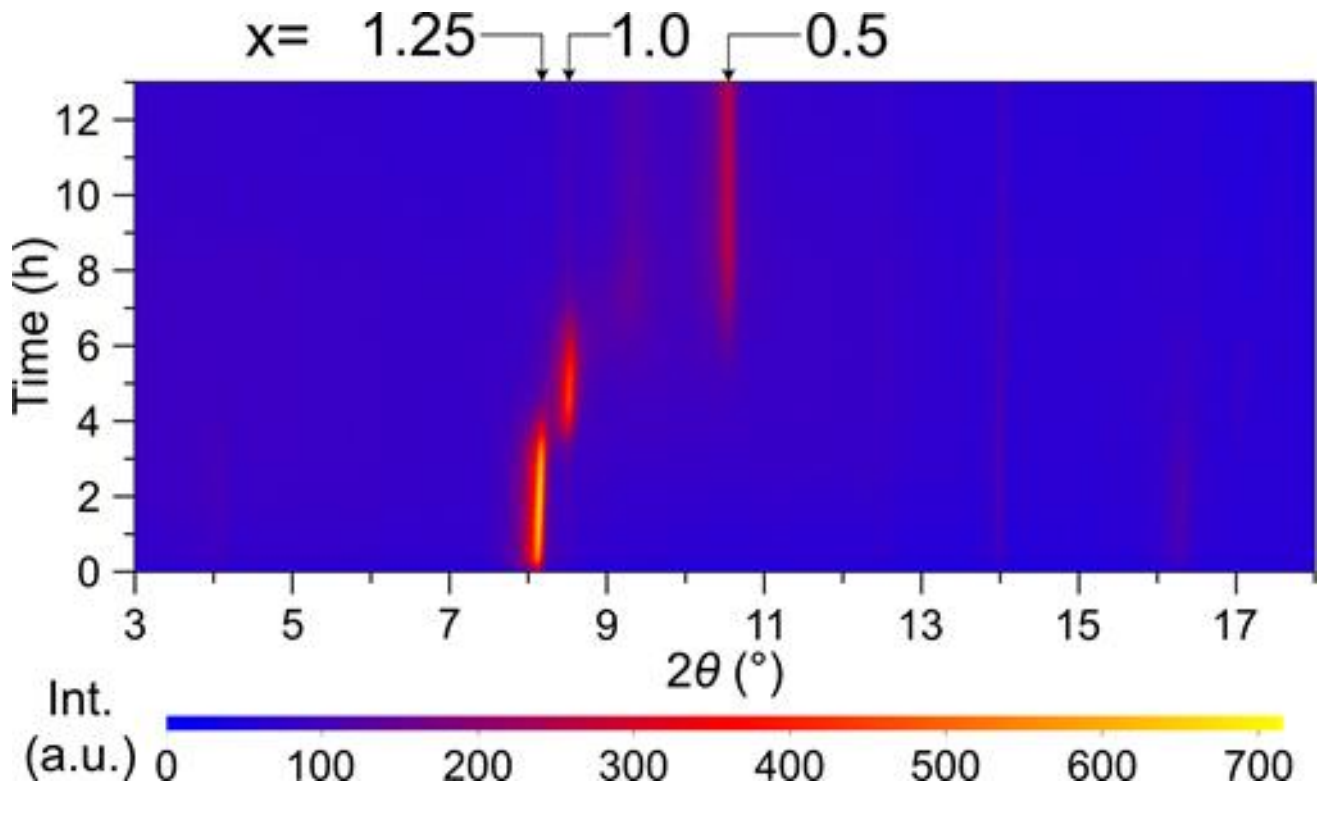

Figure 9. Time resolved PXRD analysis of $WTe_2Br_x$ showing structural evolution during bromine deintercalation with discrete phases from $x$ = 1.25 to 1.0 and 0.5 (for full 2θ range see Figure S5).

The dynamics of bromine exchange are directly captured by *in situ* PXRD studies. Starting from $WTe_2Br_{0.5}$, the bromine uptake is extremely rapid and occurs within minutes. The accompanying change in bromine content is directly reflected in the basal reflections that tracks the interlayer spacing. For the monoclinic phases, the corresponding layer-stacking reflection, indexed here as (020), shifts continuously toward lower $2\theta$ angles as the van der Waals gap expands to accommodate bromine. Upon deintercalation, time-resolved PXRD in Figure 9 reveals the reverse evolution through discrete stages, and the basal reflection shifts back toward higher $2\theta$ as bromine is released from the lattice ($2\theta$ = 8.21°, 8.52° and 10.56°).

This inherent reversibility also explains why $WTe_2Br_{1.0}$ is difficult to isolate and handle. Unless samples are stored under strictly controlled conditions, bromine readily diffuses in or out, and bromine contents of the sample near $x$ = 1.0 often relax toward $x$ = 0.5 or $x$ = 1.25. As a result, mixtures are frequently obtained in which $WTe_2Br_{1.0}$ coexists with $WTe_2Br_{0.5}$ or $WTe_2Br_{1.25}$. The $WTe_2Br_{1.0}$ phase also typically shows reduced crystallinity compared with the more stable end members, which we attribute to ongoing bromine exchange between the two stages.

Thermogravimetric analysis (TGA) of products obtained from the reaction of $WTe_2$ with excess bromine provides additional support for these composition regimes. During controlled deintercalation, distinct changes in the mass-loss rate mark transitions between the bromine-rich and bromine-poor stages. The TGA traces in Figure 10 show an initial rapid loss of excess bromine, followed by regimes of slower mass loss and a final, near-horizontal segment with only a subtle residual mass-loss rate. At 25 °C, the first clear change in slope corresponds to a bromine content consistent with $WTe_2Br_{1.25}$, whereas at 50 °C the first change in slope is shifted to the intermediate composition $WTe_2Br_{1.0}$. In both cases, the ensuing near-horizontal segment is consistent with the comparatively stable $WTe_2Br_{0.5}$ phase, which still shows a very slow residual bromine loss that becomes more pronounced at higher temperature. Accordingly, samples were stored cooled (≤ 0 °C) or kept under a sufficient bromine counter pressure at room temperature to suppress bromine evaporation. Even though phase-pure $WTe_2Br_{1.0}$ could not be isolated and structurally characterized, the established structure of $WTe_2Br_{1.25}$ shares key motifs with $WTe_2I$, which suggests that the intermediate $WTe_2Br_{1.0}$ phase may follow a closely related structural principle and has a similar structure.

Owing to the highly corrosive nature of bromine, specialized reaction and measurement setups were employed, as described in the Experimental Section and the Supporting Information.

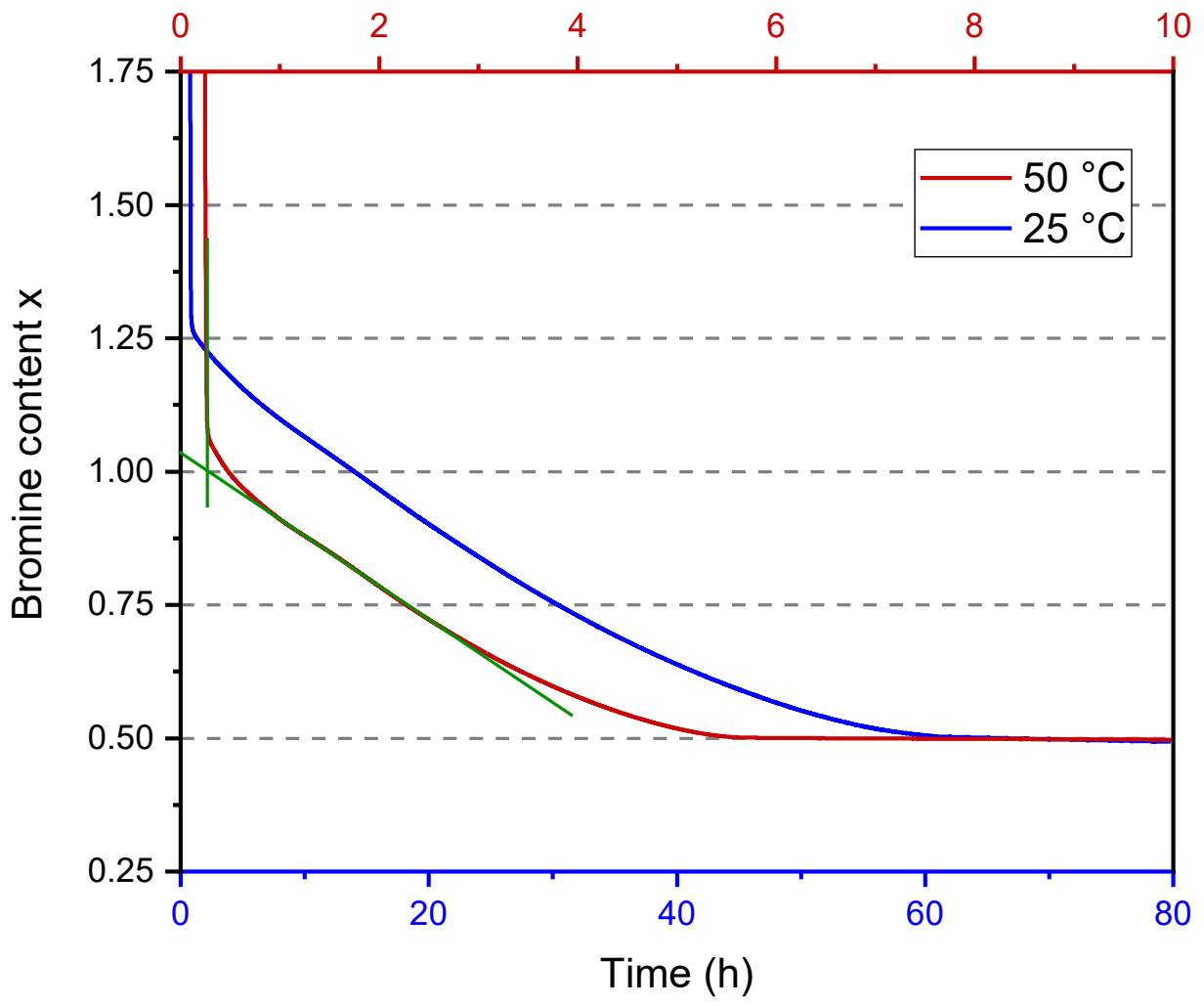

Figure 10. TGA showing the bromine loss of bromine saturated $WTe_2Br_x$ samples, showing steps at $x$ = 1.25, 1.0, and 0.5 at constant temperature of 25 °C (blue) and 50 °C (red), under constant argon flow of 320 ml·min$^{-1}$ (an onset fit for the 50 °C measurement is shown in green).

### Crystal structure of $WTe_2Br_{0.5}$

The bromine-intercalated phase $WTe_2Br_{0.5}$, representing the lowest bromine content in the $WTe_2Br_x$ series, crystallizes in the orthorhombic space group *Pmmn* with refined lattice parameters $a$ = 16.7450(4) Å, $b$ = 3.67734(3) Å, and $c$ = 6.29678(8) Å. As shown by the Rietveld plot in Figure 11, the powder diffraction pattern is well described by the refinement ($R_{Bragg}$ = 2.028%, $R_p$ = 5.438%, $R_{wp}$ = 7.104, $\chi^2$ = 1.268, Number of reflections/parameter 392/27), confirming a well-defined crystalline phase.

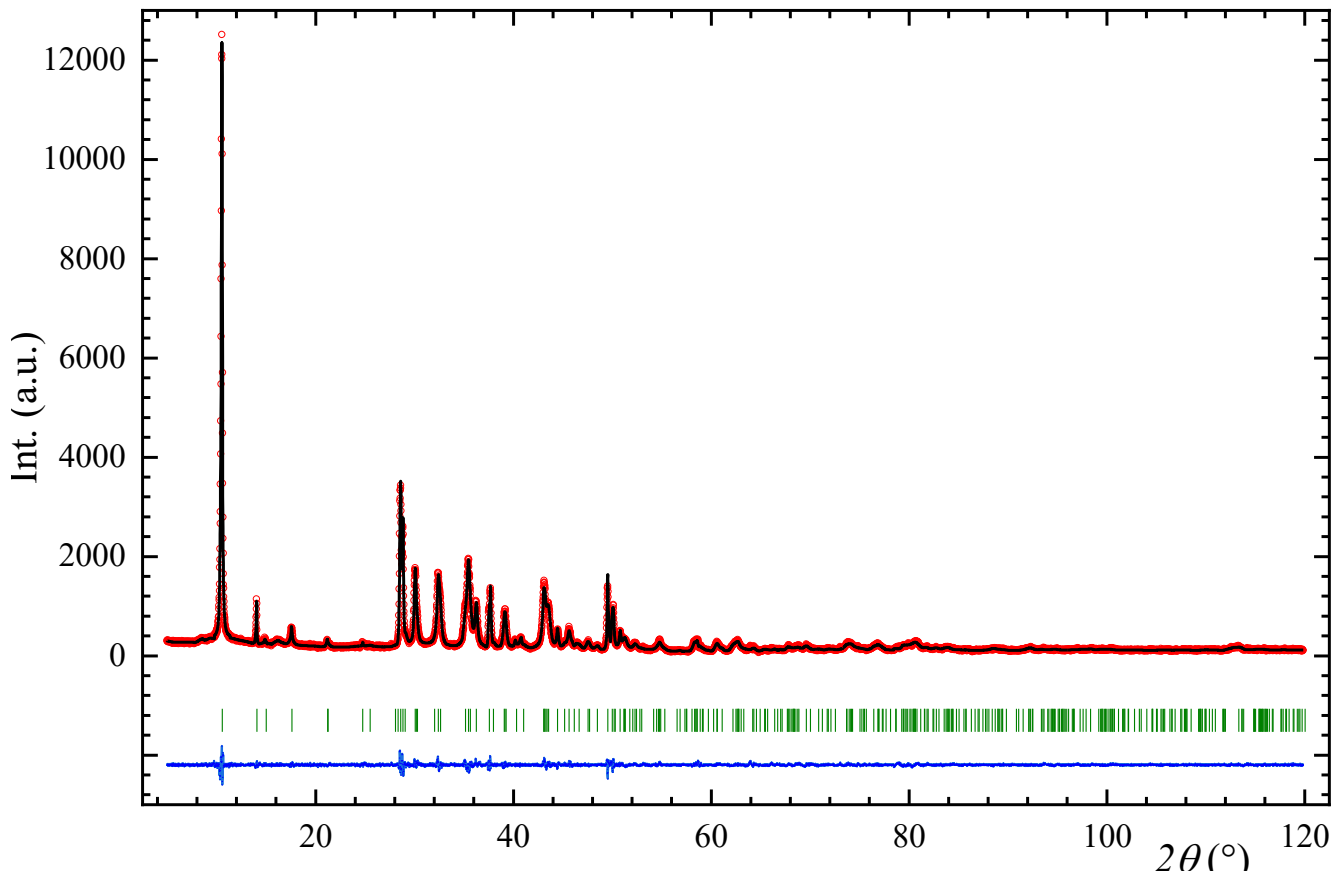

Figure 11. Rietveld PXRD structure refinement of $WTe_2Br_{0.5}$ with the space group *Pmmn* at 298 K with the experimental (red) and calculated (black) intensities. Bragg positions (green) and the difference curve (blue) are also shown.

In the chosen setting, the $a$-axis corresponds to the stacking direction. The $WTe_2$ layers adopt a stacking arrangement in which the tungsten zig-zag chains of neighboring layers are brought into eclipsed alignment along the $a$-direction, generating an enlarged interlayer void to accommodate bromine guest atoms.

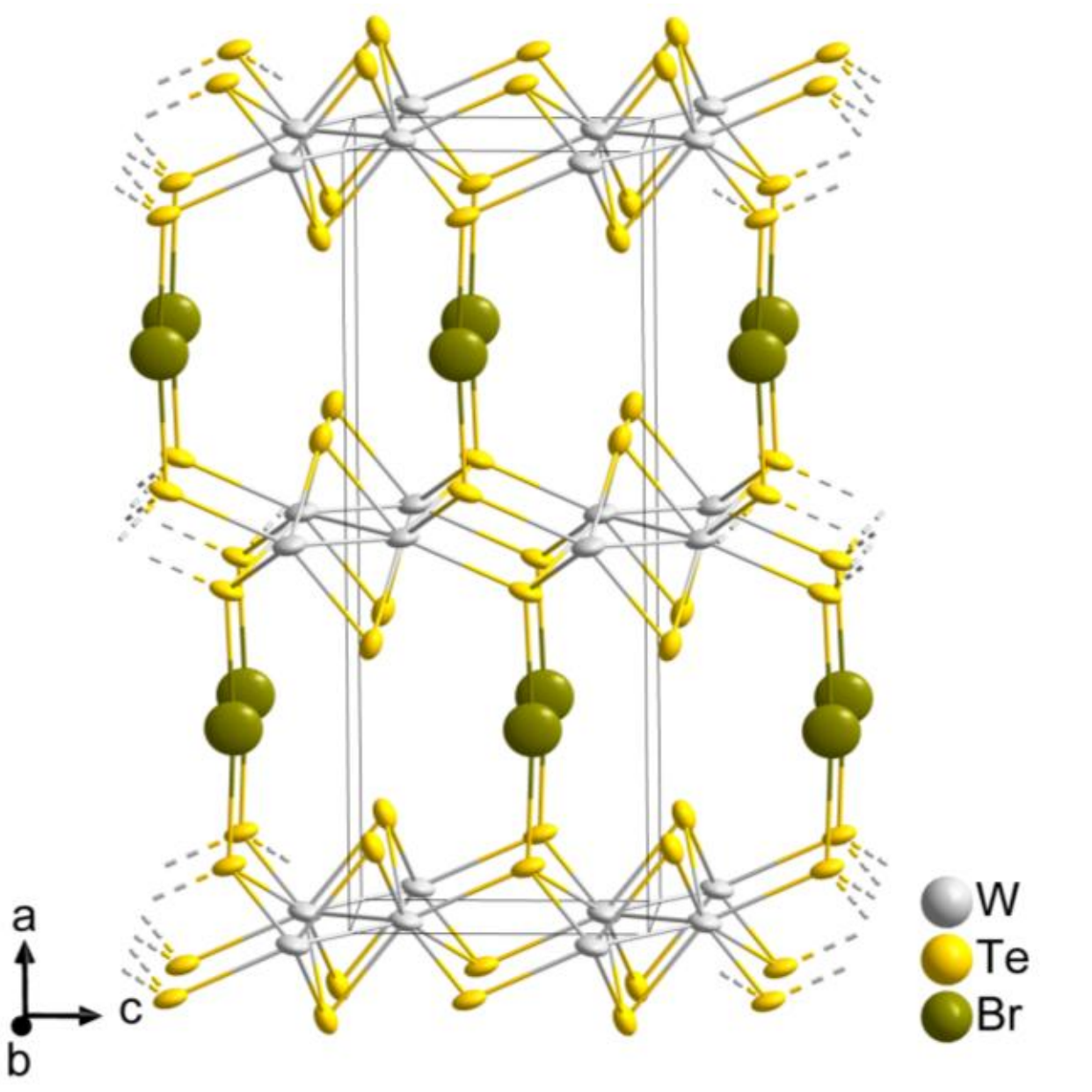

Figure 12. Section of the $WTe_2Br_{0.5}$ crystal structure highlighting the position of bromine atoms within the van der Waals gap, with their alignment toward Te1, thereby connecting adjacent $WTe_2$ layers. Atomic displacement parameters are shown as anisotropic displacement ellipsoids at the 50% probability level.

The adjacent $WTe_2$ layers are arranged in a mirror-like fashion across the bromine-containing interlayer plane. The average W-W separation of 2.909(2) Å is slightly expanded relative to pristine $WTe_2$. In contrast to $WTe_2I$, where a more continuous halogen layer separates adjacent $WTe_2$ layers, $WTe_2Br_{0.5}$ contains only half as many halogens and the bromine position is shifted such that bromine atoms align along the *a*-axis with the tungsten-chain-bridging tellurium atom Te1, giving a short Te1–Br1 contact of 2.946(2) Å that connects neighboring $WTe_2$ layers. When projected onto the *bc*-plane, the bromine positions define a rectangular net with Br···Br separations fixed by the lattice dimensions (3.6773(1) Å along *b* and 6.297(1) Å along the *c*-axis); these relatively long distances make direct Br···Br interactions inefficient.

A defining feature in the structure of $WTe_2Br_{0.5}$ is the exceptionally large anisotropic displacement of the bromine atom Br1, which occupies the Wyckoff 2b position with mm2 site symmetry. The bromine atom refines with $U_{iso}$ = 0.115(7) Å$^2$, and anisotropic atomic-displacement parameter (ADP) refinement reveals a pronounced directional character ($U_{11}$ = 0.145(9) Å$^2$, $U_{22}$ = 0.036(5) Å$^2$, $U_{33}$ = 0.163(7) Å$^2$) with the largest components along *a*- and *c*-direction and only a minor component along *b*. Attempts to improve the description by splitting the bromine position or refining the site occupancy did not yield a better model and were therefore not pursued further. These unusually large and anisotropic ADPs indicate that bromine is not sharply localized but exhibits pronounced positional disorder and/or dynamic motion within the van der Waals gap, resulting in an averaged, “smeared” electron density of bromine atoms. This crystallographic signature is consistent with the high bromine mobility inferred from the rapid and reversible bromine exchange (“breathing”) behavior observed for the $WTe_2Br_x$ system. While the present X-ray powder diffraction data are adequately described by the average *Pmmn* model, the presence of a weak superstructure or subtle incommensurate modulation cannot be fully excluded.

**Crystal structure of $WTe_2Br_{1.25}$**

At the maximum bromine content realized in this series, the compound $WTe_2Br_{1.25}$ is obtained, which crystallizes in the orthorhombic space group *Imm*2 with refined lattice parameters *a* = 10.6218(4) Å, *b* = 43.038(3) Å, and *c* = 12.6285(9) Å. The structure was determined from single-crystal X-ray diffraction at 150 K and was identified as an inversion twin (see Table 2). All measured crystals exhibited twin domains with refined Flack parameters close to 0.5. Due to multiple phase transitions during bromine uptake the resulting crystallinity is reduced and the refinement quality is modest (see Figure S6). In addition, the structural model was used for Rietveld refinement of the bulk powder sample, yielding good agreement with the experimental pattern (see Figure S7). Given the large unit cell and number of atoms, the structure refinement from PXRD data was carried out with fixed atomic coordinates, refining only lattice parameters and profile functions.

Table 2: Crystallographic data and details from single-crystal X-ray refinement for $WTe_2Br_{1.25}$.

| | |
|---|---|
| Compound | $WTe_2Br_{1.25}$ |
| CCDC | 2438541 |
| Formula | $W_{24}Te_{48}Br_{30}$ |
| Formula weight / g·mol$^{-1}$ | 12934.50 |
| Density / g·cm$^{-3}$ | 7.441 |
| *Z* | 2 |
| Crystal system | orthorhombic |
| Space group | *Imm*2 |
| *a* / Å | 10.6218(4) |
| *b* / Å | 43.038(3) |
| *c* / Å | 12.6285(9) |
| *V* / Å$^3$ | 5773.0(6) |
| *T* / K | 150.0(1) |
| Radiation type | Cu-K$_\alpha$ |
| Reflections measured | 25309 |
| Independent reflections | 4856 |
| Abs. coeff. (mm$^{-1}$) | 148.873 |
| Goof ($F^2$) | 1.046 |
| $R_{int}$ | 0.0736 |
| $R_1$ | 0.1363 |
| $wR_2$ | 0.2842 |
| Flack | 0.56(6) |

In contrast to the uniform stacking sequence in $WTe_2Br_{0.5}$, $WTe_2Br_{1.25}$ adopts a structure in which two distinct bromine layer types alternate periodically within the van der Waals (vdW) gaps between adjacent $WTe_2$ layers, leading to a doubling of the *b*-axis (see Figure 13). An interesting feature of the structure is that one vdW interlayer contains 0.5 Br, the other 0.75 Br per $WTe_2$ formula unit, corresponding with interlayer distances of 10.1573(2) Å and 11.3493(2) Å.

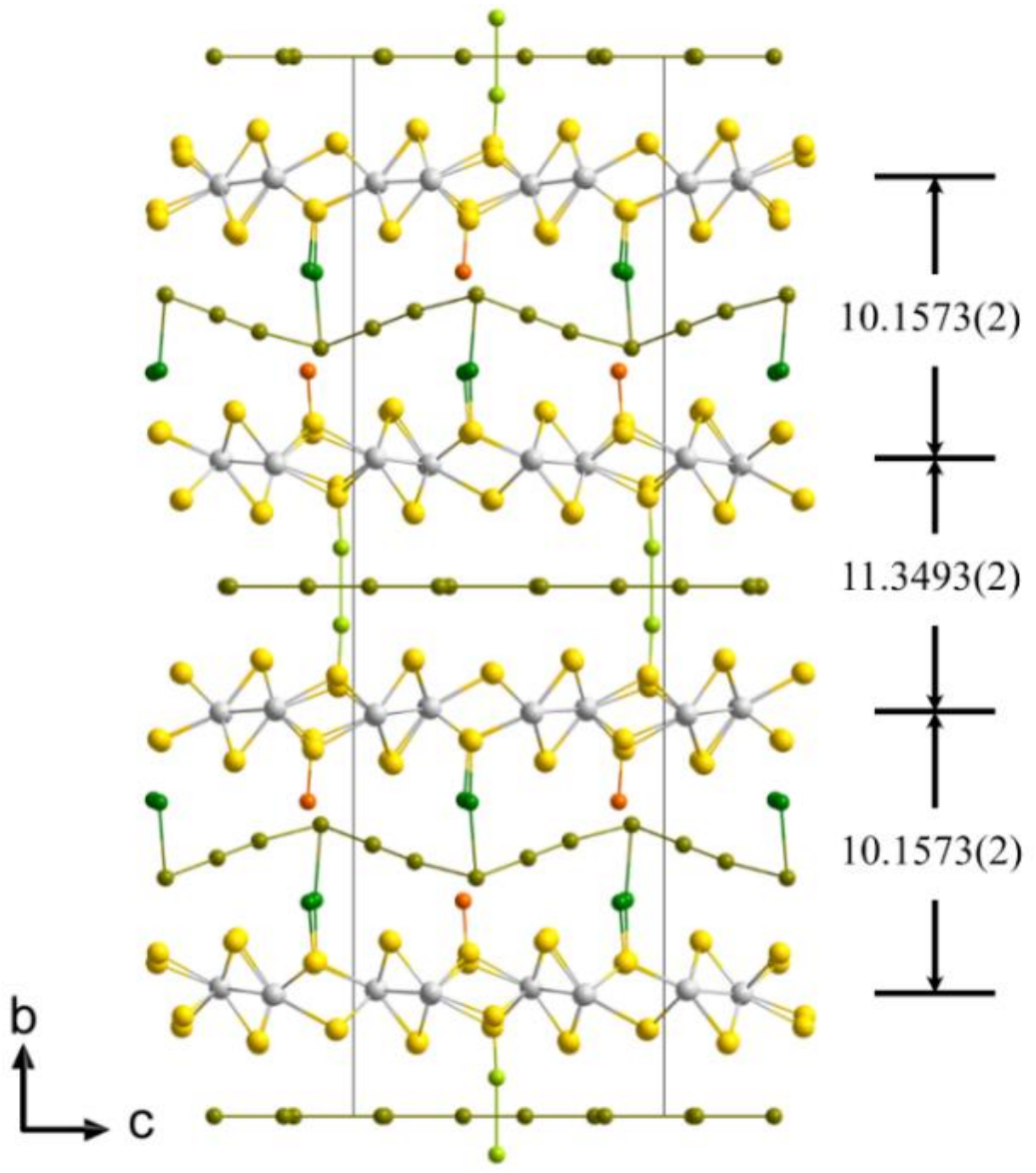

Figure 13. Section of the $WTe_2Br_{1.25}$ crystal structure viewed along the *a*-direction (W: grey, Te: yellow, Br: light green/green/olive/orange). Bromine atoms are shown in different colors to distinguish the individual bromine sites and to highlight the polybromide motifs.

The $WTe_2$ layers in $WTe_2Br_{1.25}$ are largely preserved and retain the characteristic zig-zag arrangement of tungsten atoms within a distorted $WTe_2$ layer. However, the W–W distances within the zig-zag chains are not equidistant, reflecting a distortion of the $WTe_2$ layers. Specifically, five consecutive tungsten atoms exhibit W–W distances along the crystallographic *a*-direction in the typical range of ~2.79 Å (2.771(9)–2.815(9) Å), followed by two distinctly elongated W–W contacts of approximately ~2.94 Å (2.925(9) Å and 2.954(9) Å, see Figure 14).

The bromine interlayer containing 0.5 Br atoms per $WTe_2$ formula unit adopts an essentially planar arrangement, however, containing some out of plane $Br_2$ dumbbells (Br7) aligned parallel to the *b*-axis (Figure 15, light green colored atoms), as present in the planar halogen layer observed in the commensurate $WTe_2I$ structure(Figure 8). The out of plane bromine dumbbells, with a Br–Br distance of 3.13(4) Å, establish directional contacts with Te atoms of the $WTe_2$ layers, via interlayer Te–Br contacts of 2.50(3) Å. This interaction, based on Br7, has an impact on the W–W distances along zig-zag chains (W1–W4), as visualized in Figure 14 with the longer W–W bonds shown as dashed lines.

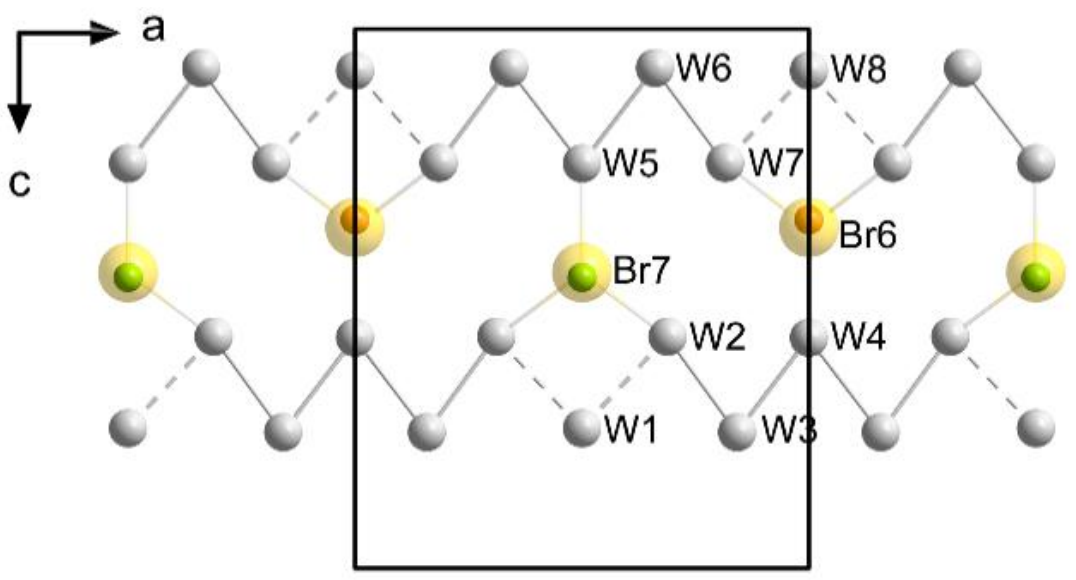

Figure 14. Section of the $WTe_2Br_{1.25}$ structure viewed along the *b*-axis with the unit cell highlighted. For clarity, only one interlayer section in the ac-plane of the tungsten zig-zag chains is shown. Tellurium atoms are omitted, except for Te atoms bonded to the $WTe_2$ layer as well as Br atoms, which are included. Note that Br7 and Br6 belong to different bromine interlayers, above and below the projected plane of W atoms. Elon-gated W–W contacts, indicating disruption of the tungsten chain, are highlighted by dashed lines (W: grey, Te: yellow, Br: light green/orange).

Bromine atoms within this layer organize into a rectangular network, with Br···Br separations ranging from 2.51(4) Å to 4.42(6) Å (Figure 15). The displacement parameters of the bromine atoms within the rectangular network (Br10 - Br13) behave strongly anisotropic, being extended in the *ac*-plane, suggesting significant positional flexibility—possibly dynamic behavior—and may indicate a tendency toward local modulation or multiple orientations of the bromine atoms.

The bromine interlayer containing 0.75 bromine per $WTe_2$ formula unit hosts a more complex arrangement involving polybromide species. Within this vdW interlayer, three distinct types of bromine motifs can be distinguished (see Figure 13 and Figure 16). First, a single bromine atom (Br6 in Figure 14, orange color) is observed in close contact with a tellurium atom in the adjacent $WTe_2$ layer, with a Br5–Te12 distance of 2.48(3) Å, which perturbs the crystallographic W–W chain (W5–W8; Figure 14). Second, a linear chain of bromine atoms is also found interacting with the $WTe_2$ layer (Figure 13 green colored atoms), but without visibly perturbing the $WTe_2$ layers. Third, a more complex, square-pyramidal arrangement of bromine atoms is located fully within the interlayer space, forming a layer through edge-sharing connections (Figure 13, olive colored atoms). These bromine atoms show no direct bonding to the surrounding $WTe_2$ framework but are connected into the polybromide chain through the apex bromine atom Br1. All Br-Br distances fall within the expected range for polybromide species (2.30–3.67 Å)[51-53], spanning from 2.36(3) Å to 3.56(2) Å, with the shortest contacts observed at the bridging $Br_2$ unit between two apex bromine atoms.

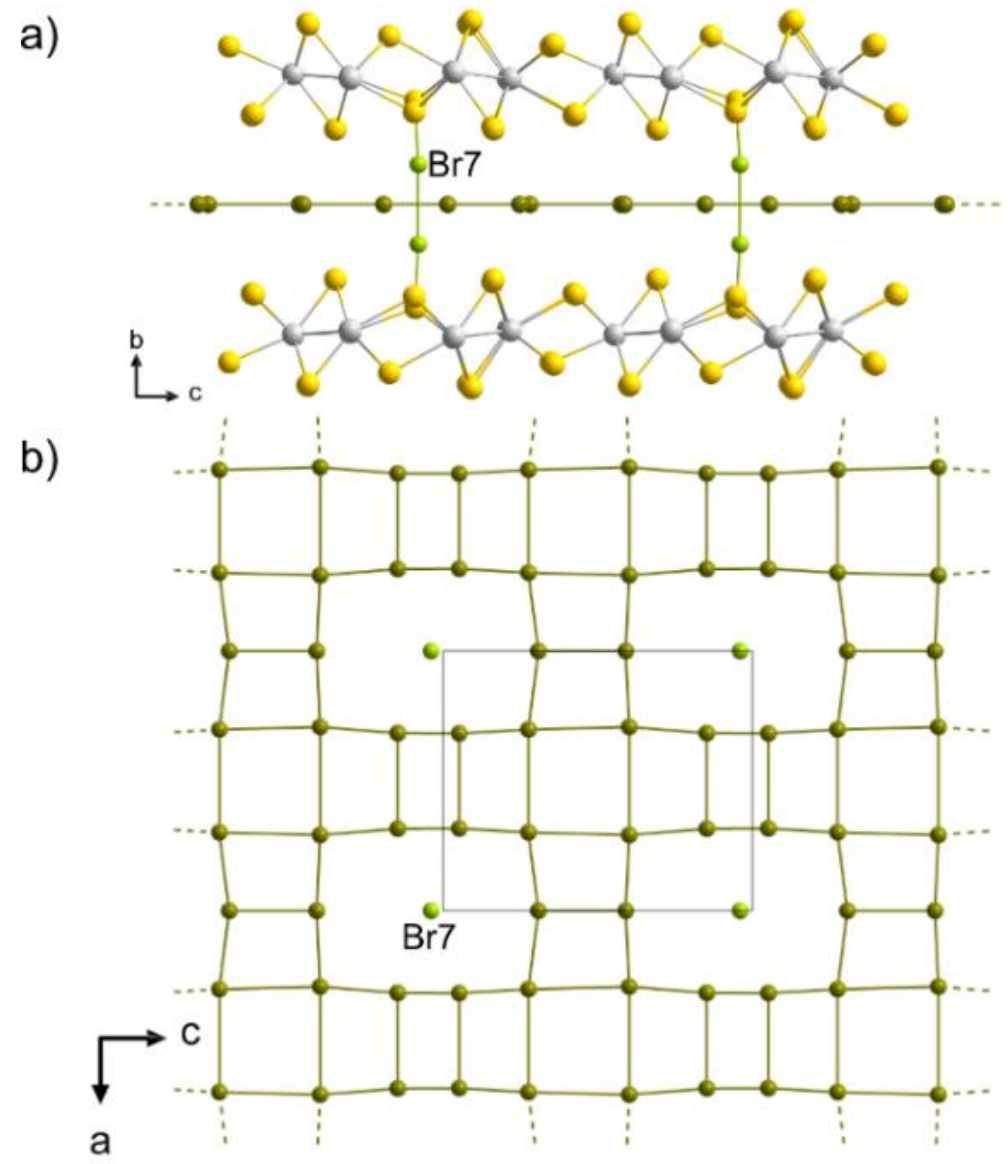

Figure 15. Section of the crystal structure of $WTe_2Br_{1.25}$ showing interlayer bromine atoms (0.5 Br per formula unit $WTe_2$) a) between adjacent $WTe_2$ layers and b) the same bromine net viewed along the *b*-direction. (W: grey, Te: yellow, Br: green/olive). The $Br_2$ dumbbells (Br7) connecting adjacent $WTe_2$ layers are aligned along the *b*-axis and highlighted green.

The atomic displacement parameters of these polybromide sites are relatively normal but slightly increased compared to those of the $WTe_2$ framework atoms. In contrast to the bromine layer forming the rectangular net, the ADPs here are less anisotropic and suggest limited flexibility or positional variation of the bromine atoms within the polybromide layer.

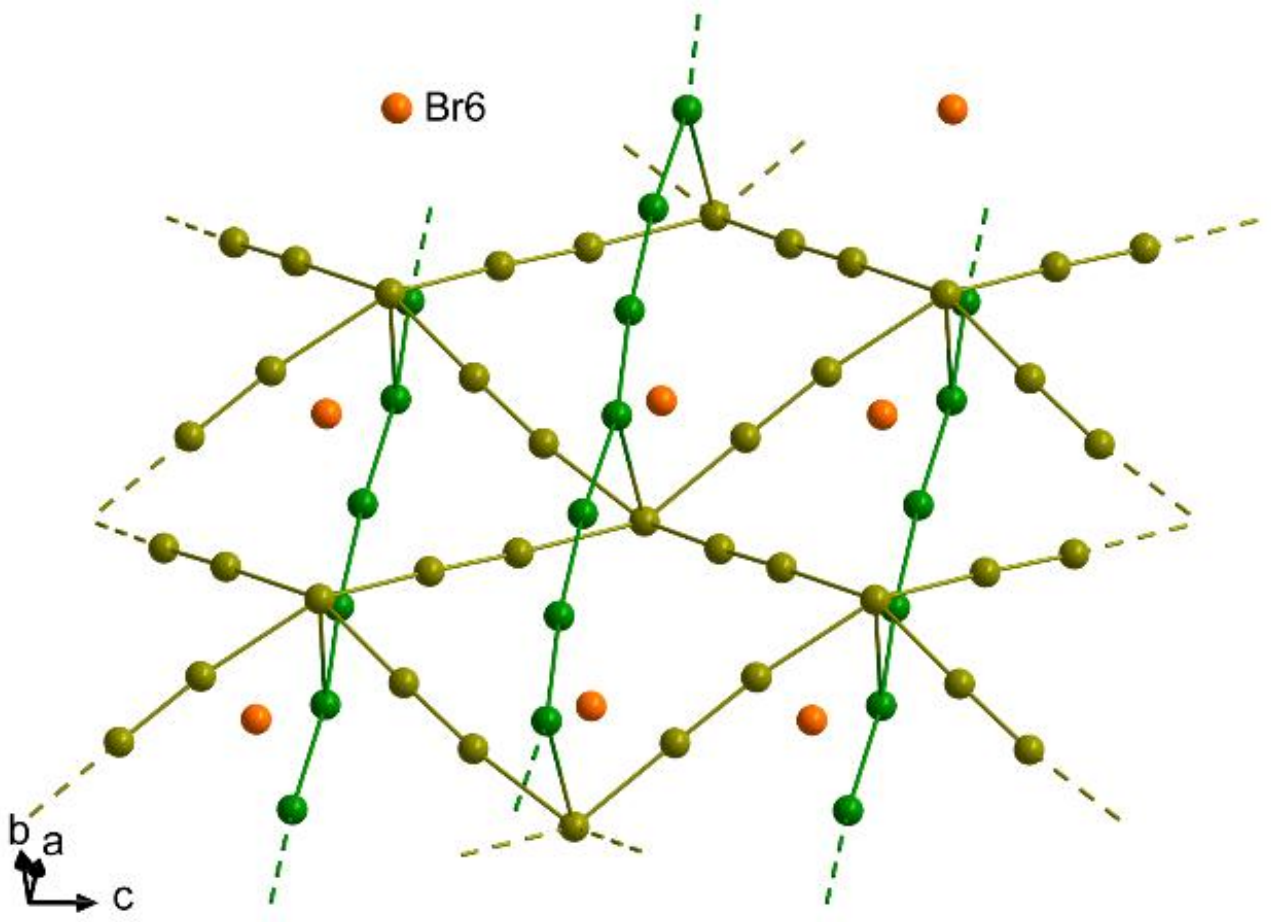

Figure 16. Section of the crystal structure of $WTe_2Br_{1.25}$ showing interlayer bromine atoms (0.75 Br per formula unit) between adjacent $WTe_2$ layers. For visual clarity, distinct bromine motifs are color-coded. The bromine atoms (Br6) connecting to adjacent $WTe_2$ layers are highlighted in orange

### Crystal structure of $WTe_2Br$

Given the structural models of $WTe_2Br_{1.25}$, $WTe_2Br_{0.5}$ and $WTe_2I$, it is reasonable to assume that the intermediate phase $WTe_2Br_{1.0}$ adopts a closely related structure. Specifically, the interlayer region is expected to feature the same bromine arrangement as observed in the 0.5 Br per formula unit bromine layer configuration of $WTe_2Br_{1.25}$, i.e., a planar halogen layer composed of $Br_2$-like units aligned with the tungsten-chain-bridging tellurium atoms. As such, $WTe_2Br_{1.0}$ likely exhibits a uniform stacking sequence with each van der Waals gap containing the same halogen motif. This would resemble the commensurate stacking found in $WTe_2I$ and represents a structurally balanced state between the alternating-filled architecture of $WTe_2Br_{0.5}$ and the polybromide-rich environment of $WTe_2Br_{1.25}$. Due to the fast and reversible intercalation behavior and high mobility of bromine in this system, isolation of pure $WTe_2Br_{1.0}$ remains challenging, and a final structural characterization has not yet been achieved.

### Electronic Band Structure of $WTe_2I$

The electronic band structure was calculated by density functional theory (DFT) for the 3D $WTe_2I$ supercell model derived from the commensurate refinement, which can be regarded as equivalent descriptions (shown in Figure 17). In comparison to the band structure of unintercalated $WTe_2$ in the $P2_1/m$ space group which is a semimetal,[34] $WTe_2I$ shows metallic behavior due to a downward shift in energy of tungsten orbitals between A (–½ 0 ½) and E (–½ ½ ½). The oxidation of the $WTe_2$ network by iodine is quite minimal, with a downward shift of the Fermi energy of only 0.04 eV.

The electronic structure of the unmodulated *Pmmn* structure of $WTe_2I$ ($a$ = 21.8967(2) Å, $b$ = 3.4759(0) Å, $c$ = 6.3270(1) Å) has been previously reported.[49, 50] The unmodulated structure is also metallic, but shows several important differences to the supercell structure: a much higher degree of oxidation of the $WTe_2$ layers, and the presence of iodine bands with high dispersion crossing the Fermi level. These crossing bands lead to nesting of the Fermi surface and charge density wave (CDW) instabilities, which create the modulated structures that we have characterized here.

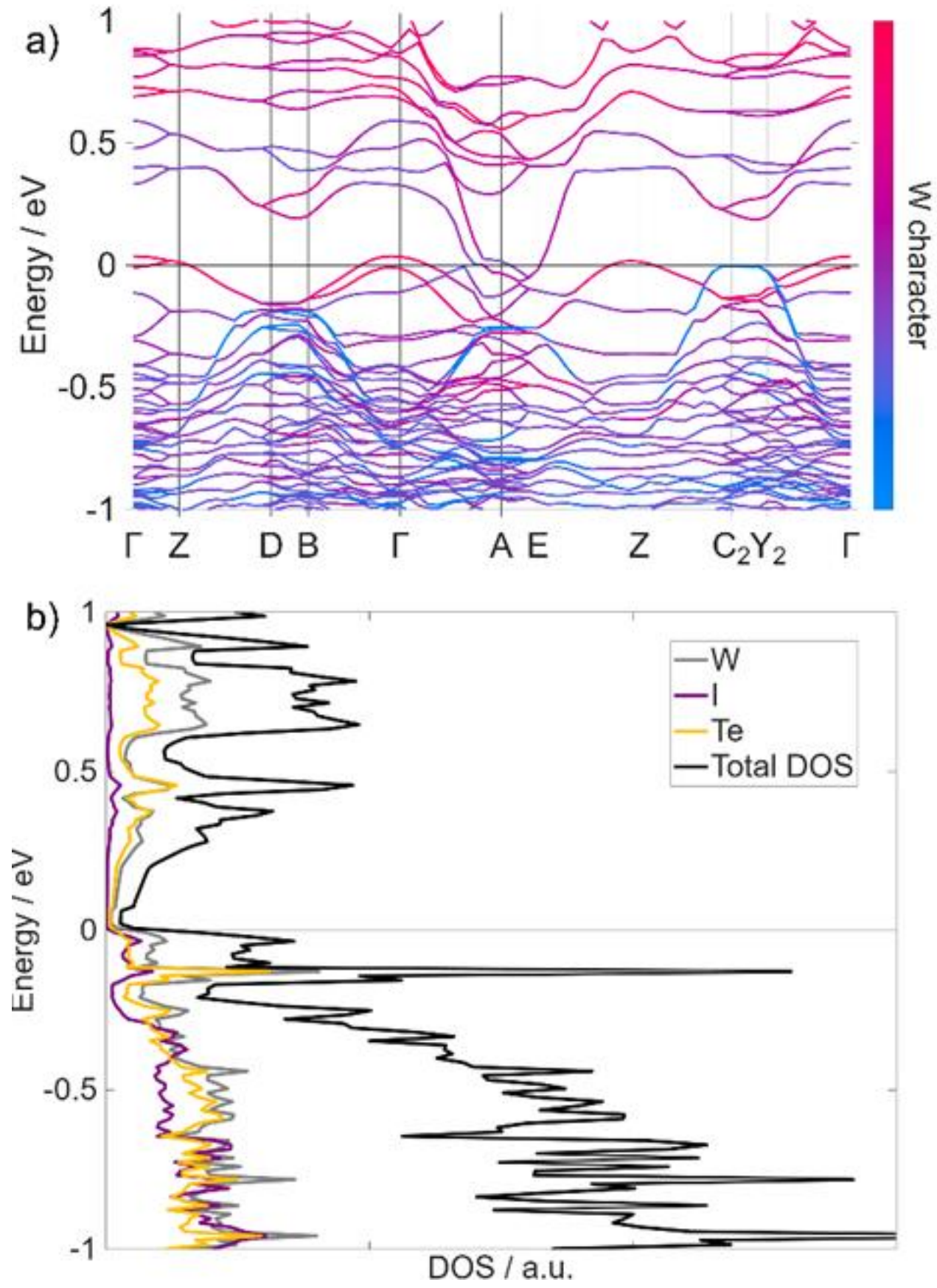

Figure 17. Calculated electronic band structure of commensurately modulated $WTe_2I$, with bands colored by their tungsten character (a), and the corresponding electronic density of states (b). Special points in and paths through reciprocal space were chosen following the literature.[54]

In supercell $WTe_2I$ structure, some iodine states form a flat band at the Fermi energy between $C_2$ (–½ ½ 0) and $Y_2$ (-½ 0 0) (shown in blue in Figure 17). The flatness of this band indicates a very high degree of electron localization. When moving away from the $C_2$–$Y_2$ line, the band takes a parabolic form with many anticrossings of the W bands, which demonstrates strong coupling between the I and W states. This band is associated with the iodine atoms I2, I8 and I5, I11 which correspond to I1 and I2 in the commensurate structure at specific internal phase coordinate $t$ (see Figure 18, highlighted in red). The overall picture is of localized metallic electrons on some portions of the iodine layer, potentially representing frozen CDWs.[49]

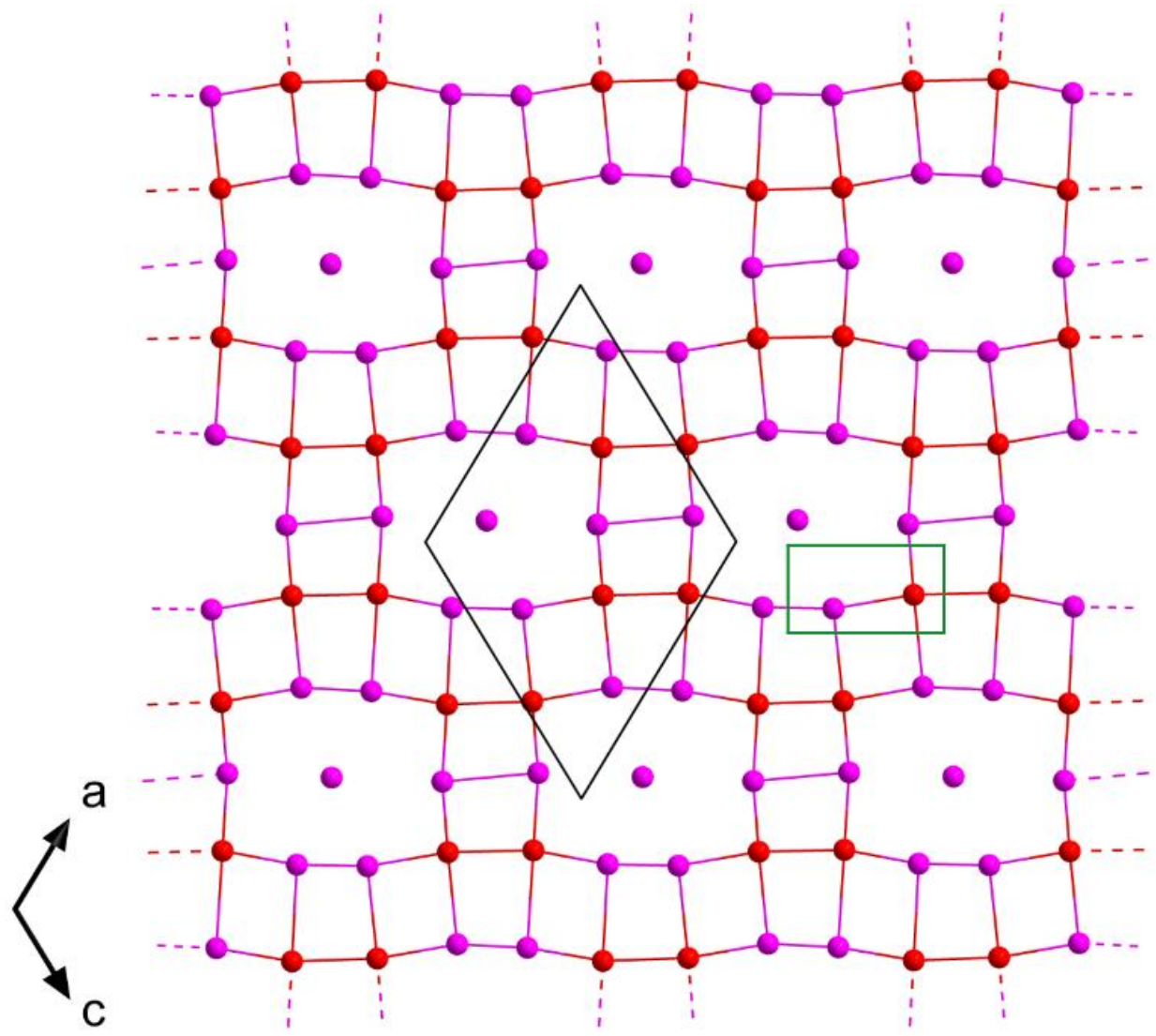

Figure 18. Section of the $WTe_2I$ supercell structure only showing one iodine (red/purple) layer with highlighted I2, I8 and I5, I11 iodines in red with viewing direction along the *b*-axis. Unit cells are shown in black (supercell) and green (commensurate).

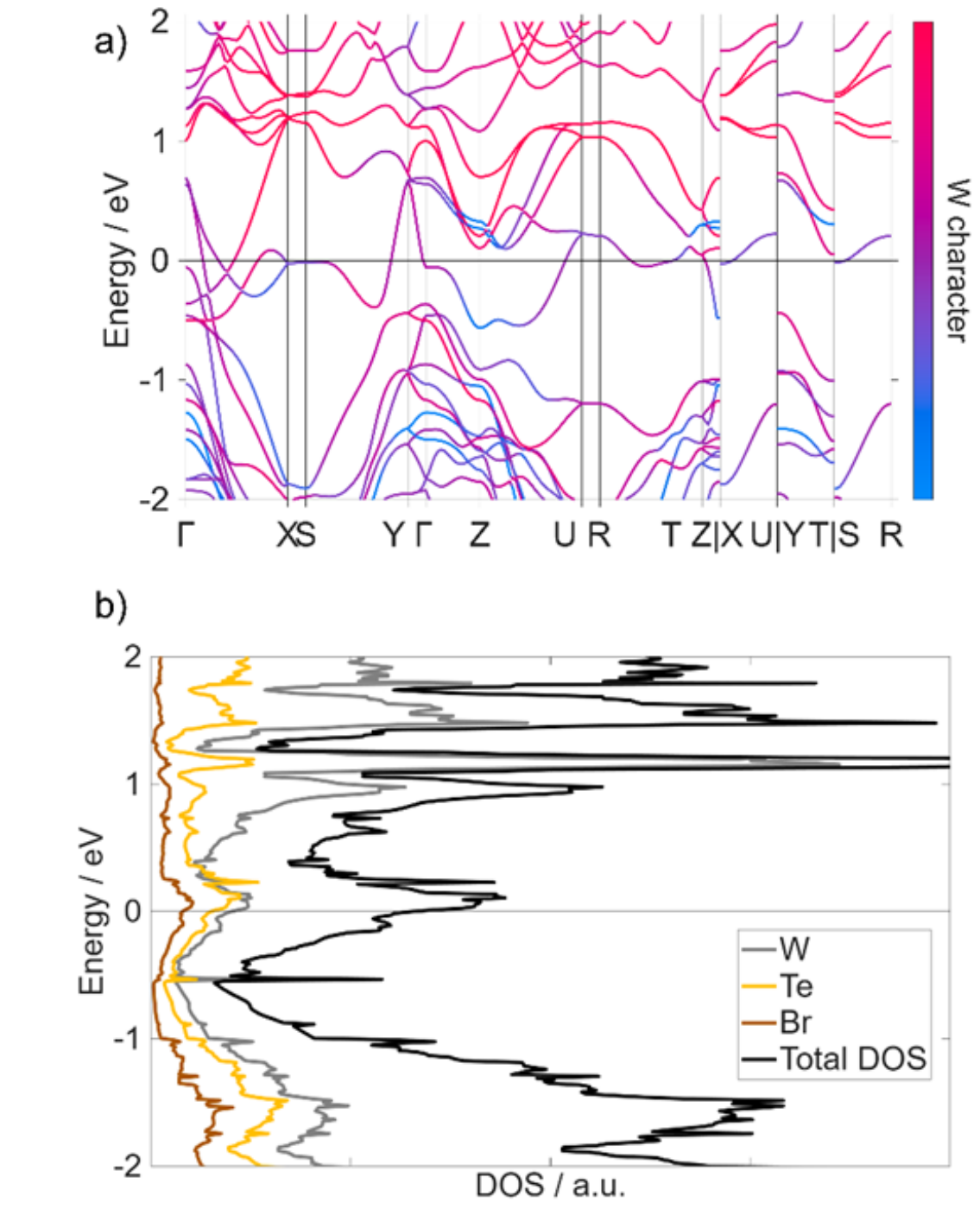

Figure 19. Calculated electronic band structure of $WTe_2Br_{0.5}$, with bands colored by their tungsten character (a), and the corresponding electronic density of states (b). Special points in and paths through reciprocal space were chosen following the literature.[54]

## Electronic Band Structure and Phonon Band Structure of $WTe_2Br_{0.5}$

The electronic band structure of $WTe_2Br_{0.5}$ was also calculated using density functional theory (DFT) (Figure 19). Like $WTe_2I$, $WTe_2Br_{0.5}$ is a metal, with additional bands appearing near the Fermi energy compared to $WTe_2$.[49] These include an interesting flat band at the Fermi energy between X (½ 0 0) and S (½ ½ 0); however, unlike in the case of $WTe_2I$, there is no contribution of the Br atoms to the states near the Fermi energy, and therefore these localized metallic electrons are located within the $WTe_2$ layers. $WTe_2Br_{0.5}$ shows a higher degree of oxidation of the $WTe_2$ layers than modulated $WTe_2I$, with a shift of the Fermi energy of 0.35 eV, indicating a more ionic interaction with the intercalant.

The phonon band structure of $WTe_2Br_{0.5}$ was also calculated; it is shown in Figure 20. Like unmodulated $WTe_2I$,[49] $WTe_2Br_{0.5}$ shows many soft modes (modes with negative energy), corresponding to instabilities of the crystal structure at 0 K. These soft modes include both displacements of the Br atoms and of the $WTe_2$ layers. Therefore, $WTe_2Br_{0.5}$, like $WTe_2I$, has a propensity towards the formation of a disordered or modulated structure.

The observation of an unmodulated structure for $WTe_2Br_{0.5}$, in comparison to the modulated structure of $WTe_2I$, can be understood by consideration of the electronic band structure. $WTe_2Br_{0.5}$ lacks the extra sheet-like bands crossing the Fermi level which lead to the formation of CDWs, CDW instabilities, and a modulated structure in $WTe_2I$.[49] Therefore, the phonon instabilities in $WTe_2Br_{0.5}$ likely cause dynamic and/or static disorder, as is seen in the large ADPs. The high density of Br states below 10 meV would also contribute thermal displacements to the ADPs, although these modes contain displacements of the Br atoms in all three Cartesian directions.

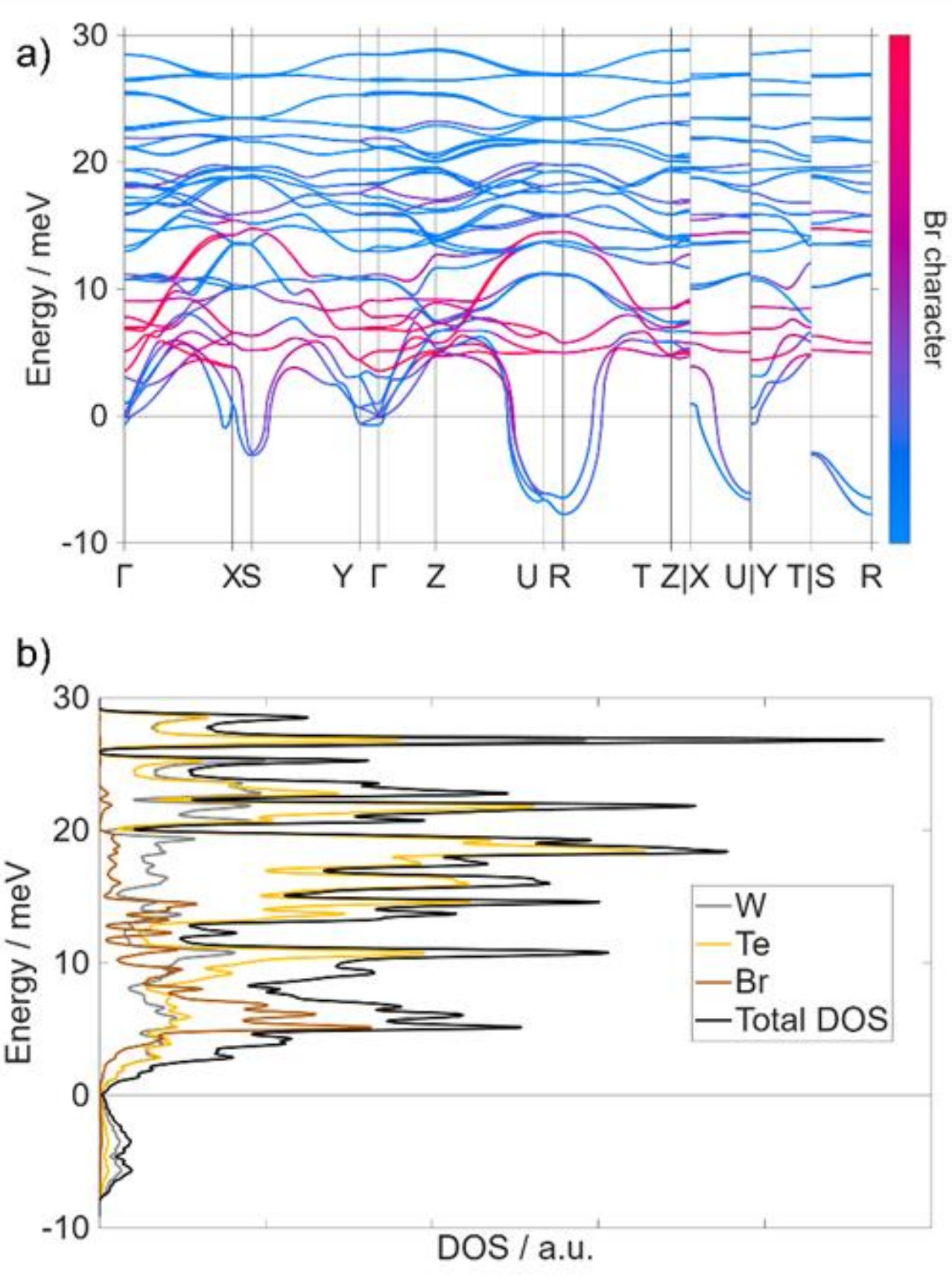

c)

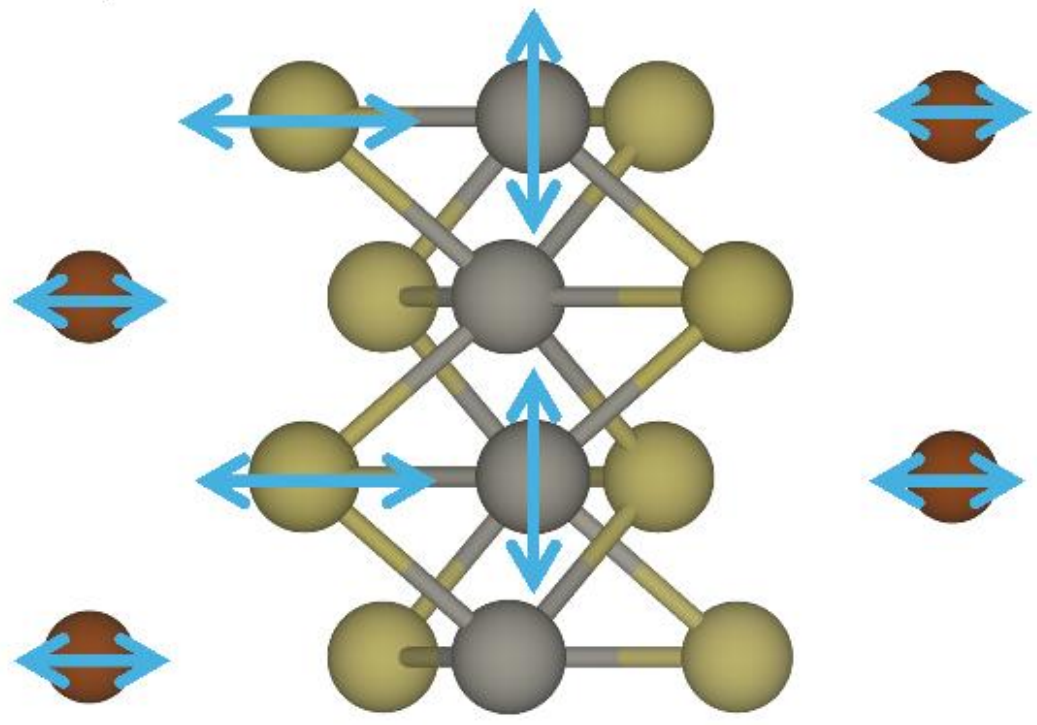

Figure 20. Calculated phonon band structure of $WTe_2Br_{0.5}$, with bands colored by their bromine character (a), and the corresponding phonon density of states (b). A cartoon view of an unstable mode at R (0.5 0.5 0.5) along *c* is shown in (c). Special points in and paths through reciprocal space were chosen following the literature. [54]

## Experimental

**Synthesis:** All material handling and manipulations were performed in an argon-filled glove box (MBraun Labmaster 130, $O_2$ < 1 ppm, $H_2O$ < 1 ppm). All synthesis steps consistently afforded yields greater than 95%.

**Powder:** $WTe_2$ was synthesized by combining tungsten powder (1256.24 mg, 6.84 mmol, ABCR GmbH, 99.95%, particle size 0.6–0.9 µm), tellurium pieces (1830.95 mg, 14.35 mmol, Evochem, 99.999%), and tungsten(VI) chloride ($WCl_6$; 135.49 mg, 0.34 mmol, Arcos, 99.9+%) in a 1:2.1:0.05 molar ratio, with $WCl_6$ serving as an oxygen getter. The mixture was sealed in a dry, evacuated silica ampoule (40 mm length, 16 mm diameter) and heated in a commercial (Carbolite) furnace to 800 °C for 6 hours, employing heating and cooling rates of 2 K/min. The resulting product was thoroughly ground, vacuum-sealed in a second silica ampoule (200 mm length, 16 mm diameter), and heated to 500 °C for 20 hours under a temperature gradient, with the opposite end of the ampoule maintained at room temperature, to facilitate the removal of excess tellurium, tungsten oxychloride impurities and unreacted $WCl_6$.

**Single crystals of $WTe_2$** was synthesized by combining tungsten powder (1256.24 mg, 6.84 mmol, ABCR GmbH, 99.95%, particle size 0.6–0.9 µm), tellurium pieces (1830.95 mg, 14.35 mmol, Evochem, 99.999%), and tungsten(VI) chloride ($WCl_6$; 135.49 mg, 0.34 mmol, Arcos, 99.9+%) in a 1:2.1:0.05 molar ratio, with $WCl_6$ serving as an oxygen getter. The mixture was sealed in a dry, evacuated silica ampoule (40 mm length, 16 mm diameter) and heated in a commercial (Carbolite) furnace to 800 °C for 6 hours, employing heating and cooling rates of 2 K/min.

**$WTe_2I$ crystals:** $WTe_2$ single-crystals (~100 mg) were loaded into a borosilicate glass screw-cap vial (5/20 mL for incommensurate/commensurate crystals) with four molar equivalents of iodine, and the vial was sealed with a custom-machined PEEK cap equipped with a PTFE liner. The reaction mixture was heated in a Simon-Müller laboratory oven at a rate of 2 $K\cdot min^{-1}$ to 120 °C, held at this temperature for 12 h, and then cooled to room temperature at the same rate.

**$WTe_2Br_x$ (*x* = 0.5 and 1.25):** Bromine (p.a.) was degassed and purified prior to use according to Brauer.[55] Due to the corrosive nature and high vapor pressure of bromine, all operations were performed either in an argon glovebox or using Schlenk techniques under dry argon (with appropriate cold traps). Typically, $WTe_2$ (powder or single crystals, ca. 100 mg) was reacted with a large excess of bromine (ca. 3 mL). $WTe_2$ powder was placed in a Schlenk tube, liquid bromine was added and the mixture was allowed to react for 3 h. Excess bromine was then removed by passing a constant argon flow over the sample. Because of the high density of bromine vapors, a needle was inserted through the septum and positioned directly above the sample to provide an efficient outlet for the $Br_2$/argon vapors and the vessel was gently shaken from time to time. For $WTe_2Br_{0.5}$, evaporation at room temperature was sufficient, whereas for $WTe_2Br_{1.25}$ the Schlenk tube was cooled to 0-5 °C during the bromine removal step and needed at least 3 h. Needle-shaped $WTe_2$ single crystals were intercalated either (i) by exposure to bromine vapor at room temperature for 4 days or (ii) by direct contact with liquid bromine at 6 °C for 2 days.

**Powder X-ray diffraction:** PXRD patterns of products were collected with a Stadi-P (STOE, Darmstadt) powder diffractometer using germanium monochromated Cu-$K_{\alpha1}$ radiation ($\lambda$ = 1.5406 Å) and a Mythen 1K detector. PXRD measurements of $WTe_2Br_x$ (*x* = 0.5, 1.0, and 1.25) were performed in transmission geometry using interchangeable window materials selected according to the experimental requirements. The windows comprised ultrathin glass (Schott AF 32® eco thin glass, 30 µm), Kapton® film (25 µm), and Mylar® film (15 µm). To prevent corrosion by bromine and to enable reliable sealing, the metal parts of the original transmission holder were replaced by custom-made PTFE components (rotor disc and counter disc) and soda-lime glass components in a redesigned holder (see Figure S8 and S9). In all experiments, $WTe_2$ powder was immobilized on the glass window using Lithelen grease to ensure a fixed sample position during measurement, and all samples were assembled and sealed under inert conditions (argon atmosphere). For $WTe_2Br_{0.5}$, a standard glass disc without a filling slit was employed; the sample was covered with a Kapton® film window that was sealed using Lithelen grease and allowed gradual pressure equilibration via bromine diffusion through the Kapton film (see Figure S8 a) and d)). For $WTe_2Br_{1.25}$, a glass disc featuring a filling slit was used; ultrathin-glass windows were bonded to both sides of the disc using cyanoacrylate adhesive (superglue). Bromine was introduced through the slit under argon, and the opening was subsequently sealed with wax and additional cyanoacrylate. For in situ PXRD experiments, Mylar film was used as the window material to facilitate faster bromine release during the measurement. The crystal structure of $WTe_2Br_{0.5}$ was solved from PXRD using EXPO2014,[56] and refined by Rietveld method in FullProf (FP)[57] with a modified Thompson-Cox-Hastings pseudo-Voigt (TCHZ) profile function. [58, 59] The instrumental resolution function (IRF) was obtained from the NIST Si640f standard[60] and fitted in WinPLOTR.[61] Additionally, multiple texturing effects, such as particle form, size, and orientation, were included in the Rietveld refinements to enhance the accuracy of the model. All intermediate and final structure models were validated with PLATON.[62]

**Single-Crystal X-Ray Diffraction (SC-XRD):** SC-XRD studies were performed using a Rigaku XtaLAB Synergy-S diffractometer equipped with a HyPix 6000HE detector using $MoK_\alpha$ ($\lambda$ = 0.71073 Å) and $CuK_\alpha$ ($\lambda$ = 1.54184 Å) radiation at 150 K. Data reduction, scaling and absorption corrections were performed using CrysAlis$^{Pro}$.[63], taking into account the crystal shape and size.[64] The supercell structure of $WTe_2I$ and $WTe_2Br_{1.25}$ were solved with the ShelXT 2018/2 solution program[65], employing dual methods and refined with ShelXL 2018/3 (Sheldrick, 2015)[66] using Olex2 1.5 (Dolomanov et al., 2009)[67] as the graphical interface. The models were refined using full matrix least squares minimization on $|F|^2$. For $WTe_2Br_{1.25}$ crystals, the needle shaped $WTe_2$ single crystals were selected directly from bromine layered with perfluoropolyalkylether (viscosity 1800 cSt; ABCR GmbH). The modulated $WTe_2I$ structures were solved using the charge-flipping algorithm SUPERFLIP[68] (as implemented in JANA2020), and subsequent refinements were carried out in JANA2020.[69-72]

**ICP-OES:** After dissolution of $WTe_2I$ in 2 wt % $NaOH/H_2O_2$, the W/Te/I ratio was determined by ICP-OES (iCAP 7400 Thermo Fisher Scientific) and yielded in averaged ratios W:Te:I of 1.0 : 1.99(3):0.49(3) for $WTe_2I$ phases.

**TXRF (Total Internal Reflection X-Ray Fluorescence) Spectroscopy:** TXRF studies were performed using a S2 Picofox (Bruker AXS Microanalysis, Berlin, Germany) equipped with a Mo X-ray tube, which was operated at 50 kV and 600 µA. The measurement period for each sample was 1000 s (live time). Fitting of the resulting spectra was done using the Spectra software (Bruker Nano GmbH) in the super byas mode (maximum stripping cycles of 2000). The average ratio for $WTe_2I$ with W:Te:I was 1.0 : 1.98(7):0.51(6).

**Bromide Content Determination (Argentometric Titration):** Approximately 100 mg of $WTe_2B_{r0.5}$ was dissolved in concentrated $HNO_3$ (5 mL) with addition of ~1 mL of $H_2O_2$ (30%). The mixture was carefully heated to ca. 80 °C for 30 min, allowed to cool to room temperature, and diluted with bidistilled $H_2O$ to a final volume of 100 ml. The bromide content was determined by argentometric titration with potentiometric end-point detection using a standardized $AgNO_3$ solution (0.01 M). Titrations were performed on a Schott TitroLine easy automatic titrator (SI Analytics) equipped with a silver indicator electrode. The equivalence point was identified from the inflection in the electrode potential (mV), corresponding to complete precipitation of $Br^-$ as AgBr. Repeated determinations gave bromide contents consistent with the expected composition of $WTe_2Br_{0.5}$ (W:Br = 1:0.49(3)).

**DFT Calculations:** Calculations were performed using the ABINIT software package with the projector augmented-wave (PAW) method and a plane-wave basis set.[73, 74] The Perdew–Burke–Ernzerhof exchange–correlation functional was used with the vdw-DFT-D3 dispersion correction.[75, 76] PAW data files were used as received from the ABINIT library.[77] Methfessel–Paxton smearing was used to determine band occupation.[78] Convergence studies were used to choose a 18 Ha ($WTe_2I$) and 20 Ha ($WTe_2Br_{0.5}$), plane wave basis set cutoff energy outside the PAW spheres; a 100 Ha cutoff was used within the spheres. The Brillouin zone was sampled with a 4×2×4 ($WTe_2I$) or 6×10×6 ($WTe_2Br_{0.5}$) grid of k-points. The calculation of the phonon band structure of $WTe_2Br_{0.5}$ was performed using density functional perturbation theory on a 3×5×3 grid of q-points. Structural relaxation was performed prior to calculation of the electronic and phononic structures.

## Conclusions

In this work, the crystal structure of the iodine intercalate $WTe_2I$ was reinvestigated and shown to occur in two closely related (3+1)D variants within $P2_1/m(\alpha 0\gamma)00$: an incommensurate form and a commensurate polytype that "locks in" to $\boldsymbol{q}$ = (1/2, 0, 1/6) and can be equivalently described as a 3D supercell. In both descriptions, the modulation is dominated by the iodine sublattice, while the $WTe_2$ host responds only by subtle, coupled distortions. A key microscopic element is a local defect motif in which an I2 unit reorients by 90° (I3 site) and is accommodated by a concomitant Te shift, consistent with localized host–guest bonding that effectively pins the modulation and periodically perturbs the tungsten chains.

Extending this chemistry from iodine to bromine, we identify a multistage series $WTe_2Br_x$ with distinct regimes at $x$ = 0.5, 1.0 and 1.25 and a striking, fast, and reversible "breathing" behavior already at room temperature. In situ PXRD directly captures the rapid evolution of the layer-spacing reflections during uptake and release, demonstrating that composition and structure can be cycled under mild conditions. Structurally, $WTe_2Br_{0.5}$ (*Pmmn*) represents a bromine-poor end member with a half-filled interlayer motif and a short Te-Br contact that links adjacent layers; its exceptionally large, anisotropic bromine displacement parameters point to pronounced positional freedom consistent with high bromine mobility. At the bromine-rich end, $WTe_2Br_{1.25}$ (*Imm*2) adopts an alternating-filled architecture with two chemically distinct interlayer environments, combining $Br_2$-like bridging motifs and more complex polybromide arrangements, while the $WTe_2$ framework remains intact but locally distorted in response to directional Te-Br interactions. In both $WTe_2I$ and $WTe_2Br_{0.5}$, the electronic structure of the $WTe_2$ layers is changed significantly by intercalation, leading to a metallic state with localized electrons in flat bands at the Fermi level. Together, these results establish $WTe_2I$ and $WTe_2Br_x$ as rare, structurally resolved examples of stoichiometric halogen intercalation in a TMDC, highlighting how a layered semimetal can tolerate strong oxidizing guests while retaining topotactic reversibility—opening a route to controllable oxidation (hole doping), staging, and modulation phenomena in van der Waals solids.

## Conflicts of interest

There are no conflicts to declare.

## Data availability

Crystallographic data have been deposited at the Cambridge Crystallographic Data Centre (CCDC) under 2520418 ($WTe_2I$, incommensurate), 2519690 ($WTe_2I$, commensurate), 2519736 ($WTe_2I$, supercell), 2503030 ($WTe_2Br_{0.5}$), 2519455 ($WTe_2Br_{1.25}$). Data are available within the article. The data that support the findings of this study are available on request from the corresponding author, H.-J. Meyer. Computational data are available at doi:10.5281/zenodo.18420902.

## Acknowledgements

The authors acknowledge support of this research by the Deutsche Forschungsgemeinschaft (Bonn) through the project ME 914/32-1 and by the state of Baden-Württemberg through bwHPC and the German Research Foundation (DFG) through grant no INST 40/575-1 FUGG (JUSTUS 2 cluster). Carl P. Romao acknowledges support from the project FerrMion of the Ministry of Education, Youth and Sports, Czech Republic, co-funded by the European Union (CZ.02.01.01/00/22_008/0004591).

# Structural Evolution during Reversible Halogen Intercalation in $WTe_2$: Commensurate–Incommensurate $WTe_2I$ and Multistage $WTe_2Br_x$ ($x$ = 0.5, 1.0 and 1.25)

Patrick Schmidt,[a] Carl P. Romao,[b] Hans-Jürgen Meyer*[a]

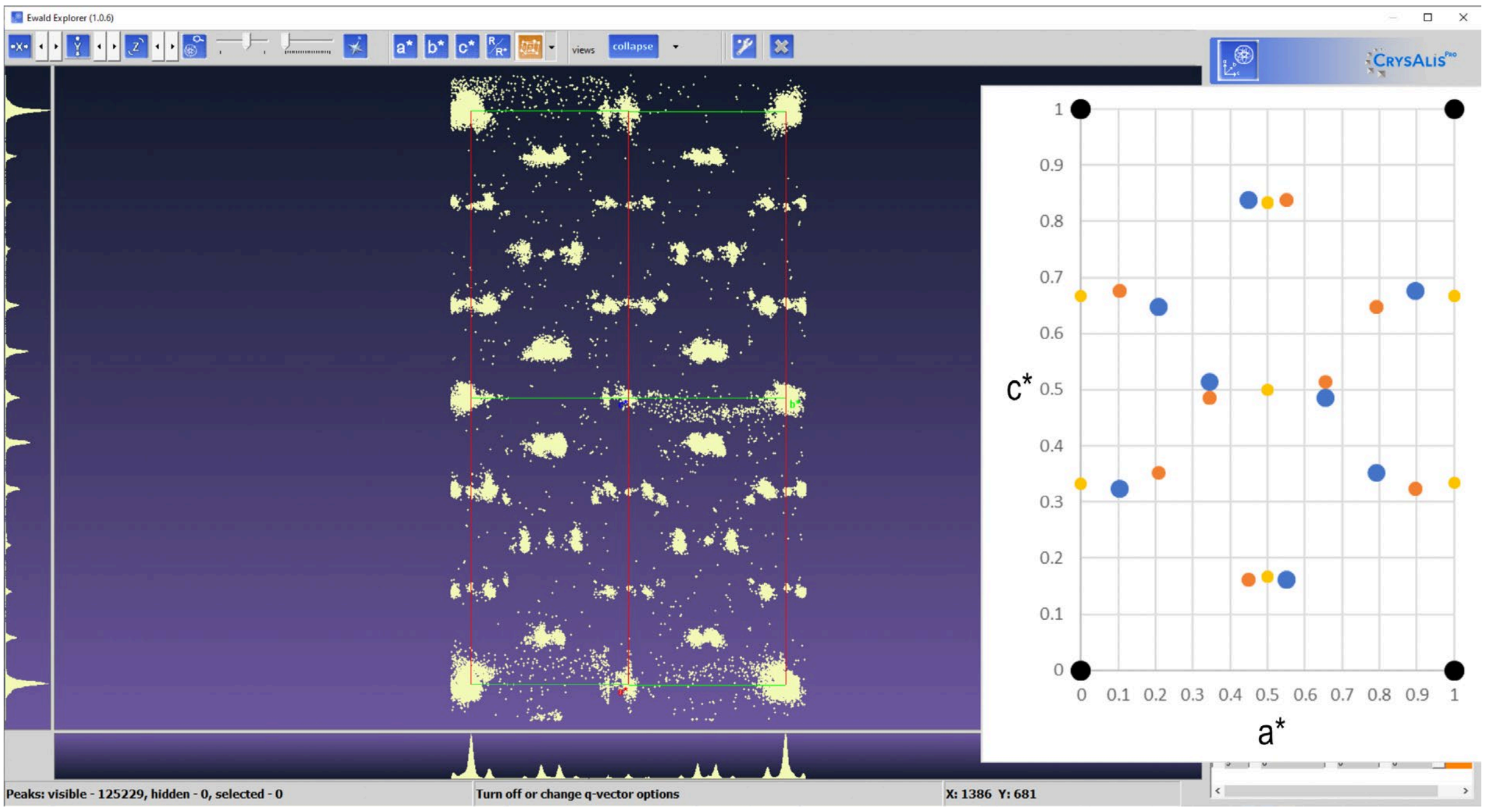

Figure S1. Screenshot from the Ewald Explorer in CrysalisPro showing the collapsed reciprocal lattice, with a simulated satellite pattern on the right (with m≤|±4|). Black dots mark the main reflections, while red and blue dots indicate incommensurate, twinned satellite reflections. Yellow dots denote the commensurate case.

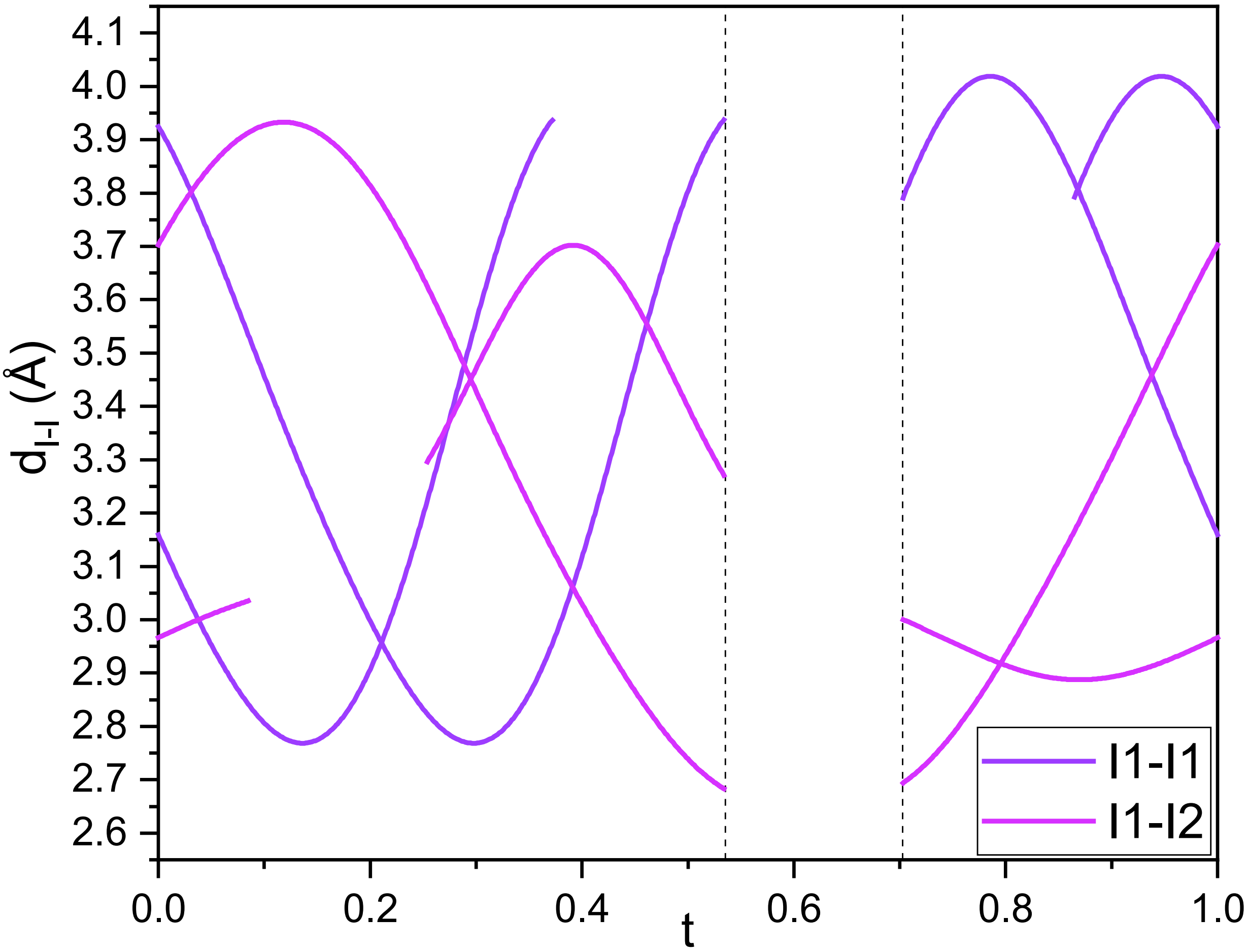

Figure S2. *t* plots of the distances for I1-I1 and I1-I2 in grey with crenel limits marked as dashed vertical line.

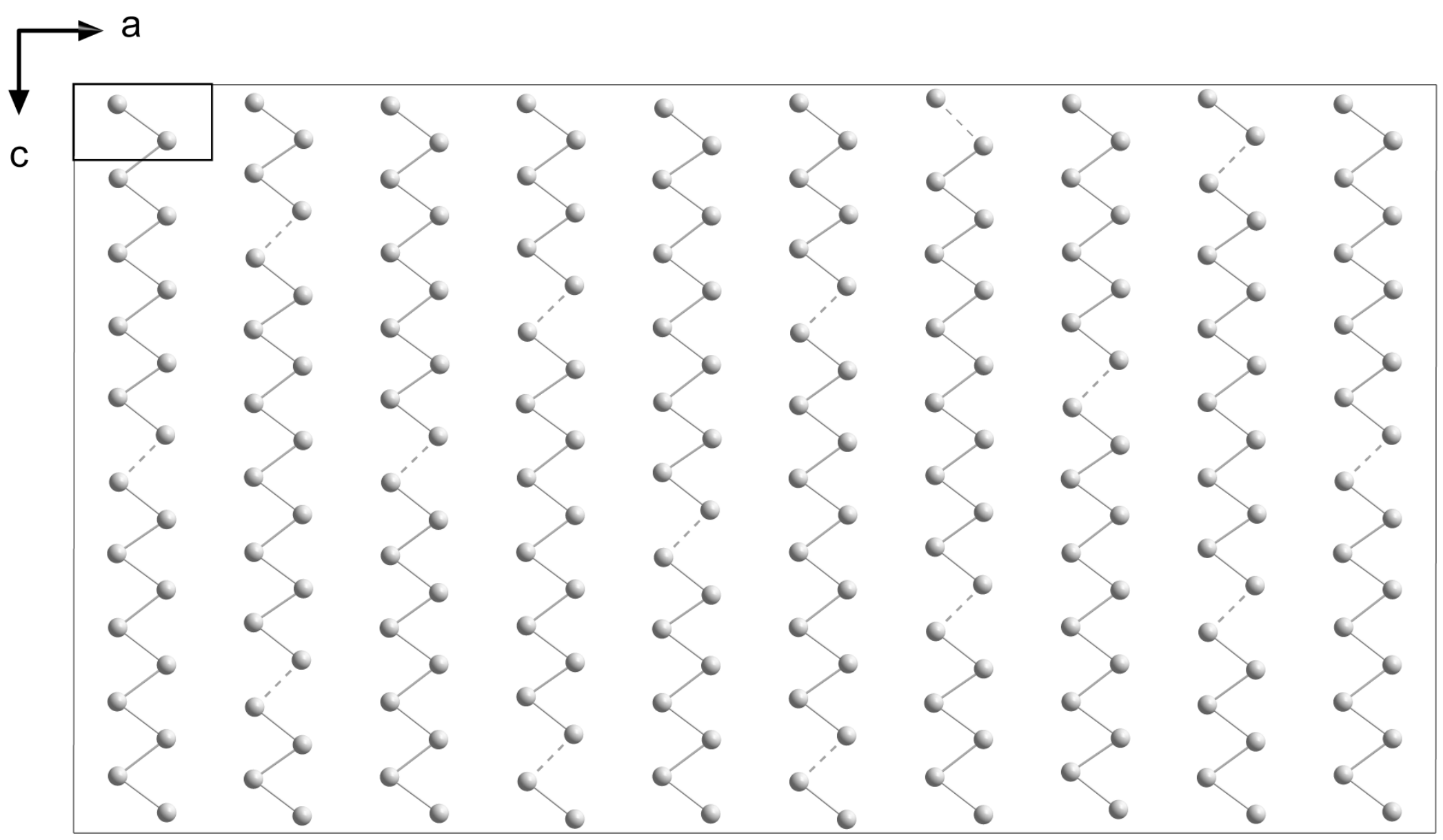

Figure S3. Section of the incommensurate $WTe_2I$ structure showing only one layer of tungsten atoms with long bonds indicated as dashed lines. The unit cell is highlighted in the top left corner.

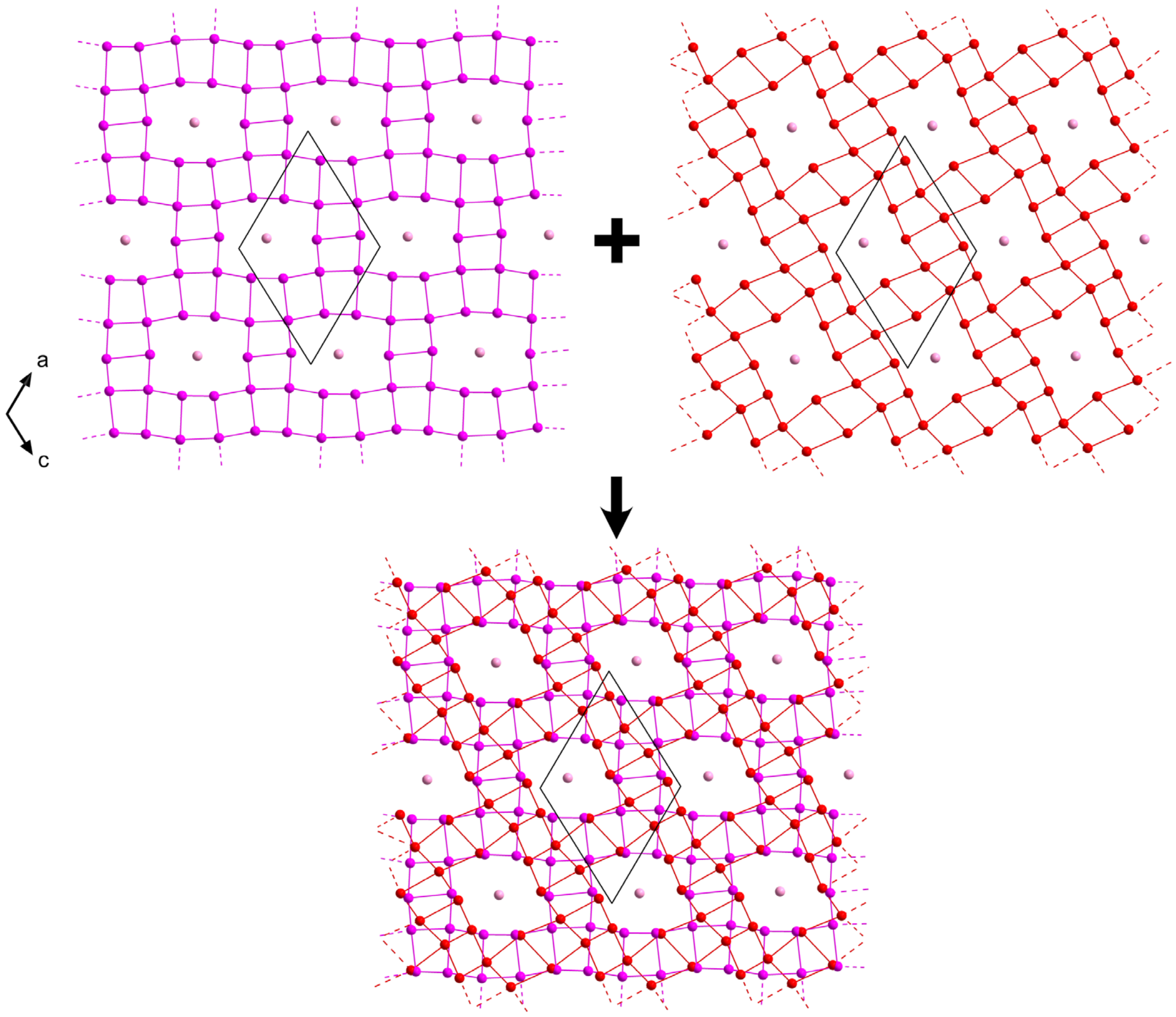

Figure S4. Sections of the $WTe_2I$ supercell structure illustrating the iodine-layer disorder model. The top panels show the two disorder components separately (part a, purple; part b, red), while the bottom panel displays the combined iodine layer. Out-of-plane iodine atoms are highlighted in light purple. The structure is viewed along the *b*-axis, with the unit cell outlined.

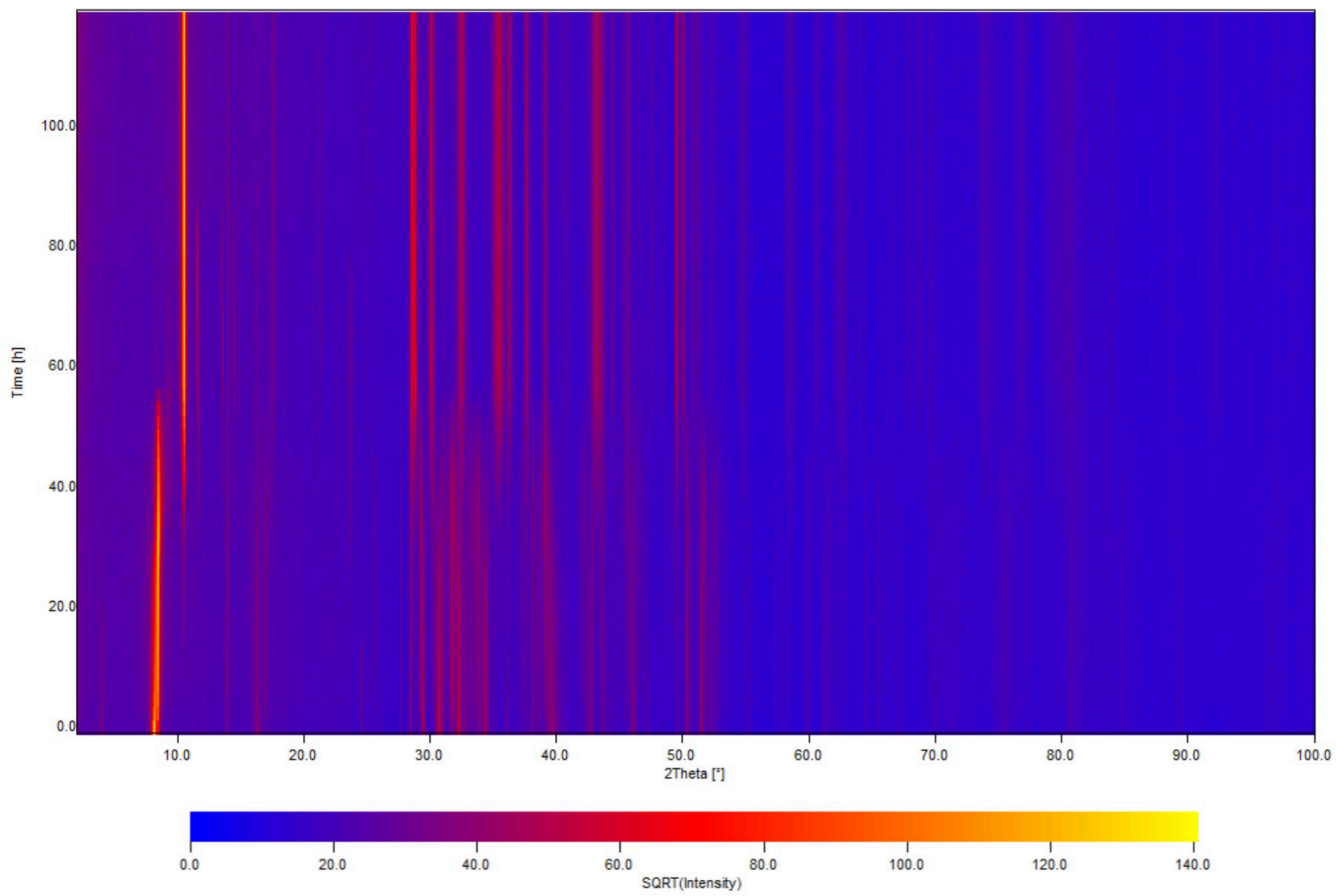

Figure S5. Time resolved PXRD analysis of $WTe_2Br_x$ showing structural evolution while bromine deintercalation.

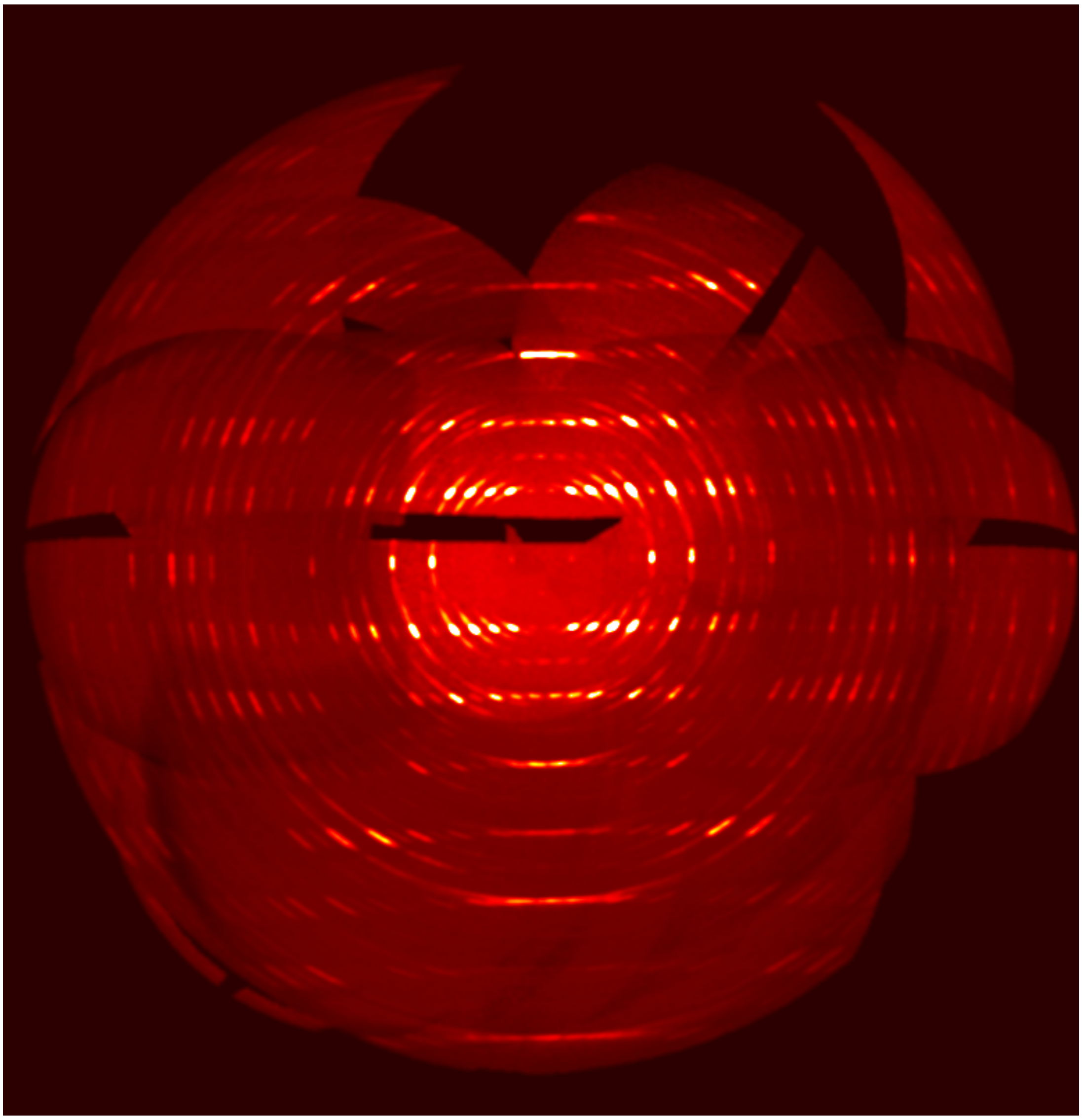

Figure S6. Reconstruction of the (-3*kn*) of the reciprocal-space plane of $WTe_2Br_{1.25}$, highlighting reduced crystallinity due to multiple phase transitions.

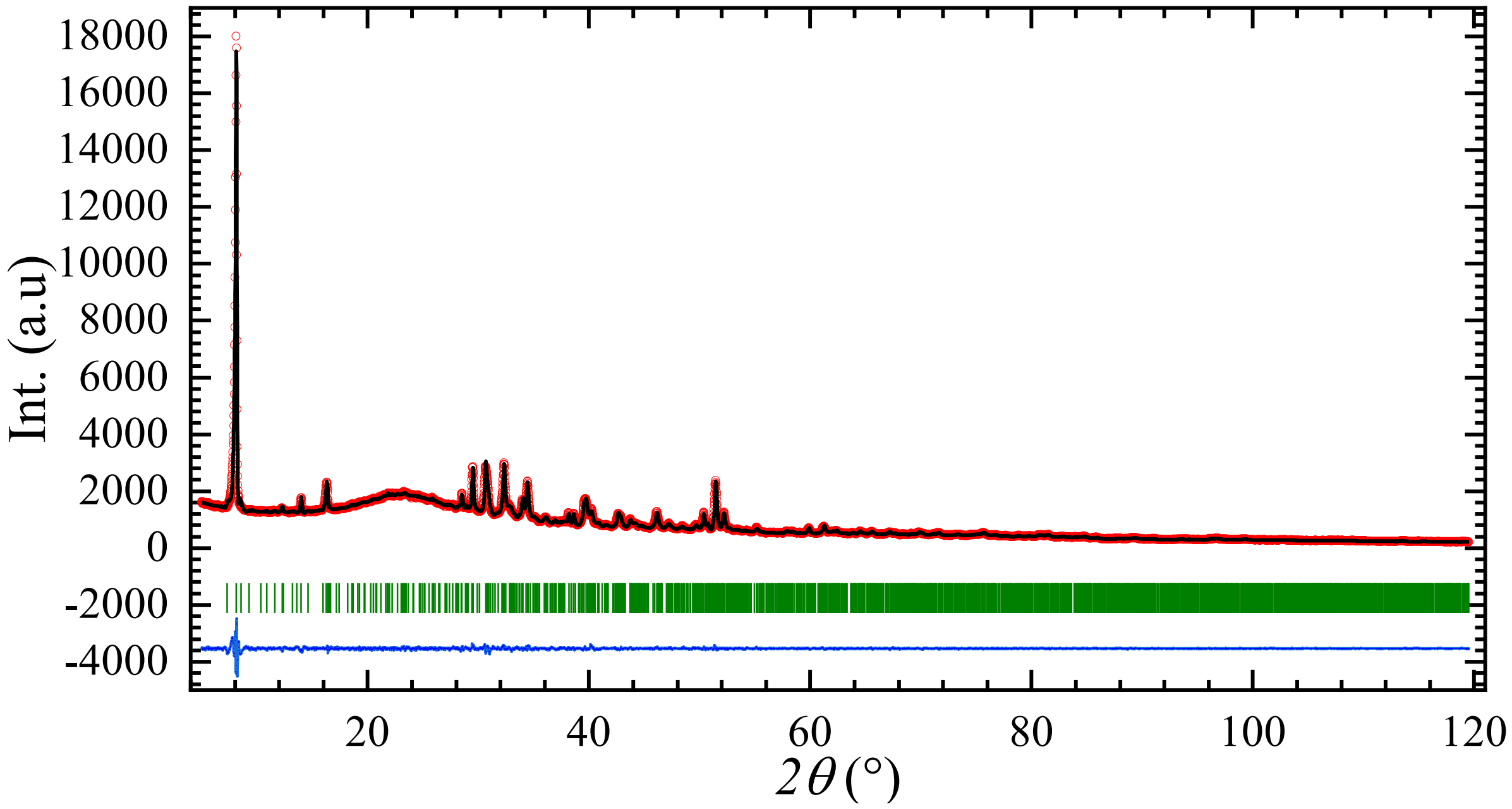

Figure S7. Rietveld refinement of $WTe_2Br_{1.25}$ using SXRD results for lattice and position parameters. The refinement shows a good agreement of PXRD and SXRD results.

## Specialized Transmission Sample Holder

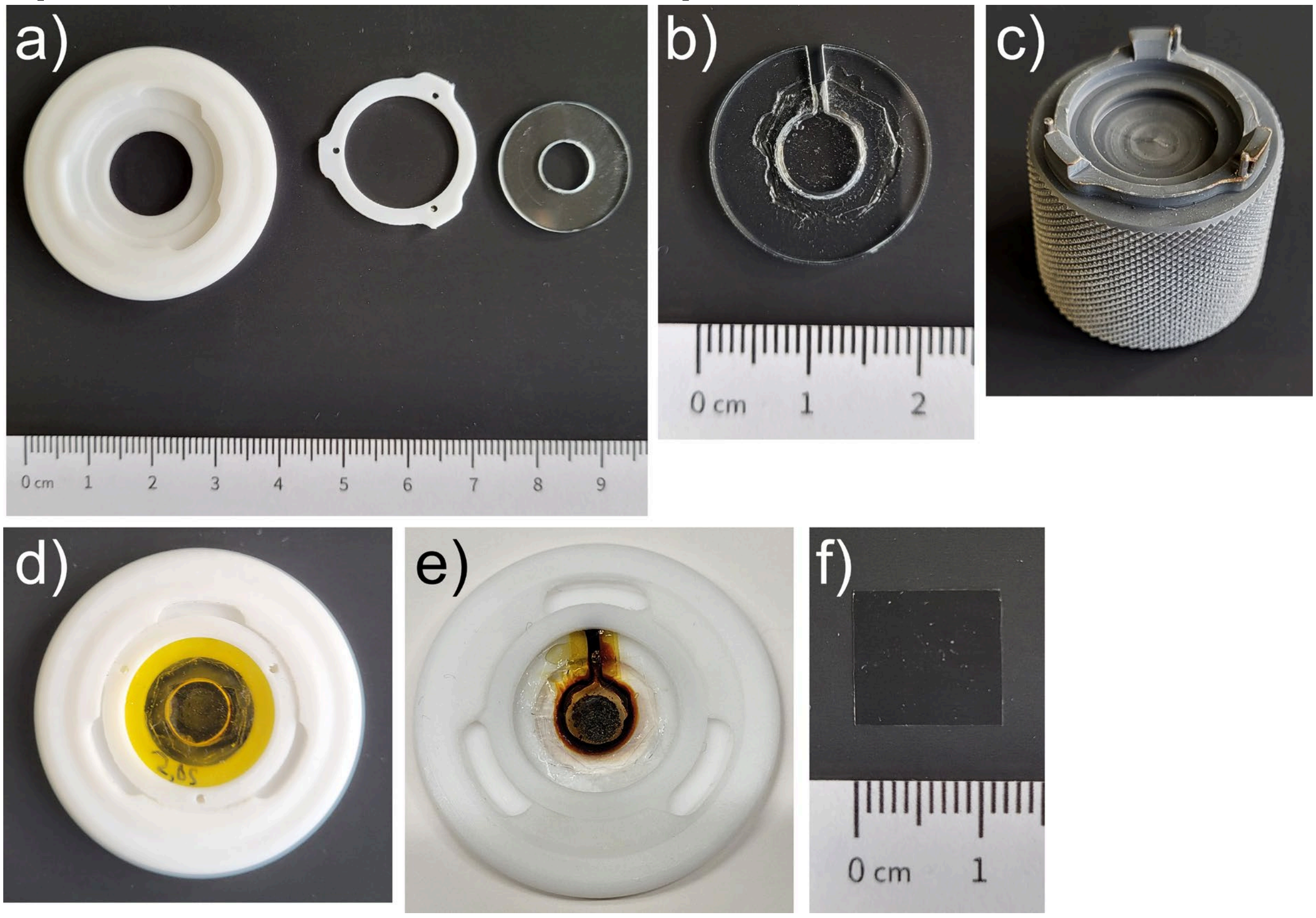

Figure S8. Components and assembly of the redesigned transmission sample holder. (a) PTFE rotor disc, PTFE counter disc, and glass disc. (b) Glass disc with an insertion slit and an ultrathin-glass window bonded to one side. (c) Assembly aid. (d) Transmission holder equipped with an assembled rotor disc (ultrathin glass on one side and Kapton on the other). (e) Examples of assembled rotor discs prepared for bromine-containing samples; for experiments with excess bromine, both sides were sealed with ultrathin glass and the perimeter was sealed with wax and cyanoacrylate under argon. (f) Ultrathin-glass window (Schott AF 32®).

## Exploded view drawing

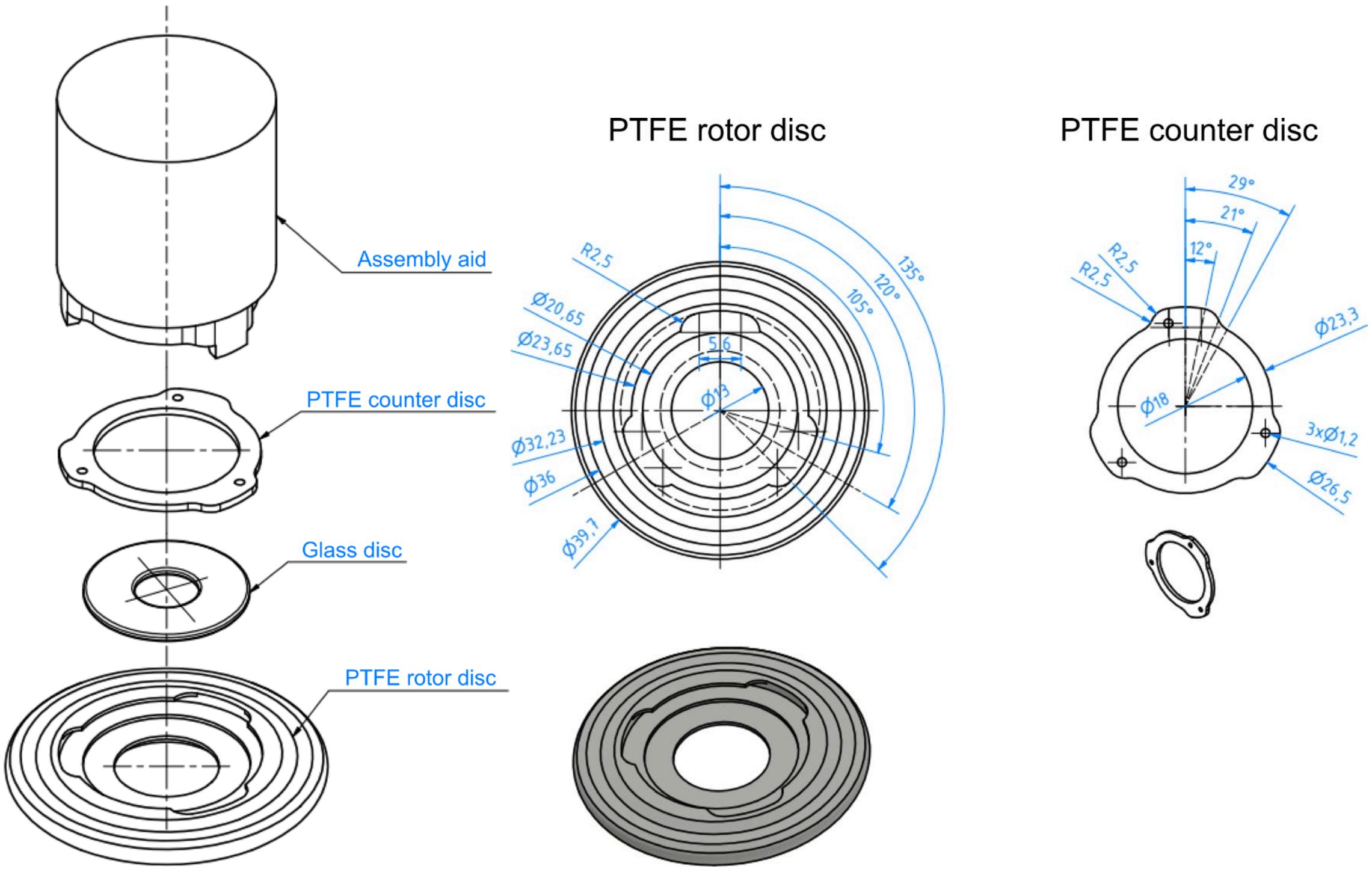

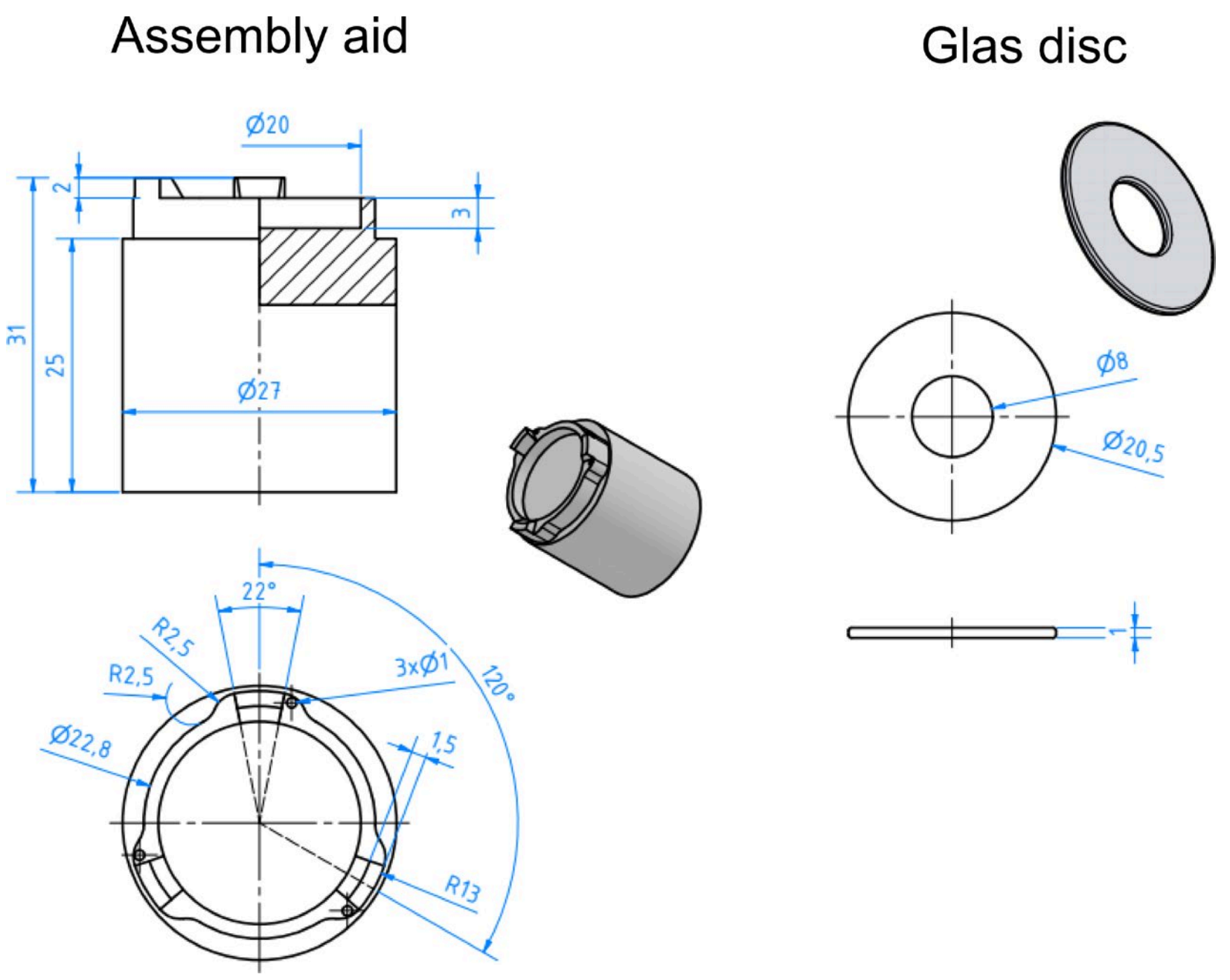

Figure S9. Engineering drawing of the redesigned transmission holder components (PTFE rotor disc, PTFE counter disc, glass disc, and assembly aid), including an exploded view and key dimensions (mm).

# Crystallographic details for $WTe_2I$ (supercell)

**Table S1**: Fractional Atomic Coordinates (×10[4]) and Equivalent Isotropic Displacement Parameters (Å[2]×10[3]) for **$WTe_2I$ (supercell)**. $U_{eq}$ is defined as 1/3 of the trace of the orthogonalised $U_{ij}$.

| Atom | x | y | z | $U_{eq}$ |
|---|---|---|---|---|
| W1 | 10340.8(7) | 10074.9(3) | 11376.4(7) | 7.57(13) |
| W2 | 12243.8(8) | 10053.4(4) | 9537.5(8) | 8.11(14) |
| W3 | 18784.5(7) | 10050.7(4) | 13031.9(8) | 7.03(13) |
| W4 | 16172.5(7) | 9956.1(4) | 12038.8(8) | 7.31(13) |
| W5 | 17157.2(7) | 10038.2(4) | 14661.4(8) | 7.08(13) |
| W6 | 14525.4(6) | 9978.7(3) | 13694.4(6) | 6.94(11) |
| Te1 | 9174.9(11) | 9090.6(5) | 10028.2(11) | 8.51(18) |
| Te2 | 12484.8(13) | 10971.8(6) | 8234.0(12) | 9.0(2) |
| Te3 | 15801.4(13) | 9057.9(6) | 13340.0(12) | 8.5(2) |
| Te4 | 14134.7(11) | 9056.6(5) | 14952.2(11) | 7.95(18) |
| Te5 | 17459.0(13) | 10958.6(6) | 13354.1(12) | 8.2(2) |
| Te6 | 9010.4(12) | 10972.9(6) | 11644.3(12) | 7.4(2) |
| Te7 | 10862.9(11) | 9332.5(6) | 13394.7(12) | 7.9(2) |
| Te8 | 20758.1(12) | 10691.9(6) | 14936.8(12) | 7.6(2) |
| Te9 | 17602.7(11) | 9298.0(5) | 16706.1(11) | 7.33(18) |
| Te10 | 14124.3(12) | 10697.7(6) | 11583.2(12) | 8.0(2) |
| Te11 | 14260.3(12) | 9310.0(6) | 10046.9(12) | 7.9(2) |
| Te12 | 12502.9(11) | 9513.6(5) | 11682.9(11) | 8.48(17) |
| I1 | 12390.5(13) | 8251.2(5) | 11550.8(13) | 17.6(2) |
| I2A | 17522(5) | 7500 | 13841(6) | 52.0(13) |
| I3A | 15352(3) | 7500 | 11151(4) | 31.9(8) |
| I4A | 13166(6) | 7500 | 8826(6) | 36.5(11) |
| I5A | 10319(5) | 7500 | 6530(4) | 40.1(10) |
| I6A | 15867(6) | 7500 | 15192(6) | 58.0(14) |
| I7A | 9118(7) | 7500 | 8110(8) | 73.7(18) |
| I8A | 14665(5) | 7500 | 16743(4) | 39.4(10) |
| I9A | 11810(6) | 7500 | 14423(6) | 40.6(12) |
| I10A | 9570(4) | 7500 | 12122(4) | 34.9(8) |
| I11A | 7449(5) | 7500 | 9458(6) | 56.5(14) |
| I2B | 13600(20) | 12500 | 10224(16) | 58(5) |
| I4B | 14321(17) | 12500 | 13021(15) | 44(4) |
| I3B | 11067(17) | 12500 | 8914(17) | 44(4) |
| I5B | 10720(20) | 12500 | 11142(12) | 50(5) |
| I6B | 14478(12) | 12500 | 15563(13) | 29(3) |
| I7B | 10810(20) | 12500 | 13717(12) | 50(5) |
| I8B | 14046(17) | 12500 | 17773(14) | 38(3) |
| I9B | 11430(20) | 12500 | 16481(16) | 51(4) |
| I10B | 12130(30) | 7500 | 14788(16) | 50(8) |
| I11B | 12890(30) | 7500 | 8436(15) | 43(7) |

**Table S2**: Anisotropic Displacement Parameters (×10$^4$) for **$WTe_2I$ (supercell)**. The anisotropic displacement factor exponent takes the form: $-2\pi^2[h^2a^{*2} \times U_{11} + \ldots + 2hka^* \times b^* \times U_{12}]$

| Atom | $U_{11}$ | $U_{22}$ | $U_{33}$ | $U_{23}$ | $U_{13}$ | $U_{12}$ |
|---|---|---|---|---|---|---|
| W1 | 5.7(3) | 10.2(3) | 6.0(3) | -0.1(2) | 2.0(2) | -0.5(2) |
| W2 | 6.4(3) | 10.8(3) | 6.0(4) | 0.4(3) | 1.9(3) | -1.0(3) |
| W3 | 4.9(3) | 10.9(3) | 5.3(3) | 0.4(3) | 2.3(2) | -1.1(3) |
| W4 | 5.1(3) | 11.3(3) | 5.7(4) | 0.5(3) | 2.7(2) | -1.2(3) |
| W5 | 4.8(3) | 11.1(3) | 5.3(4) | 0.8(3) | 2.3(2) | -1.3(3) |
| W6 | 4.7(4) | 10.6(3) | 5.4(4) | 0.9(2) | 2.3(2) | -1.4(2) |
| Te1 | 9.9(6) | 11.5(4) | 3.9(5) | 1.0(4) | 3.1(3) | -2.5(4) |
| Te2 | 10.5(6) | 11.3(6) | 6.0(6) | 0.5(4) | 4.5(5) | -1.7(4) |
| Te3 | 11.2(6) | 11.0(5) | 3.9(5) | 1.0(4) | 4.0(4) | -1.8(4) |
| Te4 | 11.0(6) | 10.6(4) | 3.3(5) | 1.0(4) | 4.3(3) | -1.7(4) |
| Te5 | 10.5(6) | 10.9(5) | 4.1(5) | 1.0(4) | 4.1(5) | -1.6(4) |
| Te6 | 8.7(5) | 10.2(5) | 5.0(5) | 0.4(4) | 4.6(4) | -1.0(4) |
| Te7 | 3.5(5) | 13.3(5) | 6.6(5) | 0.2(4) | 2.2(4) | -1.9(4) |
| Te8 | 3.2(5) | 12.1(5) | 7.8(6) | 0.5(4) | 2.8(4) | -1.7(4) |
| Te9 | 3.4(5) | 11.4(4) | 7.4(5) | 0.9(4) | 2.7(3) | -1.3(4) |
| Te10 | 3.7(5) | 11.8(5) | 8.4(6) | 0.9(4) | 2.8(4) | -1.4(4) |
| Te11 | 3.5(5) | 12.6(5) | 8.1(6) | 1.3(5) | 3.0(4) | -0.7(4) |
| Te12 | 4.2(5) | 14.5(4) | 7.7(5) | 0.0(4) | 3.6(3) | -1.5(4) |
| I1 | 20.0(6) | 12.4(5) | 20.0(7) | -1.0(5) | 9.0(4) | -0.2(4) |
| I2A | 62(3) | 26.7(18) | 92(4) | 0 | 57(3) | 0 |
| I3A | 39.1(19) | 16.2(13) | 49(2) | 0 | 27.0(17) | 0 |
| I4A | 49(3) | 21.2(17) | 44(3) | 0 | 25(3) | 0 |
| I5A | 63(3) | 17.9(14) | 49(2) | 0 | 34(2) | 0 |
| I6A | 91(4) | 17.1(14) | 79(4) | 0 | 50(3) | 0 |
| I7A | 100(5) | 16.6(16) | 110(6) | 0 | 53(4) | 0 |
| I8A | 65(3) | 19.7(14) | 49(2) | 0 | 40(2) | 0 |
| I9A | 53(3) | 22.1(18) | 52(4) | 0 | 29(3) | 0 |
| I10A | 46(2) | 17.4(13) | 54(2) | 0 | 34.1(18) | 0 |
| I11A | 64(3) | 29.3(19) | 99(4) | 0 | 58(3) | 0 |
| I2B | 91(15) | 53(10) | 29(8) | 0 | 26(9) | 0 |
| I4B | 55(10) | 29(6) | 34(7) | 0 | 9(7) | 0 |
| I3B | 56(10) | 27(7) | 51(10) | 0 | 27(8) | 0 |
| I5B | 97(13) | 11(5) | 10(5) | 0 | -2(7) | 0 |
| I6B | 31(3) | 25(3) | 26(3) | 0 | 10(2) | 0 |
| I7B | 85(14) | 37(7) | 10(5) | 0 | 6(6) | 0 |
| I8B | 67(10) | 18(5) | 33(7) | 0 | 27(7) | 0 |
| I9B | 86(14) | 43(8) | 39(8) | 0 | 42(9) | 0 |
| I10B | 73(16) | 26(8) | 7(6) | 0 | -19(7) | 0 |
| I11B | 70(15) | 19(7) | 5(7) | 0 | -13(7) | 0 |

**Table S3**: Bond Lengths in Å for **$WTe_2I$ (supercell)**.

| Atom | Atom | Length / Å |
|---|---|---|
| W1 | W1[1] | 3.0734(14) |
| W1 | W2[1] | 2.8219(10) |
| W1 | Te1 | 2.6803(13) |
| W1 | Te1[1] | 2.7458(15) |
| W1 | Te6 | 2.6604(15) |
| W1 | Te7 | 2.7622(14) |
| W1 | Te9[2] | 2.8591(13) |
| W1 | Te12 | 2.7676(14) |
| W2 | W3[3] | 2.7874(10) |
| W2 | Te1[1] | 2.7594(16) |
| W2 | Te2 | 2.6596(15) |
| W2 | Te6[1] | 2.7211(16) |
| W2 | Te10 | 2.8528(16) |
| W2 | Te11 | 2.7641(15) |
| W2 | Te12 | 2.7478(15) |
| W3 | W4 | 2.8361(11) |
| W3 | Te2[3] | 2.7492(16) |
| W3 | Te5 | 2.6997(16) |
| W3 | Te6[4] | 2.7259(15) |
| W3 | Te7[4] | 2.8329(15) |
| W3 | Te8[5] | 2.7905(15) |
| W3 | Te8 | 2.8190(15) |
| W4 | W5 | 2.8479(10) |
| W4 | Te2[3] | 2.7213(16) |
| W4 | Te3 | 2.6904(15) |
| W4 | Te5 | 2.7388(16) |
| W4 | Te10 | 2.8055(15) |
| W4 | Te11 | 2.8387(16) |
| W4 | Te11[3] | 2.8393(15) |
| W5 | W6 | 2.8572(11) |
| W5 | Te3 | 2.7292(16) |

| Atom | Atom | Length / Å |
|---|---|---|
| W5 | Te4[2] | 2.7022(16) |
| W5 | Te5 | 2.6953(15) |
| W5 | Te7[2] | 2.8304(15) |
| W5 | Te8[5] | 2.8419(15) |
| W5 | Te9 | 2.8035(14) |
| W6 | W6[2] | 2.8355(14) |
| W6 | Te3 | 2.6978(16) |
| W6 | Te4 | 2.7049(14) |
| W6 | Te4[2] | 2.7085(13) |
| W6 | Te9[2] | 2.8771(14) |
| W6 | Te10 | 2.8554(14) |
| W6 | Te12 | 2.7383(12) |
| Te12 | I1 | 2.7663(15) |
| I2A | I3A | 3.111(8) |
| I2A | I6A | 3.136(8) |
| I3A | I4A | 2.846(7) |
| I5A | I7A | 2.906(9) |
| I6A | I8A | 2.875(7) |
| I7A | I11A | 3.148(10) |
| I9A | I10A | 2.863(8) |
| I10A | I11A | 3.068(8) |
| I2B | I4B | 3.10(2) |
| I2B | I3B | 2.73(3) |
| I4B | I6B | 3.01(2) |
| I3B | I5B | 2.93(3) |
| I5B | I7B | 3.09(2) |
| I6B | I8B | 2.97(2) |
| I7B | I9B | 3.09(2) |
| I8B | I9B | 2.82(3) |

——

[1]2-x,2-y,2-z; [2]3-x,2-y,3-z; [3]3-x,2-y,2-z; [4]1+x,+y,+z; [5]4-x,2-y,3-z

Table S4: Atomic Occupancies for all atoms that are not fully occupied in **$WTe_2I$ (supercell)**.

| Atom | Occupancy |
|---|---|
| I2A | 0.803(4) |
| I3A | 0.803(4) |
| I4A | 0.803(4) |
| I5A | 0.803(4) |
| I6A | 0.803(4) |
| I7A | 0.803(4) |
| I8A | 0.803(4) |
| I9A | 0.803(4) |
| I10A | 0.803(4) |
| I11A | 0.803(4) |

| Atom | Occupancy |
|---|---|
| I2B | 0.197(4) |
| I4B | 0.197(4) |
| I3B | 0.197(4) |
| I5B | 0.197(4) |
| I6B | 0.197(4) |
| I7B | 0.197(4) |
| I8B | 0.197(4) |
| I9B | 0.197(4) |
| I10B | 0.197(4) |
| I11B | 0.197(4) |

# Crystallographic details for incommensurate $WTe_2I$

Table S5: Occupational waves for incommensurate $WTe_2I$.

| Atom | Wave/Parameter | Occ |
|---|---|---|
| Te2 | delta | 0.8318(8) |
| | x40 | 0.2710(4) |
| Te3 | delta | 0.1682(8) |
| | x40 | 0.7754(5) |
| I1 | delta | 0.8318(8) |
| | x40 | 0.3929(4) |
| I2 | delta | 0.8318(8) |
| | x40 | 0.1633(4) |
| I3 | delta | 0.1682(8) |
| | x40 | 0.7865(4) |

Table S6: Positional parameters for incommensurate $WTe_2I$.

| Atom | Occ | Wave | x | y | z | Ueq/Uiso |
|---|---|---|---|---|---|---|
| W1 | 1 | | 0.32142(3) | 0.495374(9) | 0.25174(13) | 0.00479(7) |
| | | s,1 | -0.00194(5) | 0.001783(16) | -0.0072(12) | |
| | | c,1 | 0.00190(5) | -0.001740(16) | -0.0085(10) | |
| | | s,2 | -0.00203(5) | -0.000332(16) | 0.0046(2) | |
| | | c,2 | 0.00211(5) | -0.000476(16) | 0.0091(8) | |
| | | s,3 | -0.0010(3) | 0.00067(8) | -0.0068(9) | |
| | | c,3 | 0.0032(2) | -0.00036(7) | 0.0057(11) | |
| | | s,4 | 0 | 0 | -0.067(2) | |
| | | c,4 | 0.3826(5) | 1 | 0 | |
| Te1 | 1 | | 0.57919(5) | 0.595074(13) | 0.24836(18) | 0.00651(7) |
| | | s,1 | -0.00264(8) | 0.00038(3) | 0.02596(15) | |
| | | c,1 | -0.01089(8) | -0.00028(3) | -0.01238(15) | |
| | | s,2 | 0.00422(8) | -0.00187(3) | 0.00920(15) | |
| | | c,2 | 0.00324(8) | -0.00144(3) | -0.00400(16) | |
| Te2 | 0.8318(8) | | 0.07022(7) | 0.56927(2) | 0.7492(2) | 0.00552(9) |
| | | o,1 | -0.00111(10) | 0.00106(3) | 0.0018(2) | |
| | | o,2 | 0.00011(6) | -0.00041(2) | -0.00017(12) | |
| | | o,3 | -0.00154(8) | 0.00046(3) | -0.01128(14) | |
| | | o,4 | 0.00000(13) | -0.00028(4) | 0.0047(3) | |
| Te3 | 0.1682(8) | | 0.0817(4) | 0.54812(10) | 0.7526(8) | 0.0056(4) |
| | | o,1 | -0.0013(5) | -0.00001(17) | -0.0011(13) | |
| | | o,2 | -0.001(3) | 0.0001(8) | 0.006(6) | |
| I1 | 0.8318(8) | | 0.34767(14) | 0.75 | 0.7591(4) | 0.0218(3) |
| | | o,1 | 0.02889(15) | 0 | 0.0932(3) | |
| | | o,2 | -0.01800(15) | 0 | -0.0089(4) | |
| | | o,3 | 0.0001(2) | 0 | -0.0335(4) | |
| | | o,4 | 0.00596(17) | 0 | -0.0221(4) | |
| | | o,5 | 0.0008(3) | 0 | 0.0010(5) | |
| | | o,6 | -0.0026(2) | 0 | -0.0113(5) | |
| I2 | 0.8318(8) | | -0.17524(15) | 0.75 | 0.7370(4) | 0.0254(3) |
| | | o,1 | -0.03392(15) | 0 | -0.0879(3) | |
| | | o,2 | -0.00962(16) | 0 | 0.0148(4) | |
| | | o,3 | -0.00434(19) | 0 | -0.0074(5) | |
| | | o,4 | -0.0032(2) | 0 | 0.0410(5) | |
| | | o,5 | -0.0020(2) | 0 | -0.0104(6) | |
| | | o,6 | -0.0011(3) | 0 | -0.0055(5) | |
| I3 | 0.1682(8) | | 0.1052(2) | 0.67498(7) | 0.7481(5) | 0.0151(3) |
| | | o,1 | -0.0031(8) | -0.0001(2) | 0.0282(15) | |

| | | | | |
|---|---|---|---|---|
| | o,2 | -0.004(4) | 0.0007(12) | -0.004(7) |

Table S7: ADP harmonic parameters for incommensurate $WTe_2I$.

| Atom | Wave | U11 | U22 | U33 | U12 | U13 | U23 |
|---|---|---|---|---|---|---|---|
| W1 | | 0.00241(9) | 0.00714(9) | 0.00482(17) | -0.00014(6) | -0.00143(14) | 0.00020(12) |
| | s,1 | -0.00019(12) | 0.00021(13) | -0.0004(2) | 0.00008(11) | 0.00026(12) | -0.00021(13) |
| | c,1 | 0.00005(12) | -0.00017(14) | 0.00099(17) | -0.00011(11) | -0.00007(13) | -0.00023(13) |
| Te1 | | 0.00502(12) | 0.00856(13) | 0.00596(13) | -0.00101(10) | -0.0012(2) | -0.0002(2) |
| | s,1 | 0.0010(2) | 0.0013(2) | 0.0005(2) | -0.00105(18) | -0.00049(16) | 0.00068(17) |
| | c,1 | 0.0012(2) | 0.0029(2) | 0.0015(2) | -0.00243(17) | 0.00108(17) | -0.00133(17) |
| Te2 | | 0.00326(14) | 0.00726(16) | 0.00603(15) | 0.00004(13) | -0.0015(3) | -0.0002(3) |
| | o,1 | -0.00010(18) | -0.0005(2) | 0.0018(2) | 0.00021(17) | 0.00013(15) | 0.00003(19) |
| | o,2 | -0.00001(16) | 0.00011(18) | -0.00068(18) | 0.00009(14) | -0.00018(13) | -0.00017(14) |
| Te3 | | 0.0031(4) | 0.0086(6) | 0.0050(9) | -0.0011(5) | -0.0010(5) | 0.0001(6) |
| I1 | | 0.0176(4) | 0.0156(3) | 0.0323(6) | 0 | -0.0053(4) | 0 |
| | o,1 | 0.0024(4) | -0.0014(4) | 0.0101(6) | 0 | -0.0048(4) | 0 |
| | o,2 | 0.0025(5) | -0.0015(4) | -0.0115(7) | 0 | 0.0078(5) | 0 |
| I2 | | 0.0183(4) | 0.0158(3) | 0.0419(7) | 0 | -0.0045(4) | 0 |
| | o,1 | 0.0024(4) | -0.0014(4) | 0.0123(7) | 0 | -0.0090(4) | 0 |
| | o,2 | -0.0010(5) | 0.0016(4) | 0.0149(6) | 0 | -0.0041(5) | 0 |
| I3 | | 0.0174(6) | 0.0089(5) | 0.0191(7) | -0.0002(4) | -0.0017(5) | 0.0005(5) |

Table S8: List of used orthogonalized functions for incommensurate $WTe_2I$.

| | | | | | | | | | | | | | | | | | | | | | | | | | | | | | | | | | | | | | | | |
|---|---|---|---|---|---|---|---|---|---|---|---|---|---|---|---|---|---|---|---|---|---|---|---|---|---|---|---|---|---|---|---|---|---|---|---|---|---|---|---|
| Te2 | o,1 | - | -0.309 | - | - | - | - | - | - | - | - | - | - | - | - | - | - | - | - | - | 1.616 | | | | | | | | | | | | | | | | | | |
| | o,2 | - | 0.051 | - | - | - | - | - | - | - | - | - | - | - | - | - | - | - | - | - | -0.091 | - | - | - | - | - | - | - | - | 1.315 | | | | | | | | | |
| | o,3 | - | -0.095 | - | - | - | - | - | - | - | - | - | - | - | - | - | - | - | - | - | 0.171 | - | - | - | - | - | - | - | - | -0.098 | - | - | - | - | - | - | - | 1.38 | |
| | o,4 | - | -0.448 | - | - | - | - | - | - | - | - | - | - | - | - | - | - | - | - | - | 0.853 | - | - | - | - | - | - | - | - | -0.097 | - | - | - | - | - | - | - | 0.168 | 1.708 |
| Te3 | o,1 | - | - | - | - | - | - | - | - | - | - | -0.518 | - | - | - | - | - | - | - | - | 3.414 | | | | | | | | | | | | | | | | | | |
| | o,2 | - | - | - | - | - | - | - | - | - | - | -2.531 | - | - | - | - | - | - | - | - | 3.806 | - | - | - | - | - | - | - | 3.49 | | | | | | | | | | |
| I1 | o,1 | - | -0.169 | - | - | - | - | - | - | - | - | 1.408 | | | | | | | | | | | | | | | | | | | | | | | | | | | |
| | o,2 | - | 0.263 | - | - | - | - | - | - | - | - | -0.297 | - | - | - | - | - | - | - | 1.509 | | | | | | | | | | | | | | | | | | | |
| | o,3 | - | -0.439 | - | - | - | - | - | - | - | - | 0.542 | - | - | - | - | - | - | - | -0.646 | 1.705 | | | | | | | | | | | | | | | | | | |
| | o,4 | - | 0.128 | - | - | - | - | - | - | - | - | -0.095 | - | - | - | - | - | - | - | 0.237 | -0.179 | 1.383 | | | | | | | | | | | | | | | | | |
| | o,5 | - | -0.583 | - | - | - | - | - | - | - | - | 0.652 | - | - | - | - | - | - | - | -0.916 | 0.947 | -0.332 | 1.745 | | | | | | | | | | | | | | | | |
| | o,6 | - | -0.361 | - | - | - | - | - | - | - | - | 0.51 | - | - | - | - | - | - | - | -0.481 | 0.642 | 0.04 | 0.348 | 1.523 | | | | | | | | | | | | | | | |
| I2 | o,1 | - | -0.25 | - | - | - | - | - | - | - | 1.518 | | | | | | | | | | | | | | | | | | | | | | | | | | | | |
| | o,2 | - | -0.188 | - | - | - | - | - | - | - | 0.291 | 1.401 | | | | | | | | | | | | | | | | | | | | | | | | | | | |
| | o,3 | - | 0.381 | - | - | - | - | - | - | - | -0.603 | -0.415 | 1.627 | | | | | | | | | | | | | | | | | | | | | | | | | | |
| | o,4 | - | -0.253 | - | - | - | - | - | - | - | 0.448 | 0.197 | -0.323 | 1.449 | | | | | | | | | | | | | | | | | | | | | | | | | |
| | o,5 | - | -0.034 | - | - | - | - | - | - | - | -0.005 | 0.133 | -0.15 | -0.163 | 1.439 | | | | | | | | | | | | | | | | | | | | | | | | |
| | o,6 | - | 0.685 | - | - | - | - | - | - | - | -1.13 | -0.678 | 1.033 | -0.552 | -0.046 | 1.848 | | | | | | | | | | | | | | | | | | | | | | | |

Table S9: List of distances for incommensurate $WTe_2I$.

| | average / Å | min / Å | max / Å |
|---|---|---|---|
| $W1-W1^{i}$ | 2.870(11) | 2.781(16) | 3.097(16) |
| $W1-W1^{ii}$ | 2.846(11) | 2.830(14) | 2.862(14) |
| W1-Te1 | 2.720(3) | 2.684(3) | 2.755(3) |
| $W1-Te1^{i}$ | 2.706(8) | 2.555(12) | 2.856(12) |
| $W1-Te1^{ii}$ | 2.702(8) | 2.543(12) | 2.851(12) |
| $W1-Te2^{iii}$ | 2.820(7) | 2.749(9) | 2.905(9) |
| W1-Te2 | 2.819(7) | 2.739(9) | 2.889(9) |
| $W1-Te2^{iv}$ | 2.845(3) | 2.813(3) | 2.877(3) |
| $W1-Te3^{iii}$ | 2.76(4) | 2.73(7) | 2.78(7) |
| W1-Te3 | 2.73(4) | 2.45(7) | 2.78(7) |
| $W1-Te3^{iv}$ | 2.74(3) | 2.73(6) | 2.74(6) |
| Te3-I3 | 2.77(3) | 2.76(6) | 2.79(6) |
| $I1-I1^{iii}$ | 3.415(6) | 2.769(8) | 4.018(8) |
| $I1-I1^{v}$ | 3.405(6) | 2.769(8) | 4.018(8) |
| I1-I2 | 3.341(6) | 2.688(8) | 3.933(8) |
| $I1-I2^{vi}$ | 3.189(6) | 2.887(8) | 3.702(8) |
| $I1-I2^{vii}$ | 3.341(6) | 2.688(8) | 3.933(8) |
| $I1-I2^{viii}$ | 3.189(6) | 2.887(8) | 3.702(8) |
| $I1-I3^{iii}$ | 4.084(9) | 4.011(15) | 4.255(15) |
| $I1-I3^{v}$ | 4.297(9) | 4.140(13) | 4.407(13) |
| $I1-I3^{ix}$ | 4.084(9) | 4.011(15) | 4.255(15) |
| $I1-I3^{x}$ | 4.297(9) | 4.140(13) | 4.407(13) |
| $I2-I2^{iii}$ | 3.419(7) | 2.765(9) | 4.028(9) |
| $I2-I2^{v}$ | 3.416(7) | 2.765(9) | 4.029(9) |
| $I2-I2^{ix}$ | 3.419(7) | 2.765(9) | 4.028(9) |
| $I2-I2^{vii}$ | 0 | 0 | 0 |
| $I2-I2^{x}$ | 3.416(7) | 2.765(9) | 4.029(9) |
| $I2-I3^{v}$ | 4.123(8) | 4.064(14) | 4.250(14) |
| $I2-I3^{x}$ | 4.123(8) | 4.064(14) | 4.250(14) |
| $I3-I3^{iii}$ | 3.659(15) | 3.659(15) | 3.659(15) |
| $I3-I3^{v}$ | 3.658(15) | 3.658(15) | 3.658(15) |
| $I3-I3^{vii}$ | 3.272(11) | 3.24(2) | 3.29(2) |

(i) -x+1,-y+1,-z
(ii) -x+1,-y+1,-z+1
(iii) x,y,z-1
(iv) -x,-y+1,-z+1
(v) x,y,z+1
(vi) x+1,y,z
(vii) x,-y+3/2,z
(viii) x+1,-y+3/2,z
(ix) x,-y+3/2,z-1
(x) x,-y+3/2,z+1
(xi) x-1,y,z

Table S10: List of angles for incommensurate $WTe_2I$.

| | average / ° | min / ° | max / ° |
|---|---|---|---|
| $W1^{i}$-W1-$W1^{ii}$ | 74.7(3) | 70.4(3) | 83.2(3) |
| $W1^{i}$-W1-Te1 | 57.78(14) | 55.55(17) | 59.50(17) |
| $W1^{i}$-W1-$Te1^{i}$ | 58.3(2) | 53.7(2) | 60.5(3) |
| $W1^{i}$-W1-$Te1^{ii}$ | 104.93(15) | 102.07(12) | 110.76(16) |
| $W1^{i}$-W1-$Te2^{iii}$ | 93.3(3) | 90.1(4) | 96.5(4) |
| $W1^{i}$-W1-Te2 | 140.27(13) | 137.64(16) | 143.27(12) |
| $W1^{i}$-W1-$Te2^{iv}$ | 136.1(4) | 127.6(5) | 142.3(5) |
| $W1^{i}$-W1-$Te3^{iii}$ | 82.2(14) | 81.4(14) | 89.2(14) |
| $W1^{i}$-W1-Te3 | 145.6(13) | 145.2(14) | 146.3(13) |
| $W1^{i}$-W1-$Te3^{iv}$ | 139.1(15) | 138.6(15) | 140.2(15) |
| $W1^{ii}$-W1-Te1 | 58.04(14) | 54.70(18) | 61.91(19) |
| $W1^{ii}$-W1-$Te1^{i}$ | 104.74(15) | 100.93(13) | 110.0(2) |
| $W1^{ii}$-W1-$Te1^{ii}$ | 58.6(2) | 56.8(3) | 59.9(3) |
| $W1^{ii}$-W1-$Te2^{iii}$ | 140.51(12) | 137.11(13) | 144.01(11) |
| $W1^{ii}$-W1-Te2 | 92.9(3) | 90.2(4) | 94.1(3) |
| $W1^{ii}$-W1-$Te2^{iv}$ | 137.0(4) | 133.2(6) | 141.6(5) |
| $W1^{ii}$-W1-$Te3^{iii}$ | 145.8(12) | 145.2(12) | 146.6(13) |
| $W1^{ii}$-W1-Te3 | 87.4(14) | 86.4(14) | 93.9(16) |
| $W1^{ii}$-W1-$Te3^{iv}$ | 138.3(15) | 138.0(15) | 138.7(15) |
| Te1-W1-$Te1^{i}$ | 116.1(3) | 109.3(4) | 119.9(3) |
| Te1-W1-$Te1^{ii}$ | 116.6(3) | 111.1(4) | 121.3(3) |
| Te1-W1-$Te2^{iii}$ | 83.74(15) | 81.31(14) | 86.70(13) |
| Te1-W1-Te2 | 83.82(14) | 81.73(17) | 86.06(14) |
| Te1-W1-$Te2^{iv}$ | 155.34(14) | 152.2(3) | 159.93(11) |
| Te1-W1-$Te3^{iii}$ | 85.2(12) | 83.5(12) | 87.7(13) |
| Te1-W1-Te3 | 87.0(13) | 85.8(13) | 91.8(14) |
| Te1-W1-$Te3^{iv}$ | 151.2(13) | 149.5(13) | 152.0(13) |
| $Te1^{i}$-W1-$Te1^{ii}$ | 79.86(7) | 77.14(7) | 84.58(9) |
| $Te1^{i}$-W1-$Te2^{iii}$ | 99.9(3) | 95.4(4) | 104.4(5) |
| $Te1^{i}$-W1-Te2 | 158.24(13) | 156.7(3) | 160.69(10) |
| $Te1^{i}$-W1-$Te2^{iv}$ | 81.23(16) | 79.4(2) | 84.66(17) |
| $Te1^{i}$-W1-$Te3^{iii}$ | 85.8(14) | 84.0(14) | 88.1(14) |
| $Te1^{i}$-W1-Te3 | 153.9(13) | 153.6(14) | 154.3(12) |
| $Te1^{i}$-W1-$Te3^{iv}$ | 85.3(14) | 84.7(14) | 86.3(14) |
| $Te1^{ii}$-W1-$Te2^{iii}$ | 158.39(13) | 156.05(10) | 159.76(11) |
| $Te1^{ii}$-W1-Te2 | 100.1(3) | 96.8(4) | 103.8(5) |
| $Te1^{ii}$-W1-$Te2^{iv}$ | 81.39(16) | 78.55(14) | 85.19(17) |
| $Te1^{ii}$-W1-$Te3^{iii}$ | 154.5(12) | 154.2(12) | 154.9(12) |
| $Te1^{ii}$-W1-Te3 | 86.4(14) | 84.5(14) | 98.7(16) |
| $Te1^{ii}$-W1-$Te3^{iv}$ | 85.2(14) | 84.6(14) | 86.1(14) |
| $Te2^{iii}$-W1-Te2 | 75.02(8) | 72.66(7) | 77.19(8) |
| $Te2^{iii}$-W1-$Te2^{iv}$ | 78.53(13) | 76.69(13) | 80.30(15) |
| $Te2^{iii}$-W1-Te3 | 80.7(14) | 79.8(14) | 81.2(13) |
| $Te2^{iii}$-W1-$Te3^{iv}$ | 71.7(14) | 71.4(14) | 72.2(14) |
| Te2-W1-$Te2^{iv}$ | 78.43(13) | 76.87(16) | 79.87(15) |
| Te2-W1-$Te3^{iii}$ | 79.0(14) | 78.4(14) | 80.0(14) |
| Te2-W1-$Te3^{iv}$ | 71.8(14) | 71.1(14) | 72.2(14) |
| $Te2^{iv}$-W1-$Te3^{iii}$ | 71.1(12) | 70.7(12) | 72.2(12) |
| $Te2^{iv}$-W1-Te3 | 71.9(13) | 70.7(12) | 74.8(14) |
| $Te3^{iii}$-W1-Te3 | 82.8(16) | 82.8(16) | 82.8(16) |
| W1-Te1-$W1^{i}$ | 63.9(2) | 61.8(3) | 69.3(3) |
| W1-Te1-$W1^{ii}$ | 63.3(2) | 61.3(3) | 65.4(3) |
| $W1^{i}$-Te1-$W1^{ii}$ | 79.8(3) | 75.3(3) | 89.7(3) |
| W1-Te2-$W1^{v}$ | 73.8(3) | 71.3(3) | 75.8(2) |
| W1-Te2-$W1^{iv}$ | 102.09(16) | 99.7(2) | 105.4(2) |
| $W1^{v}$-Te2-$W1^{iv}$ | 102.19(16) | 99.1(2) | 105.36(19) |
| W1-Te3-$W1^{v}$ | 89.9(10) | 81.8(5) | 90.6(17) |
| W1-Te3-W1$^{iv}$ | 110.9(11) | 110.1(17) | 114.4(5) |
| W1-Te3-I3 | 113.4(11) | 112(2) | 119.1(5) |
| W1$^{v}$-Te3-W1$^{iv}$ | 111.2(11) | 110.1(7) | 111.6(9) |
| W1$^{v}$-Te3-I3 | 114.2(11) | 112.3(5) | 115.4(7) |
| W1$^{iv}$-Te3-I3 | 114.7(10) | 113.6(4) | 115.3(19) |
| I1$^{iii}$-I1-I1$^{v}$ | 176.1(2) | 171.3(2) | 180 |
| I1$^{iii}$-I1-I2 | 89.3(2) | 76.22(19) | 107.0(2) |
| I1$^{iii}$-I1-I2$^{vi}$ | 89.7(2) | 82.1(2) | 97.28(17) |
| I1$^{iii}$-I1-I2$^{vii}$ | 89.3(2) | 76.22(19) | 107.0(2) |
| I1$^{iii}$-I1-I2$^{viii}$ | 89.7(2) | 82.1(2) | 97.28(17) |
| I1$^{iii}$-I1-I3$^{v}$ | 154.94(10) | 152.26(10) | 156.47(8) |
| I1$^{iii}$-I1-I3$^{x}$ | 154.94(10) | 152.26(10) | 156.47(8) |
| I1$^{v}$-I1-I2 | 91.0(2) | 81.1(2) | 100.33(17) |
| I1$^{v}$-I1-I2$^{vi}$ | 94.6(2) | 77.28(18) | 99.9(2) |
| I1$^{v}$-I1-I2$^{vii}$ | 91.0(2) | 81.1(2) | 100.33(17) |
| I1$^{v}$-I1-I2$^{viii}$ | 94.6(2) | 77.28(18) | 99.9(2) |
| I1$^{v}$-I1-I3$^{iii}$ | 149.64(11) | 148.82(13) | 151.36(8) |
| I1$^{v}$-I1-I3$^{ix}$ | 149.64(11) | 148.82(13) | 151.36(8) |
| I2-I1-I2$^{vi}$ | 167.5(2) | 155.8(2) | 180 |
| I2-I1-I2$^{vii}$ | 0 | 0 | 0 |
| I2-I1-I2$^{viii}$ | 167.5(2) | 155.8(2) | 180 |
| I2-I1-I3$^{iii}$ | 75.05(15) | 72.67(16) | 76.11(14) |
| I2-I1-I3$^{v}$ | 66.88(13) | 65.76(11) | 69.54(16) |
| I2-I1-I3$^{ix}$ | 75.05(15) | 72.67(16) | 76.11(14) |
| I2-I1-I3$^{x}$ | 66.88(13) | 65.76(11) | 69.54(16) |
| I2$^{vi}$-I1-I2$^{vii}$ | 167.5(2) | 155.8(2) | 180 |
| I2$^{vi}$-I1-I2$^{viii}$ | 0 | 0 | 0 |
| I2$^{vi}$-I1-I3$^{iii}$ | 102.09(16) | 101.80(18) | 102.33(15) |
| I2$^{vi}$-I1-I3$^{v}$ | 95.29(14) | 91.83(14) | 102.89(12) |
| I2$^{vi}$-I1-I3$^{ix}$ | 102.09(16) | 101.80(18) | 102.33(15) |
| I2$^{vi}$-I1-I3$^{x}$ | 95.29(14) | 91.83(14) | 102.89(12) |
| I2$^{vii}$-I1-I2$^{viii}$ | 167.5(2) | 155.8(2) | 180 |
| I2$^{vii}$-I1-I3$^{iii}$ | 75.05(15) | 72.67(16) | 76.11(14) |
| I2$^{vii}$-I1-I3$^{v}$ | 66.88(13) | 65.76(11) | 69.54(16) |
| I2$^{vii}$-I1-I3$^{ix}$ | 75.05(15) | 72.67(16) | 76.11(14) |
| I2$^{vii}$-I1-I3$^{x}$ | 66.88(13) | 65.76(11) | 69.54(16) |
| I2$^{viii}$-I1-I3$^{iii}$ | 102.09(16) | 101.80(18) | 102.33(15) |
| I2$^{viii}$-I1-I3$^{v}$ | 95.29(14) | 91.83(14) | 102.89(12) |
| I2$^{viii}$-I1-I3$^{ix}$ | 102.09(16) | 101.80(18) | 102.33(15) |
| I2$^{viii}$-I1-I3$^{x}$ | 95.29(14) | 91.83(14) | 102.89(12) |
| I3$^{iii}$-I1-I3$^{ix}$ | 47.28(14) | 44.9(2) | 48.29(11) |
| I3$^{v}$-I1-I3$^{x}$ | 44.80(13) | 43.19(11) | 46.3(2) |
| I1$^{xi}$-I2-I1 | 168.9(2) | 157.3(3) | 180 |
| I1$^{xi}$-I2-I2$^{iii}$ | 94.48(18) | 77.86(17) | 99.7(2) |
| I1$^{xi}$-I2-I2$^{v}$ | 89.74(19) | 82.2(2) | 97.1(2) |
| I1$^{xi}$-I2-I2$^{ix}$ | 94.48(18) | 77.86(17) | 99.7(2) |
| I1$^{xi}$-I2-I2$^{vii}$ | 0 | 0 | 0 |
| I1$^{xi}$-I2-I2$^{x}$ | 89.74(19) | 82.2(2) | 97.1(2) |
| I1$^{xi}$-I2-I3$^{v}$ | 105.61(14) | 104.61(18) | 106.13(11) |
| I1$^{xi}$-I2-I3$^{x}$ | 105.61(14) | 104.61(18) | 106.13(11) |
| I1-I2-I2$^{iii}$ | 90.67(18) | 82.37(19) | 99.6(2) |
| I1-I2-I2$^{v}$ | 88.91(19) | 76.01(17) | 105.7(2) |
| I1-I2-I2$^{ix}$ | 90.67(18) | 82.37(19) | 99.6(2) |
| I1-I2-I2$^{vii}$ | 0 | 0 | 0 |
| I1-I2-I2$^{x}$ | 88.91(19) | 76.01(17) | 105.7(2) |
| I1-I2-I3$^{v}$ | 73.38(13) | 71.73(14) | 74.28(12) |
| I1-I2-I3$^{x}$ | 73.38(13) | 71.73(14) | 74.28(12) |
| I2$^{iii}$-I2-I2$^{v}$ | 175.9(2) | 171.3(3) | 179.53(17) |
| I2$^{iii}$-I2-I2$^{ix}$ | 0 | 0 | 0 |
| I2$^{iii}$-I2-I2$^{vii}$ | 0 | 0 | 0 |
| $I2^{iii}$-I2-$I2^{x}$ | 175.9(2) | 171.3(3) | 179.53(17) |
| $I2^{iii}$-I2-$I3^{v}$ | 148.13(10) | 147.38(11) | 149.85(12) |
| $I2^{iii}$-I2-$I3^{x}$ | 148.13(10) | 147.38(11) | 149.85(12) |
| $I2^{v}$-I2-$I2^{ix}$ | 175.9(2) | 171.3(3) | 179.53(16) |
| $I2^{v}$-I2-$I2^{vii}$ | 0 | 0 | 0 |
| $I2^{v}$-I2-$I2^{x}$ | 0 | 0 | 0 |
| $I2^{ix}$-I2-$I2^{vii}$ | 0 | 0 | 0 |
| $I2^{ix}$-I2-$I2^{x}$ | 175.9(2) | 171.3(3) | 179.53(17) |
| $I2^{ix}$-I2-$I3^{v}$ | 148.13(10) | 147.38(11) | 149.85(12) |
| $I2^{ix}$-I2-$I3^{x}$ | 148.13(10) | 147.38(11) | 149.85(12) |
| $I2^{vii}$-I2-$I2^{x}$ | 0 | 0 | 0 |
| $I2^{vii}$-I2-$I3^{v}$ | 0 | 0 | 0 |
| $I2^{vii}$-I2-$I3^{x}$ | 0 | 0 | 0 |
| $I3^{v}$-I2-$I3^{x}$ | 46.79(13) | 44.87(12) | 47.60(14) |
| Te3-I3-$I1^{iii}$ | 113.7(8) | 112.1(4) | 114.3(14) |
| Te3-I3-$I1^{v}$ | 114.6(8) | 113.1(7) | 115.4(4) |
| Te3-I3-$I2^{iii}$ | 112.4(8) | 110.8(4) | 113.0(13) |
| Te3-I3-$I2^{ix}$ | 112.4(8) | 110.8(4) | 113.0(13) |
| Te3-I3-$I3^{iii}$ | 87.8(5) | 87.8(5) | 87.8(5) |
| Te3-I3-$I3^{v}$ | 88.2(5) | 88.2(5) | 88.2(5) |
| Te3-I3-$I3^{vii}$ | 176.6(8) | 176.4(12) | 176.9(5) |
| $I1^{iii}$-I3-$I1^{v}$ | 119.6(2) | 118.4(3) | 120.96(17) |
| $I1^{iii}$-I3-$I2^{iii}$ | 39.78(14) | 38.23(17) | 42.02(13) |
| $I1^{iii}$-I3-$I2^{ix}$ | 39.78(14) | 38.23(17) | 42.02(13) |
| $I1^{iii}$-I3-$I3^{v}$ | 152.4(2) | 152.4(2) | 152.4(2) |
| $I1^{iii}$-I3-$I3^{vii}$ | 67.62(17) | 66.9(2) | 68.42(15) |
| $I1^{v}$-I3-$I2^{iii}$ | 132.7(2) | 132.17(16) | 133.8(3) |
| $I1^{v}$-I3-$I2^{ix}$ | 132.7(2) | 132.17(16) | 133.8(3) |
| $I1^{v}$-I3-$I3^{iii}$ | 146.8(4) | 146.8(4) | 146.8(4) |
| $I1^{v}$-I3-$I3^{vii}$ | 66.38(17) | 65.86(13) | 67.6(3) |
| $I2^{iii}$-I3-$I2^{ix}$ | 0 | 0 | 0 |
| $I2^{iii}$-I3-$I3^{v}$ | 144.9(2) | 144.9(2) | 144.9(2) |
| $I2^{iii}$-I3-$I3^{vii}$ | 66.62(16) | 66.20(18) | 67.60(15) |
| $I2^{ix}$-I3-$I3^{v}$ | 144.9(2) | 144.9(2) | 144.9(2) |
| $I2^{ix}$-I3-$I3^{vii}$ | 66.62(16) | 66.20(18) | 67.60(15) |
| $I3^{iii}$-I3-$I3^{vii}$ | 89.9(4) | 89.9(4) | 89.9(4) |
| $I3^{v}$-I3-$I3^{vii}$ | 90.0(3) | 90.0(3) | 90.0(3) |

(i) -x+1,-y+1,-z
(ii) -x+1,-y+1,-z+1
(iii) x,y,z-1
(iv) -x,-y+1,-z+1
(v) x,y,z+1
(vi) x+1,y,z
(vii) x,-y+3/2,z
(viii) x+1,-y+3/2,z
(ix) x,-y+3/2,z-1
(x) x,-y+3/2,z+1
(xi) x-1,y,z

# Crystallographic details for commensurate $WTe_2I$

Table S 11: Occupational waves for commensurate $WTe_2I$.

| Atom | Wave/Parameter | Occ |
|---|---|---|
| Te2 | delta | 0.8318 |
| | x40 | 0.3211(4) |
| Te3 | delta | 0.1682 |
| | x40 | 0.8231(4) |
| I1 | delta | 0.8318 |
| | x40 | 0.4567(5) |
| I2 | delta | 0.8318 |
| | x40 | 0.1933 |
| I3 | delta | 0.1682 |
| | x40 | 0.8366(5) |

Table S12: Positional parameters for commensurate $WTe_2I$.

| Atom | Occ | Wave | x | y | z | Ueq/Uiso |
|---|---|---|---|---|---|---|
| W1 | 1 | | 0.32164(6) | 0.495299(18) | 0.2518(2) | 0.00718(10) |
| | | s,1 | -0.00066(16) | 0.00147(4) | -0.0298(10) | |
| | | c,1 | 0.00256(13) | -0.00162(3) | -0.0094(4) | |
| | | s,2 | -0.00350(11) | 0.00013(3) | -0.0044(6) | |
| | | c,2 | -0.00019(14) | 0.00012(4) | 0.0026(4) | |
| | | s,3 | 0 | 0 | -0.0949(15) | |
| | | c,3 | 0.4355 | 1 | 0 | |
| Te1 | 1 | | 0.57916(10) | 0.59501(3) | 0.2561(3) | 0.00847(15) |
| | | s,1 | -0.0039(3) | 0.00054(8) | 0.0279(3) | |
| | | c,1 | -0.00876(16) | -0.00040(5) | -0.0007(5) | |
| | | s,2 | -0.0009(2) | -0.00033(5) | 0.0108(3) | |
| | | c,2 | 0.00613(17) | -0.00235(5) | -0.0041(3) | |
| Te2 | 0.8318 | | 0.07018(16) | 0.56910(4) | 0.7592(4) | 0.00718(18) |
| | | o,1 | -0.0008(2) | 0.00093(6) | -0.0008(6) | |
| | | o,2 | 0.0002(3) | 0.00012(7) | 0.0029(4) | |
| | | o,3 | 0.0011(4) | -0.00061(11) | -0.0175(8) | |
| | | o,4 | -0.0007(3) | 0.00003(8) | -0.0030(7) | |
| Te3 | 0.1682 | | 0.0811(4) | 0.54884(10) | 0.7465(8) | 0.0077(4) |
| I1 | 0.8318 | | 0.3527(6) | 0.75 | 0.7423(14) | 0.0627(11) |
| | | o,1 | 0.0010(9) | 0 | 0.0720(8) | |
| | | o,2 | -0.0266(7) | 0 | 0.0347(12) | |
| | | o,3 | -0.0065(7) | 0 | -0.0358(10) | |
| | | o,4 | 0.0058(11) | 0 | -0.0214(14) | |
| I2 | 0.8318 | | -0.1823(6) | 0.75 | 0.7503(15) | 0.0660(12) |
| | | o,1 | -0.0280(8) | 0 | 0.0009(12) | |
| | | o,2 | -0.0161(8) | 0 | 0.0755(9) | |
| | | o,3 | -0.0021(8) | 0 | 0.0387(11) | |
| | | o,4 | 0.0110(11) | 0 | 0.0226(18) | |
| I3 | 0.1682 | | 0.1055(4) | 0.67488(9) | 0.7537(9) | 0.0168(5) |

Table S13: ADP harmonic parameters for commensurate $WTe_2I$.

| Atom | Wave | U11 | U22 | U33 | U12 | U13 | U23 |
|---|---|---|---|---|---|---|---|
| W1 | | 0.00523(16) | 0.01106(18) | 0.00526(17) | -0.00038(13) | 0.0000(4) | 0.0020(2) |
| | s,1 | 0.0003(4) | 0.0005(4) | -0.0008(3) | 0.0009(3) | -0.0001(3) | 0.0002(2) |
| | c,1 | 0.0002(3) | 0.0000(3) | 0.0004(3) | 0.0002(3) | -0.0004(3) | 0.0005(3) |
| Te1 | | 0.0080(3) | 0.0118(3) | 0.0056(3) | -0.0009(2) | -0.0037(5) | 0.0036(4) |
| Te2 | | 0.0053(3) | 0.0125(3) | 0.0037(4) | -0.0006(2) | 0.0026(5) | 0.0027(4) |
| | o,1 | 0.0004(3) | -0.0016(4) | -0.0009(4) | 0.0005(3) | 0.0006(4) | -0.0009(4) |
| | o,2 | 0.0007(5) | -0.0007(5) | 0.0002(5) | 0.0006(4) | 0.0001(2) | -0.0001(3) |
| Te3 | | 0.0058(8) | 0.0144(9) | 0.0028(7) | 0.0000(8) | 0.0027(9) | -0.0003(11) |
| I1 | | 0.102(2) | 0.0259(9) | 0.060(2) | 0 | 0.018(2) | 0 |
| | o,1 | 0.005(3) | -0.0021(13) | 0.0039(18) | 0 | -0.0057(15) | 0 |
| | o,2 | -0.018(3) | 0.0008(11) | -0.0075(17) | 0 | 0.001(2) | 0 |
| I2 | | 0.106(3) | 0.0257(9) | 0.067(2) | 0 | 0.009(2) | 0 |
| | o,1 | 0.031(3) | -0.0012(12) | 0.018(2) | 0 | -0.002(2) | 0 |
| | o,2 | 0.008(2) | 0.0039(12) | 0.0100(18) | 0 | 0.0008(16) | 0 |
| I3 | | 0.0182(9) | 0.0128(8) | 0.0194(9) | -0.0005(7) | -0.0014(11) | -0.0004(10) |

Table S14: List of used orthogonalized functions for commensurate $WTe_2I$.

| | | | | | | | | | | | | | | | | | | | | | | | | | | | | | | | |
|---|---|---|---|---|---|---|---|---|---|---|---|---|---|---|---|---|---|---|---|---|---|---|---|---|---|---|---|---|---|---|---|
| Te2 | o,1 | - | -0.269 | - | - | - | - | - | - | - | - | - | - | - | - | - | 1.548 | | | | | | | | | | | | | | |
| | o,2 | - | 0.16 | - | - | - | - | - | - | - | - | - | - | - | - | - | -0.261 | - | - | - | - | - | - | 1.373 | | | | | | | |
| | o,3 | - | -0.32 | - | - | - | - | - | - | - | - | - | - | - | - | - | 0.53 | - | - | - | - | - | - | -0.314 | - | - | - | - | - | 1.553 | |
| | o,4 | - | -0.327 | - | - | - | - | - | - | - | - | - | - | - | - | - | 0.589 | - | - | - | - | - | - | -0.223 | - | - | - | - | - | 0.367 | 1.518 |
| I1 | o,1 | - | -0.069 | - | - | - | - | - | - | 1.326 | | | | | | | | | | | | | | | | | | | | | |
| | o,2 | - | 0.305 | - | - | - | - | - | - | -0.149 | - | - | - | - | - | 1.603 | | | | | | | | | | | | | | | |
| | o,3 | - | -0.197 | - | - | - | - | - | - | 0.168 | - | - | - | - | - | -0.345 | 1.437 | | | | | | | | | | | | | | |
| | o,4 | - | 0.413 | - | - | - | - | - | - | -0.181 | - | - | - | - | - | 0.772 | -0.308 | 1.64 | | | | | | | | | | | | | |
| I2 | o,1 | - | -0.284 | - | - | - | - | - | 1.573 | | | | | | | | | | | | | | | | | | | | | | |
| | o,2 | - | -0.131 | - | - | - | - | - | 0.223 | 1.351 | | | | | | | | | | | | | | | | | | | | | |
| | o,3 | - | 0.257 | - | - | - | - | - | -0.44 | -0.231 | 1.488 | | | | | | | | | | | | | | | | | | | | |
| | o,4 | - | -0.379 | - | - | - | - | - | 0.695 | 0.213 | -0.356 | 1.584 | | | | | | | | | | | | | | | | | | | |

Table S15: List of distances for commensurate $WTe_2I$.

| | average / Å | min / Å | max / Å |
|---|---|---|---|
| $W1-W1^{i}$ | 2.843(6) | 2.825(7) | 2.862(7) |
| $W1-W1^{ii}$ | 2.859(6) | 2.775(9) | 3.080(9) |
| W1-Te1 | 2.722(3) | 2.689(3) | 2.746(3) |
| $W1-Te1^{i}$ | 2.715(5) | 2.649(6) | 2.794(6) |
| $W1-Te1^{ii}$ | 2.696(5) | 2.623(7) | 2.761(7) |
| $W1-Te2^{iii}$ | 2.802(6) | 2.744(8) | 2.863(8) |
| W1-Te2 | 2.837(5) | 2.787(7) | 2.905(7) |
| $W1-Te2^{iv}$ | 2.839(4) | 2.814(6) | 2.860(6) |
| $W1-Te3^{iii}$ | 2.755(5) | 2.755(5) | 2.755(5) |
| W1-Te3 | 2.766(7) | 2.766(7) | 2.766(7) |
| $W1-Te3^{iv}$ | 2.739(3) | 2.739(3) | 2.739(3) |
| Te3-I3 | 2.762(3) | 2.762(3) | 2.762(3) |
| $I1-I1^{iii}$ | 3.416(11) | 2.900(12) | 3.915(12) |
| $I1-I1^{v}$ | 3.416(11) | 2.900(12) | 3.915(12) |
| I1-I2 | 3.374(14) | 2.966(18) | 4.015(18) |
| $I1-I2^{vi}$ | 3.143(14) | 3.051(17) | 3.223(17) |
| $I1-I3^{v}$ | 4.164(8) | 4.164(8) | 4.164(8) |
| $I1-I3^{vii}$ | 4.164(8) | 4.164(8) | 4.164(8) |
| $I2-I2^{iii}$ | 3.443(13) | 2.967(16) | 4.002(16) |
| $I2-I2^{v}$ | 3.443(13) | 2.967(16) | 4.002(16) |
| $I2-I3^{iii}$ | 4.148(11) | 4.148(11) | 4.148(11) |
| $I2-I3^{viii}$ | 4.148(11) | 4.148(11) | 4.148(11) |
| $I3-I3^{ix}$ | 3.287(3) | 3.287(3) | 3.287(3) |

(i) -x+1,-y+1,-z
(ii) -x+1,-y+1,-z+1
(iii) x,y,z-1
(iv) -x,-y+1,-z+1
(v) x,y,z+1
(vi) x+1,y,z
(vii) x,-y+3/2,z+1
(viii) x,-y+3/2,z-1
(ix) x,-y+3/2,z
(x) x-1,y,z

Table S16: List of angles for commensurate $WTe_2I$.

| | average / ° | min / ° | max / ° |
|---|---|---|---|
| $W1^{i}-W1-W1^{ii}$ | 74.97(16) | 71.72(19) | 82.50(17) |
| $W1^{i}-W1-Te1$ | 58.38(10) | 57.09(11) | 60.83(11) |
| $W1^{i}-W1-Te1^{i}$ | 58.59(12) | 57.18(14) | 59.34(15) |
| $W1^{i}-W1-Te1^{ii}$ | 104.87(11) | 102.29(11) | 109.22(11) |
| $W1^{i}-W1-Te2^{iii}$ | 93.0(2) | 91.9(2) | 93.50(19) |
| $W1^{i}-W1-Te2$ | 140.47(16) | 137.74(16) | 142.44(15) |
| $W1^{i}-W1-Te2^{iv}$ | 136.3(3) | 133.8(3) | 139.1(3) |
| $W1^{i}-W1-Te3^{iii}$ | 87.1(2) | 87.1(2) | 87.1(2) |
| $W1^{i}-W1-Te3$ | 145.68(10) | 145.68(10) | 145.68(10) |
| $W1^{i}-W1-Te3^{iv}$ | 138.87(18) | 138.87(18) | 138.87(18) |
| $W1^{ii}-W1-Te1$ | 57.67(10) | 56.70(12) | 58.57(10) |
| $W1^{ii}-W1-Te1^{i}$ | 105.25(11) | 102.55(10) | 109.73(11) |
| $W1^{ii}-W1-Te1^{ii}$ | 58.63(12) | 54.50(15) | 61.06(13) |
| $W1^{ii}-W1-Te2^{iii}$ | 140.04(16) | 138.02(16) | 142.81(15) |
| $W1^{ii}-W1-Te2$ | 92.9(2) | 91.6(2) | 94.6(2) |
| $W1^{ii}-W1-Te2^{iv}$ | 136.6(3) | 129.4(3) | 141.9(3) |
| $W1^{ii}-W1-Te3^{iii}$ | 145.00(12) | 145.00(12) | 145.00(12) |
| $W1^{ii}-W1-Te3$ | 82.1(2) | 82.1(2) | 82.1(2) |
| W1$^{ii}$-W1-Te3$^{iv}$ | 138.71(19) | 138.71(19) | 138.71(19) |
| Te1-W1-Te1$^{i}$ | 116.87(16) | 114.04(17) | 119.9(2) |
| Te1-W1-Te1$^{ii}$ | 116.29(17) | 111.2(2) | 119.63(17) |
| Te1-W1-Te2$^{iii}$ | 83.75(17) | 81.88(17) | 85.62(16) |
| Te1-W1-Te2 | 83.27(17) | 81.75(17) | 85.50(16) |
| Te1-W1-Te2$^{iv}$ | 155.39(17) | 153.53(17) | 157.29(16) |
| Te1-W1-Te3$^{iii}$ | 86.47(12) | 86.47(12) | 86.47(12) |
| Te1-W1-Te3 | 85.13(15) | 85.13(15) | 85.13(15) |
| Te1-W1-Te3$^{iv}$ | 150.85(9) | 150.85(9) | 150.85(9) |
| Te1$^{i}$-W1-Te1$^{ii}$ | 79.83(9) | 77.81(9) | 83.73(9) |
| Te1$^{i}$-W1-Te2$^{iii}$ | 100.4(2) | 97.8(2) | 103.4(3) |
| Te1$^{i}$-W1-Te2 | 158.47(17) | 157.59(17) | 159.61(17) |
| Te1$^{i}$-W1-Te2$^{iv}$ | 80.83(18) | 78.93(18) | 83.49(19) |
| Te1$^{i}$-W1-Te3$^{iii}$ | 85.48(18) | 85.48(18) | 85.48(18) |
| Te1$^{i}$-W1-Te3 | 154.93(12) | 154.93(12) | 154.93(12) |
| Te1$^{i}$-W1-Te3$^{iv}$ | 85.84(11) | 85.84(11) | 85.84(11) |
| Te1$^{ii}$-W1-Te2$^{iii}$ | 158.18(18) | 156.56(16) | 160.50(18) |
| Te1$^{ii}$-W1-Te2 | 99.9(2) | 96.6(2) | 103.3(3) |
| Te1$^{ii}$-W1-Te2$^{iv}$ | 81.40(18) | 80.4(2) | 83.30(18) |
| Te1$^{ii}$-W1-Te3$^{iii}$ | 153.84(12) | 153.84(12) | 153.84(12) |
| Te1$^{ii}$-W1-Te3 | 85.9(2) | 85.9(2) | 85.9(2) |
| Te1$^{ii}$-W1-Te3$^{iv}$ | 84.74(10) | 84.74(10) | 84.74(10) |
| Te2$^{iii}$-W1-Te2 | 75.03(19) | 73.37(17) | 77.4(2) |
| Te2$^{iii}$-W1-Te2$^{iv}$ | 78.28(18) | 77.24(17) | 79.01(17) |
| Te2$^{iii}$-W1-Te3 | 78.46(15) | 78.46(15) | 78.46(15) |
| Te2$^{iii}$-W1-Te3$^{iv}$ | 71.84(13) | 71.84(13) | 71.84(13) |
| Te2-W1-Te2$^{iv}$ | 78.91(18) | 77.6(2) | 80.1(2) |
| Te2-W1-Te3$^{iii}$ | 80.69(8) | 80.69(8) | 80.69(8) |
| Te2-W1-Te3$^{iv}$ | 71.95(13) | 71.95(13) | 71.95(13) |
| Te2$^{iv}$-W1-Te3$^{iii}$ | 71.85(15) | 71.85(15) | 71.85(15) |
| Te2$^{iv}$-W1-Te3 | 72.28(16) | 72.28(16) | 72.28(16) |
| W1-Te1-W1$^{i}$ | 63.04(13) | 62.00(16) | 64.18(11) |
| W1-Te1-W1$^{ii}$ | 63.70(13) | 61.85(13) | 68.81(17) |
| W1$^{i}$-Te1-W1$^{ii}$ | 79.82(17) | 76.3(2) | 89.14(18) |
| W1-Te2-W1$^{v}$ | 73.77(18) | 71.83(16) | 75.17(19) |
| W1-Te2-W1$^{iv}$ | 101.63(16) | 99.85(10) | 103.8(2) |
| W1$^{v}$-Te2-W1$^{iv}$ | 102.28(16) | 100.6(2) | 104.6(2) |
| W1-Te3-W1$^{v}$ | 89.8(2) | 89.8(2) | 89.8(2) |
| W1-Te3-W1$^{iv}$ | 111.06(14) | 111.06(14) | 111.06(14) |
| W1-Te3-I3 | 114.59(12) | 114.59(12) | 114.59(12) |
| W1$^{v}$-Te3-W1$^{iv}$ | 110.62(14) | 110.62(14) | 110.62(14) |
| W1$^{v}$-Te3-I3 | 113.31(12) | 113.31(12) | 113.31(12) |
| W1$^{iv}$-Te3-I3 | 114.85(10) | 114.85(10) | 114.85(10) |
| I1$^{iii}$-I1-I1$^{v}$ | 176.2(4) | 170.9(4) | 179.9(5) |
| I1$^{iii}$-I1-I2 | 90.7(4) | 87.0(4) | 96.3(4) |
| I1$^{iii}$-I1-I2$^{vi}$ | 92.1(4) | 83.7(4) | 98.3(4) |
| I1$^{iii}$-I1-I3$^{v}$ | 149.9(2) | 149.9(2) | 149.9(2) |
| I1$^{iii}$-I1-I3$^{vii}$ | 149.9(2) | 149.9(2) | 149.9(2) |
| I1$^{v}$-I1-I2 | 89.7(4) | 81.3(3) | 98.6(4) |
| I1$^{v}$-I1-I2$^{vi}$ | 91.9(4) | 83.5(4) | 96.2(4) |
| I2-I1-I2$^{vi}$ | 169.3(4) | 165.3(5) | 171.8(4) |
| I2-I1-I3$^{v}$ | 72.03(19) | 72.03(19) | 72.03(19) |
| I2-I1-I3$^{vii}$ | 72.03(19) | 72.03(19) | 72.03(19) |
| I2$^{vi}$-I1-I3$^{v}$ | 100.5(2) | 100.5(2) | 100.5(2) |
| I2$^{vi}$-I1-I3$^{vii}$ | 100.5(2) | 100.5(2) | 100.5(2) |
| I3$^{v}$-I1-I3$^{vii}$ | 46.49(9) | 46.49(9) | 46.49(9) |
| I1$^{x}$-I2-I1 | 170.3(4) | 166.8(3) | 175.2(5) |
| I1$^{x}$-I2-I2$^{iii}$ | 91.0(4) | 82.6(3) | 95.9(3) |
| I1$^{x}$-I2-I2$^{v}$ | 91.7(4) | 84.3(4) | 95.7(4) |
| I1$^{x}$-I2-I3$^{iii}$ | 105.3(3) | 105.3(3) | 105.3(3) |
| I1$^{x}$-I2-I3$^{viii}$ | 105.3(3) | 105.3(3) | 105.3(3) |
| I1-I2-I2$^{iii}$ | 89.3(4) | 80.4(4) | 97.3(3) |
| I1-I2-I2$^{v}$ | 90.2(4) | 86.9(4) | 96.9(4) |
| I1-I2-I3$^{iii}$ | 70.3(3) | 70.3(3) | 70.3(3) |
| I1-I2-I3$^{viii}$ | 70.3(3) | 70.3(3) | 70.3(3) |
| I2$^{iii}$-I2-I2$^{v}$ | 174.9(5) | 168.3(6) | 178.7(6) |
| I2$^{v}$-I2-I3$^{iii}$ | 148.1(2) | 148.1(2) | 148.1(2) |
| I2$^{v}$-I2-I3$^{viii}$ | 148.1(2) | 148.1(2) | 148.1(2) |
| I3$^{iii}$-I2-I3$^{viii}$ | 46.69(14) | 46.69(14) | 46.69(14) |
| Te3-I3-I1$^{iii}$ | 113.92(12) | 113.92(12) | 113.92(12) |
| Te3-I3-I2$^{v}$ | 112.40(13) | 112.40(13) | 112.40(13) |
| Te3-I3-I3$^{ix}$ | 176.75(11) | 176.75(11) | 176.75(11) |
| I1$^{iii}$-I3-I2$^{v}$ | 133.28(11) | 133.28(11) | 133.28(11) |
| I1$^{iii}$-I3-I3$^{ix}$ | 66.75(8) | 66.75(8) | 66.75(8) |
| I2$^{v}$-I3-I3$^{ix}$ | 66.65(10) | 66.65(10) | 66.65(10) |

(i) -x+1,-y+1,-z
(ii) -x+1,-y+1,-z+1
(iii) x,y,z-1
(iv) -x,-y+1,-z+1
(v) x,y,z+1
(vi) x+1,y,z
(vii) x,-y+3/2,z+1
(viii) x,-y+3/2,z-1
(ix) x,-y+3/2,z
(x) x-1,y,z

# Crystallographic details for $WTe_2Br_{0.5}$

Table S17: Fractional atomic coordinates and isotropic or equivalent isotropic displacement parameters (Å$^2$) for $WTe_2Br_{0.5}$.

| | x | y | z | Uiso*/Ueq |
|---|---|---|---|---|
| W1 | −0.00516 (15) | 0.25 | −0.1785 (3) | 0.0527 (15) |
| Te1 | −0.07421 (12) | 0.25 | −0.5896 (5) | 0.042 (2) |
| Te2 | 0.12012 (14) | 0.25 | 0.0912 (5) | 0.039 (2) |
| Br1 | −0.2500 | 0.25 | −0.6083 (13) | 0.115 (7) |

Table S18: for $WTe_2Br_{0.5}$.

| | $U^{11}$ | $U^{22}$ | $U^{33}$ | $U^{12}$ | $U^{13}$ | $U^{23}$ |
|---|---|---|---|---|---|---|
| W1 | 0.0192 (14) | 0.0821 (16) | 0.0568 (14) | 0 | −0.0072 (17) | 0 |
| Te1 | 0.020 (3) | 0.048 (3) | 0.0574 (18) | 0 | 0.013 (2) | 0 |
| Te2 | 0.047 (3) | 0.044 (3) | 0.0274 (17) | 0 | −0.010 (3) | 0 |
| Br1 | 0.145 (9) | 0.036 (5) | 0.163 (7) | 0 | 0 | 0 |

Table S19: Atomic displacement parameters (Å$^2$) for $WTe_2Br_{0.5}$.

| | | | |
|---|---|---|---|
| W1—W1$^{i}$ | 3.6773 (1) | Te1—Te2$^{vi}$ | 3.735 (4) |
| W1—W1$^{ii}$ | 3.6773 (1) | Te1—Te2$^{iii}$ | 3.718 (4) |
| W1—W1$^{iii}$ | 2.909 (2) | Te1—Te2$^{vii}$ | 3.735 (4) |
| W1—W1$^{iv}$ | 2.909 (2) | Te1—Te2$^{iv}$ | 3.718 (4) |
| W1—Te1 | 2.835 (4) | Te1—Br1 | 2.946 (2) |
| W1—Te1$^{v}$ | 3.884 (4) | Te2—W1 | 2.699 (4) |
| W1—Te1$^{vi}$ | 2.698 (3) | Te2—W1$^{iii}$ | 2.718 (3) |
| W1—Te1$^{vii}$ | 2.698 (3) | Te2—W1$^{iv}$ | 2.718 (3) |
| W1—Te2 | 2.699 (4) | Te2—Te1$^{v}$ | 3.825 (4) |
| W1—Te2$^{iii}$ | 2.718 (3) | Te2—Te1$^{vi}$ | 3.735 (4) |
| W1—Te2$^{iv}$ | 2.718 (3) | Te2—Te1$^{iii}$ | 3.718 (4) |
| Te1—W1$^{viii}$ | 3.884 (4) | Te2—Te1$^{vii}$ | 3.735 (4) |
| Te1—W1 | 2.835 (4) | Te2—Te1$^{iv}$ | 3.718 (4) |
| Te1—W1$^{vi}$ | 2.698 (3) | Te2—Te2$^{i}$ | 3.6773 (1) |
| Te1—W1$^{vii}$ | 2.698 (3) | Te2—Te2$^{ii}$ | 3.6773 (1) |
| Te1—Te1$^{i}$ | 3.6773 (1) | Br1—Te1 | 2.946 (2) |
| Te1—Te1$^{ii}$ | 3.6773 (1) | Br1—Te1$^{ix}$ | 2.946 (2) |
| Te1—Te1$^{vi}$ | 3.291 (3) | Br1—Br1$^{i}$ | 3.6773 (1) |
| Te1—Te1$^{vii}$ | 3.291 (3) | Br1—Br1$^{ii}$ | 3.6773 (1) |
| Te1—Te2$^{viii}$ | 3.825 (4) | | |

Symmetry codes: (i) *x*, *y*−1, *z*; (ii) *x*, *y*+1, *z*; (iii) −*x*, *y*−1/2, −*z*; (iv) −*x*, *y*+1/2, −*z*; (v) *x*, *y*, *z*+1; (vi) −*x*, *y*−1/2, −*z*−1; (vii) −*x*, *y*+1/2, −*z*−1; (viii) *x*, *y*, *z*−1; (ix) −*x*−1/2, −*y*+1/2, *z*.

Table S20: for $WTe_2Br_{0.5}$.

| | | | |
|---|---|---|---|
| W1$^{i}$—W1—W1$^{ii}$ | 180 | W1$^{vii}$—Te1—Br1 | 118.07 (13) |
| W1$^{i}$—W1—W1$^{iii}$ | 50.80 (6) | Te1$^{i}$—Te1—Te1$^{ii}$ | 180 |
| W1$^{i}$—W1—W1$^{iv}$ | 129.20 (6) | Te1$^{i}$—Te1—Te1$^{vi}$ | 56.03 (7) |
| W1$^{i}$—W1—Te1 | 90.00 (10) | Te1$^{i}$—Te1—Te1$^{vii}$ | 123.97 (7) |
| W1$^{i}$—W1—Te1$^{v}$ | 90.00 (8) | Te1$^{i}$—Te1— | 90.00 (8) |
| W1$^{i}$—W1—Te1$^{vi}$ | 47.04 (7) | Te1$^{i}$—Te1—Te2$^{vi}$ | 60.51 (9) |
| W1$^{i}$—W1—Te1$^{vii}$ | 132.96 (7) | Te1$^{i}$—Te1—Te2$^{iii}$ | 60.36 (9) |
| W1$^{i}$—W1—Te2 | 90.00 (10) | Te1$^{i}$—Te1—Te2$^{vii}$ | 119.49 (9) |
| W1$^{i}$—W1—Te2$^{iii}$ | 47.43 (7) | Te1$^{i}$—Te1—Te2$^{iv}$ | 119.64 (9) |
| W1$^{i}$—W1—Te2$^{iv}$ | 132.57 (7) | Te1$^{i}$—Te1—Br1 | 90.00 (5) |
| W1$^{ii}$—W1—W1$^{iii}$ | 129.20 (6) | Te1$^{ii}$—Te1—Te1$^{vi}$ | 123.97 (7) |
| W1$^{ii}$—W1—W1$^{iv}$ | 50.80 (6) | Te1$^{ii}$—Te1— | 56.03 (7) |
| W1$^{ii}$—W1—Te1 | 90.00 (10) | Te1$^{ii}$—Te1— | 90.00 (8) |
| W1$^{ii}$—W1—Te1$^{v}$ | 90.00 (8) | Te1$^{ii}$—Te1—Te2$^{vi}$ | 119.49 (9) |
| W1$^{ii}$—W1—Te1$^{vi}$ | 132.96 (7) | Te1$^{ii}$—Te1—Te2$^{iii}$ | 119.64 (9) |
| W1$^{ii}$—W1—Te1$^{vii}$ | 47.04 (7) | Te1$^{ii}$—Te1— | 60.51 (9) |
| W1$^{ii}$—W1—Te2 | 90.00 (10) | Te1$^{ii}$—Te1—Te2$^{iv}$ | 60.36 (9) |
| W1$^{ii}$—W1—Te2$^{iii}$ | 132.57 (7) | Te1$^{ii}$—Te1—Br1 | 90.00 (5) |
| W1$^{ii}$—W1—Te2$^{iv}$ | 47.43 (7) | Te1$^{vi}$—Te1— | 67.93 (7) |
| W1$^{iii}$—W1—W1$^{iv}$ | 78.40 (6) | Te1$^{vi}$—Te1— | 62.46 (10) |
| W1$^{iii}$—W1—Te1 | 136.86 (16) | Te1$^{vi}$—Te1— | 99.81 (13) |
| W1$^{iii}$—W1—Te1$^{vi}$ | 87.61 (13) | Te1$^{vi}$—Te1— | 65.82 (10) |
| W1$^{iii}$—W1—Te1$^{vii}$ | 145.05 (13) | Te1$^{vi}$—Te1— | 136.09 (13) |
| W1$^{iii}$—W1—Te2 | 57.84 (10) | Te1$^{vi}$—Te1— | 98.22 (10) |
| W1$^{iii}$—W1—Te2$^{iii}$ | 57.20 (10) | Te1$^{vi}$—Te1—Br1 | 140.20 (11) |
| W1$^{iii}$—W1—Te2$^{iv}$ | 108.26 (10) | Te1$^{vii}$—Te1— | 62.46 (10) |
| W1$^{iv}$—W1—Te1 | 136.86 (16) | Te1$^{vii}$—Te1— | 136.09 (13) |
| W1$^{iv}$—W1—Te1$^{vi}$ | 145.05 (13) | Te1$^{vii}$—Te1— | 98.22 (10) |
| W1$^{iv}$—W1—Te1$^{vii}$ | 87.61 (13) | Te1$^{vii}$—Te1— | 99.81 (13) |
| W1$^{iv}$—W1—Te2 | 57.84 (10) | Te1$^{vii}$—Te1— | 65.82 (10) |
| W1$^{iv}$—W1—Te2$^{iii}$ | 108.26 (10) | Te1$^{vii}$—Te1—Br1 | 140.20 (11) |
| W1$^{iv}$—W1—Te2$^{iv}$ | 57.20 (10) | Te2$^{viii}$—Te1— | 74.38 (10) |
| Te1—W1—Te1$^{v}$ | 138.61 (18) | Te2$^{viii}$—Te1— | 128.28 (14) |
| Te1—W1—Te1$^{vi}$ | 72.95 (12) | Te2$^{viii}$—Te1— | 74.38 (10) |
| Te1—W1—Te1$^{vii}$ | 72.95 (12) | Te2$^{viii}$—Te1— | 128.28 (14) |
| Te1—W1—Te2 | 153.1 (2) | Te2$^{viii}$—Te1—Br1 | 146.01 (13) |
| Te1—W1—Te2$^{iii}$ | 84.02 (14) | Te2$^{vi}$—Te1— | 115.33 (15) |
| Te1—W1—Te2$^{iv}$ | 84.02 (14) | Te2$^{vi}$—Te1— | 58.99 (9) |
| Te1$^{v}$—W1—Te1$^{vi}$ | 131.55 (16) | Te2$^{vi}$—Te1— | 156.19 (15) |
| Te1$^{v}$—W1—Te1$^{vii}$ | 131.55 (16) | Te2$^{vi}$—Te1—Br1 | 76.14 (16) |
| Te1$^{v}$—W1—Te2 | 68.33 (12) | Te2$^{iii}$—Te1— | 156.19 (15) |
| Te1$^{v}$—W1—Te2$^{iii}$ | 66.18 (11) | Te2$^{iii}$—Te1— | 59.28 (9) |
| Te1$^{v}$—W1—Te2$^{iv}$ | 66.18 (11) | Te2$^{iii}$—Te1—Br1 | 80.04 (17) |
| Te1$^{vi}$—W1— | 85.92 (8) | Te2$^{vii}$—Te1— | 115.33 (15) |
| Te1$^{vi}$—W1—Te2 | 87.57 (15) | Te2$^{vii}$—Te1—Br1 | 76.14 (16) |
| Te1$^{vi}$—W1—Te2$^{iii}$ | 89.85 (13) | Te2$^{iv}$—Te1—Br1 | 80.04 (17) |
| Te1$^{vi}$—W1—Te2$^{iv}$ | 156.82 (13) | W1—Te2—W1$^{iii}$ | 64.96 (11) |
| Te1$^{vii}$—W1—Te2 | 87.57 (15) | W1—Te2—W1$^{iv}$ | 64.96 (11) |
| Te1$^{vii}$—W1—Te2$^{iii}$ | 156.82 (13) | W1—Te2—Te1$^{v}$ | 70.69 (12) |
| Te1$^{vii}$—W1— | 89.85 (13) | W1—Te2—Te1$^{vi}$ | 46.20 (11) |
| Te2—W1—Te2$^{iii}$ | 115.04 (15) | W1—Te2—Te1$^{iii}$ | 111.74 (15) |
| Te2—W1—Te2$^{iv}$ | 115.04 (15) | W1—Te2—Te1$^{vii}$ | 46.20 (11) |
| Te2$^{iii}$—W1—Te2$^{iv}$ | 85.13 (8) | W1—Te2—Te1$^{iv}$ | 111.74 (15) |
| W1$^{viii}$—Te1—W1 | 138.61 (13) | W1—Te2—Te2$^{i}$ | 90.00 (10) |
| W1$^{viii}$—Te1—W1$^{vi}$ | 48.45 (9) | W1—Te2—Te2$^{ii}$ | 90.00 (10) |
| W1$^{viii}$—Te1— | 48.45 (9) | W1$^{iii}$—Te2—W1$^{iv}$ | 85.13 (8) |
| W1$^{viii}$—Te1—Te1$^{i}$ | 90.00 (8) | W1$^{iii}$—Te2—Te1$^{vi}$ | 72.07 (12) |
| W1$^{viii}$—Te1—Te1$^{ii}$ | 90.00 (8) | W1$^{iii}$—Te2—Te1$^{iii}$ | 49.33 (10) |
| W1$^{viii}$—Te1— | 95.88 (11) | W1$^{iii}$—Te2—Te1$^{vii}$ | 110.99 (12) |
| W1$^{viii}$—Te1— | 95.88 (11) | W1$^{iii}$—Te2—Te1$^{iv}$ | 91.00 (10) |
| W1$^{viii}$—Te1— | 150.16 (13) | W1$^{iii}$—Te2—Te2$^{i}$ | 47.43 (7) |
| W1$^{viii}$—Te1— | 150.16 (13) | W1$^{iii}$—Te2—Te2$^{ii}$ | 132.57 (7) |
| W1$^{viii}$—Te1—Br1 | 105.03 (18) | W1$^{iv}$—Te2—Te1$^{vi}$ | 110.99 (12) |
| W1—Te1—W1$^{vi}$ | 107.05 (13) | W1$^{iv}$—Te2—Te1$^{iii}$ | 91.00 (10) |
| W1—Te1—W1$^{vii}$ | 107.05 (13) | W1$^{iv}$—Te2— | 72.07 (12) |
| W1—Te1—Te1$^{i}$ | 90.00 (10) | W1$^{iv}$—Te2—Te1$^{iv}$ | 49.33 (10) |
| W1—Te1—Te1$^{ii}$ | 90.00 (10) | W1$^{iv}$—Te2—Te2$^{i}$ | 132.57 (7) |
| W1—Te1—Te1$^{vi}$ | 51.61 (10) | W1$^{iv}$—Te2—Te2$^{ii}$ | 47.43 (7) |
| W1—Te1—Te1$^{vii}$ | 51.61 (10) | Te1$^{v}$—Te2—Te1$^{vi}$ | 105.62 (13) |
| W1—Te1—Te2$^{viii}$ | 97.63 (14) | Te1$^{v}$—Te2—Te1$^{iii}$ | 51.72 (9) |
| W1—Te1—Te2$^{vi}$ | 148.89 (16) | Te1$^{v}$—Te2— | 105.62 (13) |
| W1—Te1—Te2$^{iii}$ | 46.65 (10) | Te1$^{v}$—Te2—Te1$^{iv}$ | 51.72 (9) |
| W1—Te1—Te2$^{vii}$ | 148.89 (16) | Te1$^{v}$—Te2—Te2$^{i}$ | 90.00 (8) |
| W1—Te1—Te2$^{iv}$ | 46.65 (10) | Te1$^{v}$—Te2—Te2$^{ii}$ | 90.00 (8) |
| W1—Te1—Br1 | 116.36 (19) | Te1$^{vi}$—Te2— | 115.33 (15) |
| W1$^{vi}$—Te1—W1$^{vii}$ | 85.92 (8) | Te1$^{vi}$—Te2— | 58.99 (9) |
| W1$^{vi}$—Te1—Te1$^{i}$ | 47.04 (7) | Te1$^{vi}$—Te2— | 156.19 (15) |
| W1$^{vi}$—Te1—Te1$^{ii}$ | 132.96 (7) | Te1$^{vi}$—Te2—Te2$^{i}$ | 60.51 (9) |
| W1$^{vi}$—Te1—Te1$^{vi}$ | 55.45 (10) | Te1$^{vi}$—Te2—Te2$^{ii}$ | 119.49 (9) |
| W1$^{vi}$—Te1— | 101.20 (10) | Te1$^{iii}$—Te2— | 156.19 (15) |
| W1$^{vi}$—Te1— | 45.29 (9) | Te1$^{iii}$—Te2— | 59.28 (9) |
| W1$^{vi}$—Te1—Te2$^{vi}$ | 46.22 (10) | Te1$^{iii}$—Te2—Te2$^{i}$ | 60.36 (9) |
| W1$^{vi}$—Te1—Te2$^{iii}$ | 102.81 (14) | Te1$^{iii}$—Te2—Te2$^{ii}$ | 119.64 (9) |
| W1$^{vi}$—Te1— | 88.81 (10) | Te1$^{vii}$—Te2— | 115.33 (15) |
| W1$^{vi}$—Te1—Te2$^{iv}$ | 153.61 (14) | Te1$^{vii}$—Te2—Te2$^{i}$ | 119.49 (9) |
| W1$^{vi}$—Te1—Br1 | 118.07 (13) | Te1$^{vii}$—Te2— | 60.51 (9) |
| W1$^{vii}$—Te1—Te1$^{i}$ | 132.96 (7) | Te1$^{iv}$—Te2—Te2$^{i}$ | 119.64 (9) |
| W1$^{vii}$—Te1—Te1$^{ii}$ | 47.04 (7) | Te1$^{iv}$—Te2—Te2$^{ii}$ | 60.36 (9) |
| W1$^{vii}$—Te1— | 101.20 (10) | Te2$^{i}$—Te2—Te2$^{ii}$ | 180 |
| W1$^{vii}$—Te1— | 55.45 (10) | Te1—Br1—Te1$^{ix}$ | 175.42 (13) |
| W1$^{vii}$—Te1— | 45.29 (9) | Te1—Br1—Br1$^{i}$ | 90.00 (5) |
| W1$^{vii}$—Te1— | 88.81 (10) | Te1—Br1—Br1$^{ii}$ | 90.00 (5) |
| W1$^{vii}$—Te1—Te2$^{iii}$ | 153.61 (14) | Te1$^{ix}$—Br1—Br1$^{i}$ | 90.00 (5) |
| W1$^{vii}$—Te1— | 46.22 (10) | Te1$^{ix}$—Br1—Br1$^{ii}$ | 90.00 (5) |
| W1$^{vii}$—Te1— | 102.81 (14) | Br1$^{i}$—Br1—Br1$^{ii}$ | 180 |

Symmetry codes: (i) *x*, *y*−1, *z*; (ii) *x*, *y*+1, *z*; (iii) −*x*, *y*−1/2, −*z*; (iv) −*x*, *y*+1/2, −*z*; (v) *x*, *y*, *z*+1; (vi) −*x*, *y*−1/2, −*z*−1; (vii) −*x*, *y*+1/2, −*z*−1; (viii) *x*, *y*, *z*−1; (ix) −*x*−1/2, −*y*+1/2, *z*.

# Crystallographic details for $WTe_2Br_{1.25}$

Table S 21: Fractional Atomic Coordinates (×10[4]) and Equivalent Isotropic Displacement Parameters (Å[2]×10[3]) for **$WTe_2Br_{1.25}$**. $U_{eq}$ is defined as 1/3 of the trace of the orthogonalised $U_{ij}$.

| Atom | x | y | z | $U_{eq}$ |
|---|---|---|---|---|
| W4 | 5000 | 3794(2) | 723(8) | 55(2) |
| W8 | 5000 | 3803(2) | 5768(8) | 57(2) |
| W6 | 1607(5) | 3809.8(15) | 5719(5) | 56(2) |
| W3 | 3417(5) | 3849.9(16) | 2475(5) | 58(2) |
| W2 | 1886(4) | 3782.2(15) | 703(5) | 57(2) |
| W5 | 0 | 3832(2) | 7490(7) | 55(2) |
| W7 | 3155(5) | 3850.8(15) | 7480(5) | 58(2) |
| W1 | 0 | 3858(2) | 2403(8) | 64(3) |
| Te1 | 0 | 3357(3) | 1159(9) | 53(3) |
| Te3 | 3411(6) | 3323(2) | 1265(6) | 52(2) |
| Te13 | 0 | 4055(3) | 9515(10) | 55(3) |
| Te12 | 5000 | 3538(3) | 3811(9) | 57(3) |
| Te15 | 3329(6) | 4157(2) | -525(7) | 55(2) |
| Te14 | 1712(6) | 3461(2) | 8708(7) | 61(3) |
| Te10 | 1651(6) | 3557(2) | 3727(7) | 60(3) |
| Te8 | 5000 | 4294(3) | 7073(10) | 55(3) |
| Te6 | 1594(7) | 4313.7(17) | 7032(7) | 52(2) |
| Te16 | 5000 | 3555(3) | 8679(10) | 60(3) |
| Te9 | 0 | 4186(3) | 4440(11) | 57(3) |
| Te11 | 3342(6) | 4177.7(19) | 4490(7) | 55(2) |
| Te5 | 0 | 3347(3) | 6167(10) | 59(3) |
| Te7 | 3283(7) | 3357(2) | 6187(7) | 58(2) |
| Te4 | 5000 | 4313(3) | 1945(12) | 63(3) |
| Te2 | 1726(7) | 4291(3) | 1916(8) | 71(3) |
| Br5 | 5000 | 2961(5) | 3756(16) | 63(5) |
| Br6 | 5000 | 2969(5) | 8534(18) | 64(5) |
| Br3 | 1794(15) | 2597(4) | 6973(16) | 78(4) |
| Br4 | 1674(12) | 2984(4) | 3571(14) | 71(4) |
| Br7 | 0 | 4636(6) | 9607(19) | 71(6) |
| Br8 | 3170(20) | 5000 | 10520(20) | 74(6) |
| Br1 | 5000 | 2249(6) | 3950(20) | 74(6) |
| Br9 | 3160(20) | 5000 | 8530(20) | 75(6) |
| Br10 | 0 | 5000 | 5910(60) | 190(40) |
| Br2 | 3175(12) | 2438(5) | 5615(18) | 86(5) |
| Br12 | 2920(40) | 5000 | 6020(80) | 290(50) |
| Br13 | 3020(60) | 5000 | 2760(100) | 350(80) |
| Br11 | 0 | 5000 | 3080(70) | 240(60) |

Table S22: Anisotropic Displacement Parameters (×$10^4$) for **$WTe_2Br_{1.25}$**. The anisotropic displacement factor exponent takes the form: $-2\pi^2[h^2a^{*2} \times U_{11} + ... + 2hka^* \times b^* \times U_{12}]$

| Atom | $U_{11}$ | $U_{22}$ | $U_{33}$ | $U_{23}$ | $U_{13}$ | $U_{12}$ |
|---|---|---|---|---|---|---|
| W4 | 31(3) | 80(5) | 55(5) | 7(4) | 0 | 0 |
| W8 | 37(4) | 79(5) | 53(5) | -3(4) | 0 | 0 |
| W6 | 34(3) | 77(4) | 55(4) | -5(3) | 1(3) | -1(2) |
| W3 | 32(3) | 91(4) | 51(3) | 0(3) | -1(3) | 0(2) |
| W2 | 31(3) | 81(4) | 60(4) | 5(3) | -1(3) | 0(2) |
| W5 | 35(4) | 76(5) | 55(5) | 0(4) | 0 | 0 |
| W7 | 42(3) | 80(4) | 52(4) | 1(3) | -2(3) | 0(2) |
| W1 | 69(6) | 70(5) | 55(6) | 1(4) | 0 | 0 |
| Te1 | 26(4) | 76(7) | 55(7) | -5(5) | 0 | 0 |
| Te3 | 31(3) | 81(5) | 45(4) | -2(3) | 1(3) | -1(3) |
| Te13 | 23(4) | 95(8) | 49(6) | 4(6) | 0 | 0 |
| Te12 | 24(4) | 109(9) | 37(6) | -5(6) | 0 | 0 |
| Te15 | 33(3) | 85(5) | 46(4) | 0(4) | 1(3) | 2(3) |
| Te14 | 28(3) | 104(6) | 49(5) | -4(5) | 2(3) | -1(3) |
| Te10 | 25(3) | 100(6) | 55(5) | -7(4) | 1(3) | -2(3) |
| Te8 | 35(5) | 73(7) | 57(7) | -3(5) | 0 | 0 |
| Te6 | 43(4) | 55(4) | 59(5) | -5(3) | -2(4) | -1(3) |
| Te16 | 33(5) | 93(8) | 56(7) | -1(6) | 0 | 0 |
| Te9 | 36(5) | 77(7) | 58(7) | 0(6) | 0 | 0 |
| Te11 | 32(3) | 76(5) | 57(4) | -2(4) | 0(3) | 0(3) |
| Te5 | 29(5) | 95(9) | 52(7) | -1(6) | 0 | 0 |
| Te7 | 31(3) | 84(5) | 58(5) | 4(4) | 1(3) | 0(3) |
| Te4 | 24(4) | 91(8) | 74(8) | 5(7) | 0 | 0 |
| Te2 | 20(3) | 126(8) | 66(6) | -2(5) | -4(3) | 3(4) |
| Br5 | 20(6) | 118(15) | 51(11) | 3(10) | 0 | 0 |
| Br6 | 60(10) | 65(11) | 67(13) | -8(9) | 0 | 0 |
| Br3 | 57(8) | 84(9) | 92(11) | -2(8) | 21(8) | 4(7) |
| Br4 | 41(6) | 87(9) | 86(11) | 8(8) | 1(7) | 1(6) |
| Br7 | 29(7) | 124(18) | 58(12) | -3(11) | 0 | 0 |
| Br8 | 64(11) | 65(11) | 94(16) | 0 | 10(12) | 0 |
| Br1 | 41(9) | 102(15) | 78(14) | 0(12) | 0 | 0 |
| Br9 | 54(10) | 92(14) | 79(15) | 0 | 13(10) | 0 |
| Br10 | 370(130) | 60(20) | 150(60) | 0 | 0 | 0 |
| Br2 | 36(6) | 118(13) | 104(13) | 27(11) | 9(8) | 2(6) |
| Br12 | 60(20) | 450(130) | 370(110) | 0 | 110(40) | 0 |
| Br13 | 190(50) | 64(18) | 800(200) | 0 | -300(90) | 0 |
| Br11 | 500(200) | 50(20) | 170(70) | 0 | 0 | 0 |

Table S23: Bond Lengths in Å for **$WTe_2Br_{1.25}$**.

| Atom | Atom | Length/Å |
|---|---|---|
| W4 | W3 | 2.789(9) |
| W4 | W3[1] | 2.789(9) |
| W4 | Te3 | 2.725(11) |
| W4 | Te3[1] | 2.725(11) |
| W4 | Te15 | 2.843(11) |
| W4 | Te15[1] | 2.843(11) |
| W4 | Te16[2] | 2.778(16) |
| W4 | Te4 | 2.716(17) |
| W8 | W7[1] | 2.925(9) |
| W8 | W7 | 2.925(9) |
| W8 | Te12 | 2.723(15) |
| W8 | Te8 | 2.680(15) |
| W8 | Te11 | 2.883(11) |
| W8 | Te11[1] | 2.883(11) |
| W8 | Te7 | 2.701(11) |
| W8 | Te7[1] | 2.701(11) |
| W6 | W5 | 2.815(9) |
| W6 | W7 | 2.772(9) |
| W6 | Te10 | 2.741(11) |
| W6 | Te6 | 2.730(10) |
| W6 | Te9 | 2.852(12) |
| W6 | Te11 | 2.883(11) |
| W6 | Te5 | 2.682(13) |
| W6 | Te7 | 2.706(10) |
| W3 | W2 | 2.781(8) |
| W3 | Te3 | 2.736(11) |
| W3 | Te12 | 2.735(12) |
| W3 | Te10 | 2.758(10) |
| W3 | Te11 | 2.910(11) |
| W3 | Te4 | 2.692(13) |
| W3 | Te2 | 2.706(12) |
| W2 | W1 | 2.955(9) |
| W2 | Te1 | 2.775(11) |

| Atom | Atom | Length/Å |
|---|---|---|
| W2 | Te3 | 2.653(10) |
| W2 | Te13[2] | 2.763(11) |
| W2 | Te15 | 2.713(10) |
| W2 | Te14[2] | 2.879(12) |
| W2 | Te2 | 2.676(13) |
| W5 | Te13 | 2.731(16) |
| W5 | Te14[3] | 2.866(11) |
| W5 | Te14 | 2.866(11) |
| W5 | Te6 | 2.738(10) |
| W5 | Te6[3] | 2.738(10) |
| W5 | Te5 | 2.672(16) |
| W7 | Te15[4] | 2.849(11) |
| W7 | Te14 | 2.750(11) |
| W7 | Te8 | 2.782(11) |
| W7 | Te6 | 2.653(10) |
| W7 | Te16 | 2.783(11) |
| W7 | Te7 | 2.685(12) |
| W1 | Te1 | 2.668(15) |
| W1 | Te10 | 2.748(12) |
| W1 | Te10[3] | 2.748(12) |
| W1 | Te9 | 2.933(17) |
| W1 | Te2[3] | 2.685(12) |
| W1 | Te2 | 2.685(12) |
| Te13 | Br7 | 2.50(3) |
| Te12 | Br5 | 2.48(3) |
| Te10 | Br4 | 2.473(19) |
| Te16 | Br6 | 2.53(2) |
| Br3 | Br2 | 2.36(2) |
| Br8 | Br9 | 2.51(3) |

——

[1]1-x,+y,+z; [2]+x,+y,-1+z; [3]-x,+y,+z; [4]+x,+y,1+z

Table S 24: Bond Angles in ° for **$WTe_2Br_{1.25}$**.

| Atom | Atom | Atom | Angle/° |
|---|---|---|---|
| W3[1] | W4 | W3 | 74.2(3) |
| W3[1] | W4 | Te15 | 140.2(4) |
| W3 | W4 | Te15[1] | 140.2(4) |
| W3[1] | W4 | Te15[1] | 90.9(2) |
| W3 | W4 | Te15 | 90.9(2) |
| Te3[1] | W4 | W3[1] | 59.5(2) |
| Te3 | W4 | W3 | 59.5(2) |
| Te3 | W4 | W3[1] | 103.8(4) |
| Te3[1] | W4 | W3 | 103.8(4) |
| Te3[1] | W4 | Te3 | 76.6(4) |
| Te3[1] | W4 | Te15 | 159.3(5) |
| Te3 | W4 | Te15 | 99.3(2) |
| Te3 | W4 | Te15[1] | 159.3(5) |
| Te3[1] | W4 | Te15[1] | 99.3(2) |
| Te3 | W4 | Te16[2] | 87.6(4) |
| Te3[1] | W4 | Te16[2] | 87.6(4) |
| Te15[1] | W4 | Te15 | 77.3(4) |
| Te16[2] | W4 | W3[1] | 140.3(2) |
| Te16[2] | W4 | W3 | 140.3(2) |

| Atom | Atom | Atom | Angle/° |
|---|---|---|---|
| Te16[2] | W4 | Te15 | 71.8(4) |
| Te16[2] | W4 | Te15[1] | 71.8(4) |
| Te4 | W4 | W3 | 58.5(3) |
| Te4 | W4 | W3[1] | 58.5(3) |
| Te4 | W4 | Te3[1] | 118.0(4) |
| Te4 | W4 | Te3 | 118.0(4) |
| Te4 | W4 | Te15 | 82.1(4) |
| Te4 | W4 | Te15[1] | 82.1(4) |
| Te4 | W4 | Te16[2] | 146.3(6) |
| W7 | W8 | W7[1] | 84.1(3) |
| Te12 | W8 | W7[1] | 134.5(2) |
| Te12 | W8 | W7 | 134.5(2) |
| Te12 | W8 | Te11 | 74.1(4) |
| Te12 | W8 | Te11[1] | 74.1(4) |
| Te8 | W8 | W7[1] | 59.3(3) |
| Te8 | W8 | W7 | 59.3(3) |
| Te8 | W8 | Te12 | 152.7(6) |
| Te8 | W8 | Te11[1] | 84.4(4) |
| Te8 | W8 | Te11 | 84.4(4) |

| Atom | Atom | Atom | Angle/° |
|---|---|---|---|
| Te8 | W8 | Te7 | 116.1(4) |
| Te8 | W8 | Te7$^{1}$ | 116.1(4) |
| Te11$^{1}$ | W8 | W7$^{1}$ | 88.0(2) |
| Te11 | W8 | W7 | 88.0(2) |
| Te11$^{1}$ | W8 | W7 | 141.6(4) |
| Te11 | W8 | W7$^{1}$ | 141.6(4) |
| Te11 | W8 | Te11$^{1}$ | 75.3(4) |
| Te7 | W8 | W7$^{1}$ | 111.0(4) |
| Te7 | W8 | W7 | 56.8(3) |
| Te7$^{1}$ | W8 | W7 | 111.0(4) |
| Te7$^{1}$ | W8 | W7$^{1}$ | 56.8(3) |
| Te7$^{1}$ | W8 | Te12 | 83.1(4) |
| Te7 | W8 | Te12 | 83.1(4) |
| Te7 | W8 | Te11 | 95.4(3) |
| Te7 | W8 | Te11$^{1}$ | 156.9(4) |
| Te7$^{1}$ | W8 | Te11$^{1}$ | 95.5(3) |
| Te7$^{1}$ | W8 | Te11 | 156.9(4) |
| Te7 | W8 | Te7$^{1}$ | 84.9(5) |
| W5 | W6 | Te9 | 93.9(3) |
| W5 | W6 | Te11 | 142.8(3) |
| W7 | W6 | W5 | 73.7(2) |
| W7 | W6 | Te9 | 140.7(4) |
| W7 | W6 | Te11 | 91.0(3) |
| Te10 | W6 | W5 | 138.9(3) |
| Te10 | W6 | W7 | 138.7(3) |
| Te10 | W6 | Te9 | 73.5(4) |
| Te10 | W6 | Te11 | 73.3(3) |
| Te6 | W6 | W5 | 59.2(3) |
| Te6 | W6 | W7 | 57.7(2) |
| Te6 | W6 | Te10 | 150.8(4) |
| Te6 | W6 | Te9 | 83.7(3) |
| Te6 | W6 | Te11 | 83.9(3) |
| Te9 | W6 | Te11 | 76.5(3) |
| Te5 | W6 | W5 | 58.1(3) |
| Te5 | W6 | W7 | 104.8(3) |
| Te5 | W6 | Te10 | 84.8(4) |
| Te5 | W6 | Te6 | 117.3(4) |
| Te5 | W6 | Te9 | 99.2(3) |
| Te5 | W6 | Te11 | 158.1(4) |
| Te5 | W6 | Te7 | 80.7(3) |
| Te7 | W6 | W5 | 104.5(3) |
| Te7 | W6 | W7 | 58.7(3) |
| Te7 | W6 | Te10 | 84.4(3) |
| Te7 | W6 | Te6 | 116.3(3) |
| Te7 | W6 | Te9 | 157.8(4) |
| Te7 | W6 | Te11 | 95.3(3) |
| W4 | W3 | Te11 | 138.8(3) |
| W2 | W3 | W4 | 72.8(2) |
| W2 | W3 | Te11 | 137.6(3) |
| Te3 | W3 | W4 | 59.1(3) |
| Te3 | W3 | W2 | 57.5(2) |
| Te3 | W3 | Te10 | 86.5(3) |
| Te3 | W3 | Te11 | 152.9(4) |
| Te12 | W3 | W4 | 94.4(3) |
| Te12 | W3 | W2 | 143.5(4) |
| Te12 | W3 | Te3 | 86.5(4) |
| Te12 | W3 | Te10 | 80.8(3) |
| Te12 | W3 | Te11 | 73.5(3) |
| Te10 | W3 | W4 | 145.6(4) |
| Te10 | W3 | W2 | 90.9(3) |
| Te10 | W3 | Te11 | 72.6(3) |
| Te4 | W3 | W4 | 59.4(4) |

| Atom | Atom | Atom | Angle/° |
|---|---|---|---|
| Te4 | W3 | W2 | 104.0(4) |
| Te4 | W3 | Te3 | 118.5(4) |
| Te4 | W3 | Te12 | 97.6(4) |
| Te4 | W3 | Te10 | 154.9(5) |
| Te4 | W3 | Te11 | 82.9(4) |
| Te4 | W3 | Te2 | 80.3(3) |
| Te2 | W3 | W4 | 104.7(3) |
| Te2 | W3 | W2 | 58.4(3) |
| Te2 | W3 | Te3 | 115.7(4) |
| Te2 | W3 | Te12 | 156.0(4) |
| Te2 | W3 | Te10 | 91.1(3) |
| Te2 | W3 | Te11 | 82.5(3) |
| W3 | W2 | W1 | 78.5(2) |
| W3 | W2 | Te14$^{2}$ | 142.4(3) |
| Te1 | W2 | W3 | 108.9(3) |
| Te1 | W2 | W1 | 55.4(3) |
| Te1 | W2 | Te14$^{2}$ | 79.6(3) |
| Te3 | W2 | W3 | 60.4(3) |
| Te3 | W2 | W1 | 107.5(3) |
| Te3 | W2 | Te1 | 83.9(3) |
| Te3 | W2 | Te13$^{2}$ | 154.7(4) |
| Te3 | W2 | Te15 | 104.6(3) |
| Te3 | W2 | Te14$^{2}$ | 85.2(3) |
| Te3 | W2 | Te2 | 119.7(4) |
| Te13$^{2}$ | W2 | W3 | 144.7(4) |
| Te13$^{2}$ | W2 | W1 | 81.7(3) |
| Te13$^{2}$ | W2 | Te1 | 82.5(3) |
| Te13$^{2}$ | W2 | Te14$^{2}$ | 71.5(4) |
| Te15 | W2 | W3 | 93.9(3) |
| Te15 | W2 | W1 | 137.1(4) |
| Te15 | W2 | Te1 | 156.8(4) |
| Te15 | W2 | Te13$^{2}$ | 81.2(3) |
| Te15 | W2 | Te14$^{2}$ | 79.7(3) |
| Te14$^{2}$ | W2 | W1 | 130.2(3) |
| Te2 | W2 | W3 | 59.4(3) |
| Te2 | W2 | W1 | 56.7(3) |
| Te2 | W2 | Te1 | 112.0(4) |
| Te2 | W2 | Te13$^{2}$ | 85.3(4) |
| Te2 | W2 | Te15 | 82.9(3) |
| Te2 | W2 | Te14$^{2}$ | 152.7(4) |
| W6$^{3}$ | W5 | W6 | 74.6(3) |
| W6 | W5 | Te14$^{3}$ | 142.4(4) |
| W6$^{3}$ | W5 | Te14$^{3}$ | 91.3(2) |
| W6 | W5 | Te14 | 91.3(2) |
| W6$^{3}$ | W5 | Te14 | 142.4(4) |
| Te13 | W5 | W6$^{3}$ | 139.1(2) |
| Te13 | W5 | W6 | 139.1(2) |
| Te13 | W5 | Te14 | 72.1(4) |
| Te13 | W5 | Te14$^{3}$ | 72.1(4) |
| Te13 | W5 | Te6 | 86.1(4) |
| Te13 | W5 | Te6$^{3}$ | 86.1(4) |
| Te14$^{3}$ | W5 | Te14 | 78.8(4) |
| Te6$^{3}$ | W5 | W6 | 103.4(4) |
| Te6 | W5 | W6 | 58.9(3) |
| Te6$^{3}$ | W5 | W6$^{3}$ | 58.9(3) |
| Te6 | W5 | W6$^{3}$ | 103.4(4) |
| Te6$^{3}$ | W5 | Te14$^{3}$ | 98.2(3) |
| Te6 | W5 | Te14$^{3}$ | 158.0(5) |
| Te6$^{3}$ | W5 | Te14 | 158.0(5) |
| Te6 | W5 | Te14 | 98.2(3) |
| Te6 | W5 | Te6$^{3}$ | 76.4(4) |
| Te5 | W5 | W6 | 58.4(3) |

| Atom | Atom | Atom | Angle/° |
|---|---|---|---|
| Te5 | W5 | W6$^{3}$ | 58.4(3) |
| Te5 | W5 | Te13 | 149.2(6) |
| Te5 | W5 | Te14$^{3}$ | 84.3(4) |
| Te5 | W5 | Te14 | 84.3(4) |
| Te5 | W5 | Te6$^{3}$ | 117.3(4) |
| Te5 | W5 | Te6 | 117.3(4) |
| W6 | W7 | W8 | 78.5(2) |
| W6 | W7 | Te15$^{4}$ | 141.0(3) |
| W6 | W7 | Te8 | 108.2(3) |
| W6 | W7 | Te16 | 145.6(4) |
| Te15$^{4}$ | W7 | W8 | 130.0(3) |
| Te14 | W7 | W8 | 138.3(4) |
| Te14 | W7 | W6 | 94.7(3) |
| Te14 | W7 | Te15$^{4}$ | 79.6(3) |
| Te14 | W7 | Te8 | 156.2(4) |
| Te14 | W7 | Te16 | 78.9(3) |
| Te8 | W7 | W8 | 55.9(3) |
| Te8 | W7 | Te15$^{4}$ | 78.5(3) |
| Te8 | W7 | Te16 | 85.3(3) |
| Te6 | W7 | W8 | 108.3(3) |
| Te6 | W7 | W6 | 60.4(2) |
| Te6 | W7 | Te15$^{4}$ | 83.2(3) |
| Te6 | W7 | Te14 | 103.3(3) |
| Te6 | W7 | Te8 | 83.5(3) |
| Te6 | W7 | Te16 | 154.0(4) |
| Te6 | W7 | Te7 | 119.8(3) |
| Te16 | W7 | W8 | 84.2(3) |
| Te16 | W7 | Te15$^{4}$ | 71.6(3) |
| Te7 | W7 | W8 | 57.4(3) |
| Te7 | W7 | W6 | 59.4(2) |
| Te7 | W7 | Te15$^{4}$ | 154.3(4) |
| Te7 | W7 | Te14 | 83.6(3) |
| Te7 | W7 | Te8 | 113.3(4) |
| Te7 | W7 | Te16 | 86.2(4) |
| W2$^{3}$ | W1 | W2 | 85.4(3) |
| Te1 | W1 | W2$^{3}$ | 58.9(3) |
| Te1 | W1 | W2 | 58.9(3) |
| Te1 | W1 | Te10 | 88.7(4) |
| Te1 | W1 | Te10$^{3}$ | 88.7(4) |
| Te1 | W1 | Te9 | 154.8(6) |
| Te1 | W1 | Te2$^{3}$ | 115.2(4) |
| Te1 | W1 | Te2 | 115.2(4) |
| Te10 | W1 | W2$^{3}$ | 145.4(4) |
| Te10 | W1 | W2 | 87.6(2) |
| Te10$^{3}$ | W1 | W2 | 145.4(4) |
| Te10$^{3}$ | W1 | W2$^{3}$ | 87.6(2) |
| Te10$^{3}$ | W1 | Te10 | 79.3(4) |
| Te10$^{3}$ | W1 | Te9 | 72.1(3) |
| Te10 | W1 | Te9 | 72.1(3) |
| Te9 | W1 | W2 | 133.6(2) |
| Te9 | W1 | W2$^{3}$ | 133.6(2) |
| Te2$^{3}$ | W1 | W2$^{3}$ | 56.4(3) |
| Te2 | W1 | W2$^{3}$ | 111.9(4) |
| Te2 | W1 | W2 | 56.4(3) |
| Te2$^{3}$ | W1 | W2 | 111.9(4) |
| Te2$^{3}$ | W1 | Te10$^{3}$ | 91.8(3) |
| Te2 | W1 | Te10$^{3}$ | 154.5(5) |
| Te2 | W1 | Te10 | 91.8(3) |
| Te2$^{3}$ | W1 | Te10 | 154.5(5) |
| Te2 | W1 | Te9 | 82.4(4) |
| Te2$^{3}$ | W1 | Te9 | 82.4(4) |
| Te2 | W1 | Te2$^{3}$ | 86.1(5) |

| Atom | Atom | Atom | Angle/° |
|---|---|---|---|
| W2 | Te1 | W2$^{3}$ | 92.4(4) |
| W1 | Te1 | W2 | 65.7(3) |
| W1 | Te1 | W2$^{3}$ | 65.7(3) |
| W4 | Te3 | W3 | 61.4(3) |
| W2 | Te3 | W4 | 75.9(3) |
| W2 | Te3 | W3 | 62.1(3) |
| W2$^{4}$ | Te13 | W2$^{5}$ | 92.9(4) |
| W5 | Te13 | W2$^{5}$ | 111.1(4) |
| W5 | Te13 | W2$^{4}$ | 111.1(4) |
| Br7 | Te13 | W2$^{5}$ | 113.5(5) |
| Br7 | Te13 | W2$^{4}$ | 113.5(5) |
| Br7 | Te13 | W5 | 113.2(7) |
| W8 | Te12 | W3 | 110.7(5) |
| W8 | Te12 | W3$^{1}$ | 110.7(5) |
| W3 | Te12 | W3$^{1}$ | 75.9(4) |
| Br5 | Te12 | W8 | 116.4(7) |
| Br5 | Te12 | W3$^{1}$ | 118.3(5) |
| Br5 | Te12 | W3 | 118.3(5) |
| W4 | Te15 | W7$^{2}$ | 106.0(4) |
| W2 | Te15 | W4 | 73.0(3) |
| W2 | Te15 | W7$^{2}$ | 101.1(3) |
| W5 | Te14 | W2$^{4}$ | 104.1(4) |
| W7 | Te14 | W2$^{4}$ | 99.5(3) |
| W7 | Te14 | W5 | 73.3(3) |
| W6 | Te10 | W3 | 110.9(3) |
| W6 | Te10 | W1 | 111.1(4) |
| W1 | Te10 | W3 | 82.5(3) |
| Br4 | Te10 | W6 | 118.0(6) |
| Br4 | Te10 | W3 | 113.8(5) |
| Br4 | Te10 | W1 | 115.3(5) |
| W8 | Te8 | W7$^{1}$ | 64.7(3) |
| W8 | Te8 | W7 | 64.7(3) |
| W7$^{1}$ | Te8 | W7 | 89.5(4) |
| W6 | Te6 | W5 | 62.0(3) |
| W7 | Te6 | W6 | 62.0(2) |
| W7 | Te6 | W5 | 76.9(3) |
| W4$^{4}$ | Te16 | W7$^{1}$ | 109.7(4) |
| W4$^{4}$ | Te16 | W7 | 109.7(4) |
| W7 | Te16 | W7$^{1}$ | 89.5(5) |
| Br6 | Te16 | W4$^{4}$ | 115.8(7) |
| Br6 | Te16 | W7 | 114.6(5) |
| Br6 | Te16 | W7$^{1}$ | 114.6(5) |
| W6 | Te9 | W6$^{3}$ | 73.5(4) |
| W6 | Te9 | W1 | 102.9(4) |
| W6$^{3}$ | Te9 | W1 | 102.9(4) |
| W8 | Te11 | W3 | 101.6(3) |
| W6 | Te11 | W8 | 77.4(3) |
| W6 | Te11 | W3 | 102.8(3) |
| W6$^{3}$ | Te5 | W6 | 79.0(5) |
| W5 | Te5 | W6$^{3}$ | 63.4(4) |
| W5 | Te5 | W6 | 63.4(4) |
| W8 | Te7 | W6 | 83.6(3) |
| W7 | Te7 | W8 | 65.8(3) |
| W7 | Te7 | W6 | 61.9(3) |
| W3 | Te4 | W4 | 62.1(3) |
| W3$^{1}$ | Te4 | W4 | 62.1(3) |
| W3$^{1}$ | Te4 | W3 | 77.3(5) |
| W2 | Te2 | W3 | 62.2(3) |
| W2 | Te2 | W1 | 66.9(3) |
| W1 | Te2 | W3 | 84.7(4) |

[1]1-x,+y,+z; [2]+x,+y,-1+z; [3]-x,+y,+z; [4]+x,+y,1+z; [5]-x,+y,1+z